**Hidden stochastic, quantum and dynamic information of markov diffusion process and its evaluation by an entropy integral measure under the impulse control's actions, applied to information observer**

**Vladimir S. Lerner, USA**

## Abstract

We study emergence of quantum, dynamic information, and arising an information observer under impulse interactive actions on Markov diffusion process modeling an interactive random environment.

The impulse discrete yes-no action cuts the Markov process correlations revealing a Bit of hidden information, connected the process correlated states covered by entropy of cutting correlation.

*Information appears as phenomenon of interaction cutting correlations carrying entropy.*

Each inter-action models Kronicker impulse while delta-impulse models interaction between the Kronicker's impulses along the process. Each impulse step-down action cuts a maximum of impulse' minimal entropy, and the impulse' step-up action transits the cutting minimal entropy to each step-up action of the delta function which overlaps merging with it. The delta function' step-down action kills the delivering entropy producing equivalent minimax information. The merging action initiates quantum microprocess. The multiple cutting entropy of the process is sequentially converting to information micro-macroprocess.

The cutting impulse entropy integrates entropy functional (EF) along trajectories of the multi-dimensional diffusion process. The information which delivers the ending states of each impulse integrates information path functional (IPF) along the process trajectories.

The hidden information evaluates Feller kernel whose minimal path transforms Markov transition probability to probability of Brownian diffusion at each cutoff. Each transitive transformation virtually observes the origin of hidden information of the probabilities correlated states (events). The IPF integrates the observing Bits along the minimal path of interacting process assembling the information Observer.

The multiple impulses' minimax entropy-information imposes variation principle-a law on both the EF and IPF, whose extreme equations describes optimal observing micro and mcroprocess.

The observing microprocess appears similar to the considered quantum microproces, while the information macroprocess describes classical irreversible thermodynamics.

The hidden information, in addition to Bit, curries free information frozen from the correlated connections. The free information of the multiple hidden information contributions binds the observing micro-macro processes in information macrodynamics (IMD). Each IMD three information units' free information composes doublets-triplets information structures. The free information of each three structural triplets assembles their sequence in information network (IN). The triple INs' free information cooperate structure of the information Observer.

**Contents**

## Introduction

Uncovering and evaluation quantum and dynamic information hidden in correlation of stochastic process under an interactive impulse presents new and essential topic in understanding the process of interaction an observer with its random environment.

Conventional information science generally considers an information process, but traditionally uses the probability measure for the random events and Shannon's entropy measure as uncertainty function of the states [1, 2, 3].

Revealing hidden information, covering a *process*' *inner connections* between its states holding hidden statistical and quantum dependencies *along the process*, is important problem in information theory [4, 5].

The process' interactive impulses, cutting the states correlation, carry conditional entropy of the correlated events and the time intervals of random process presenting a natural source of hidden information.

In a Markov diffusion process, considered a formal and traditional model of a random non-stationary interactive process, such hidden information absorbs a Feller's kernel [6, 7], whose a minimal Markov path is formed at transformation of the Markov transition probability to the probability of Brownian diffusion.

The kernel hidden information measures an invariant information unit comparable with hidden information in other processes.

To find the hidden information we use method of cutting the process's inter-states correlations, by an impulse control modeling the impulse interactions.

The cutoff entropy along the process trajectories integrates Entropy Functional (EF) [8, 9].

The impulse control's jump action on Markov process "killing its drift" provides transformation to minimal Markov path, selecting the Feller kernel's measure [10-14] holding the hidden information.

Multiple Feller kernels' minimal paths at these transformations along the impulse cutoff integrate Information Path Functional (IPF).

Other source of a hidden information is Schrödinger's bridge of quantum process [15], which originates from a Brownian path for reversible probability densities, represented through the quantum conjugated wave functions [16, 17,18,19,20,21].

The common Brownian path for both Feller kernel and Schrödinger's bridge in the Markov diffusion process [22, 23, 24, 25] opens a possibility of their joint information evaluation during the cutting Markovian fractions.

The impulse controls, applied along a multi-dimensional Markov diffusion process, provide the cut-off transformations which *concurrently* produce both Feller's kernel and initiate Schrödinger's bridge with quantum entanglement, generating quantum information dynamics within the cutting impulse.

At the above transformations arise an additive functional on the Markov process trajectories [26, 27, 28, 29, 14], which entropy functional EF averages [8].

The probability density, being common for the Brownian path in both Feller's kernel and Schrödinger's bridge, defines the related multiplicative functional at this transformation [29, 30].

Since both additive and multiplicative functionals are defined along the trajectories of the Markov process as the solutions of Ito's controllable stochastic deferential equation, the impulse controls, applied to this equation, can model the cut-off transformation for these functionals. That allows measuring the EF and IPF under the impulse controls actions concurrently with the cutoff.

Instead of interacting particles, we study probabilities of their interactions creating an information microprocess, emerging from the process hidden uncertainties within each cut-off.

The interactive impulse, transforming each cutting minimum of a priori probability to maximum of a posteriori probability, delivers minimax law for extracting information hidden in the cutting correlation. That law formalizes the minimax variation principle for the entropy functional, whose variation equations determine the structure of the information dynamics arising at this transformation.

Integration the cutoff information units generates information macrodynamic process, which the IPF measures.

Since cutting random process is time-space distributed, the quantum entanglement, taking place at an inner locality of the distributed interacting impulses, can be non-local, which we estimate by a minimum of a maximal difference in the time between the non-local entanglements along the distributions, or a minimax distance between them.

This paper results relate to some other publications in this field [31- 34] but also distinguish by the specifics, directed on revealing hidden information, creating an information observer during the impulse interactive observations.

Cutting correlation cuts impulse entropy and time, which creates information, "surprising" an observer by revealing its new connections with previous one being accumulated and memorized.

The paper is organized *in* seven sections in part **I** and seven sections in part **II**.

Formulas in each section start with section number (1.1), (2.1), independently of the part number, while reference to these formulas from other parts and sections adds the part number:(1.1.1),(2. 1.3), (3.3.2), etc.

**Sec.1** introduces an entropy functional (EF) on trajectories of the controllable Markov diffusion process, whose functions of drift and diffusion of Ito's stochastic Eq. determine the process' additive functional, allowing also measuring the EF time intervals.

The EF is defined at transforming the Markov process (with non-zeros drift and diffusion) to Brownian movement (with zero drift) under the process' cutoff controls.

**Sec. 2** applies the impulse control, cutting the process' time-length fraction, whose amount measures the additive functional contribution during the cutoff. The control dissolves the correlations between the process' cutting states and transfers the Markov process to Brownian movement, while operator of this transformation produces a Feller's kernel and provides the entropy's functional measure during the impulse' cutoff intervals.

Specifically, the impulse step-down control action transfers Markov process to Brownian movement, and its step-up control action transfers the Brownian movement to a renewed Markov process. The step-down control extracts *maximum* of the process entropy during the minimal cutting time. The step-up control transfers the *minimum* of this *maximum* to a next impulse.

The impulse controls, sequential transforming the maximum to minimum and then from minimum to maximum along the controllable Markov process, set up the maxmin-minimax principle. The minimax is a dual complimentary variation principle, defining variation equations of information dynamics. The EF fraction, generated during the cutoff, estimates an entropy path, hidden between the process states, being connected by their correlations. The IPF measure integrates information of multiple cutting units.

**Sec.3** considers the evolution equations for both Markov's transition probabilities and the entropy functional at above transformations. This leads to equations for information quantum complex conjugated wave functions with probability density, commonly shared by the Markov diffusion and quantum-conjugated wave functions, which evaluate the equivalent entropy functional measure. Applying this probability for a class of reciprocal Markovian diffusion, we find the condition of forming a *Schrödinger's process with its bridge* for a reversible probability density, and evaluate *quantity of information* of the Schrödinger's bridge.

We also find the conditions of forming a *Schrödinger's bridge* for more general case, when Markov diffusion is not reciprocal process. In both cases, entropy for *Schrödinger's bridge* exceeds the entropy of potential cutoff by the impulse. This opens a possibility of forming the cutting unstable bridge, while the non-cutting bridge could only be stable.

**Sec.4** evaluates the *quantum information of* Schrödinger's bridge at an *entanglement* of the wave functions. Analyzing both local and non-local entanglements, we evaluate their information distances and the conditions of disentanglement.

Condition for unstable entanglement leads to possibility of self-disentanglement (self-destruction) during the interactive killing actions.

This effect, published in [35, 36, 37, and 38] is called Entanglement Sudden Death (ESD) during finite time interval for both local and the Bell's non-local entanglements.

We estimate the entropy in the ESD evolution, when it's starting probability decreases down to the moment near the moment of entanglement death, approximating this evolution time intervals.

**Sec.5** presents solutions of minimax variation problem (VP) for the EF through the extremals and VP's constraint, imposed on the solutions, which minimizes the functional Hamiltonian. These results lead to the extreme solutions for the entropy's function of action of wave function and for the probability density–both for the minimal entropy functional.

Considering the diffusion process at a locality of the correlating boundary states, formed by the impulse controls cutoff actions, we express the VP constraint through an operator of the transition equations for the transformations of the cutting off Markov process.

Applying the operator proves that the impulse controls cutoff action implements the VP at the locality of these states in the form of *minimax* and *maxmin*, depending on the impulse's step-down and step-up actions accordingly.

The solutions of the operator equations also allow classifying the boundary states at the locality on the attracting and repealing random states. The attractive dynamics boundaries carry *hidden dynamic* connections between the process' cutting states boundaries.

**Sec.6** uses the operator forms of the evolution equations to express both *Schrödinger's Equation for the wave functions* and the *entropy's functions of actions* through the *Hamiltonian* of the VP, defined by the minimum of the entropy's functional. Applying the VP solutions, we identify a maximal frequency of the information wave, whose energy spectrum is limited by Plank constant. Bringing an *information equivalent of Plank constant* to Hamiltonian form of the information Schrödinger's Equations on the extremals, allows specifying these equations for information quantum microprocess. We proved that these equations, following from imposing the VP constraint, lead to *conditions of an entanglement* of the information wave functions at each moment of the constraint imposing. The conditions of *minimizing the entropy functional* on extremals, imposing the constraint, lead to the entanglement at applying the impulse controls cutoff. By applying these conditions to the probability density and probabilities of Markov diffusion we get a minimal path for the probabilities along the minimal EF in Markov diffusion process. The minimal path forms a *Schrödinger's* process holding a mixture of Brownian's bridges. Concurrently with the Schrödinger Brownian's bridges, the same cutoff action on the additive functional of Markov diffusion, killing its drift, *selects* the Feller's information measure of the kernel. Whereas the IPF *measures the Feller kernel information cutting from entire current Markov movement.*

**Sec.7** analyzes jointly the cutting Markov diffusion and the quantum information microprocess, having the equivalent probability's densities (measured by the EF on the trajectories of both processes) and starting simultaneously under the same control cut-off time interval. This allows evaluate together the quantity of information for both Feller's kernel and Schrödinger's bridge simultaneously during both processes' current time and finding the mutual relations of these information quantities. The forming Feller's kernel of the Markov process and Schrödinger's bridge of the *quantum process* have the same life-time. A maximum of this time interval, determined by the maximal bridge's path, estimates a *maximal difference* in the time between the non-local entanglements. A minimum of this time interval estimates a unit of instance where elementary hidden information might be generated.

The information interactive dynamics, applying to information observer [9], include both stochastic and quantum dynamics, producing information for the related kernel, bridge, and entanglement concurrently.

The analysis shows that these phenomena occur at a close locality of the cutting edges, creating a "window" where observer consecutively gets the information unit, which emerges at each natural interactive cutoff killing its entropy and the time instance. Both amount of the instance and the unit information estimate the observer's quantity of receiving hidden information in part **II**.

**Sec. 1** reviews the published physical and information results related to the Observer and provides the formal principles that explain how an interactive Observer self-creates Information.

This is accomplished by unifying different interactions, independent of origin, and focusing on observation emanating from the probabilities of a random field. The field triad initiates an interactive random process whose observation is the source of a potential Observer. That is, the Observer acquires a natural emergent feature of Information processes. The random process models a Markov diffusion process emanating from the field. Diffusion encloses a set of events that hold correlations.

Observation of the Markov process begins a sequence of Kolmogorov's 0-1 law probabilities that the field generates. The sequence of the probabilistic 0-1 (No-Yes) impulses, acting on the Markov diffusion process, initiates a sequence of discrete Bayes probabilities.

These probabilistic impulses virtually observe the Markov diffusion by cutting the diffusion correlation. Entropy-uncertainty hidden in the cutting correlation is released.

Sequential cutting the entropy of correlation decreases uncertainty of the Markov diffusion and increases the probabilities of the observing process. Each impulse $\downarrow\uparrow$ action $\downarrow$ cutting the maximal probability opens a path to certainty, while the following interaction $\uparrow$ cutting maximal entropy carries the equivalent

Information unit. Such interactive impulses↓↑enclose the certainty of real observation, bringing the energy field to an actual interacting cut. This *converts* the maximal entropy to the Information units. Natural (real) interactions convert this entropy to Information. Information becomes a phenomenon of interactions. The Bayes sequence of probabilistic 0-1-0-1… interactive actions conveys the probabilistic logic of the maximin-minimax principle, which at conversion to certainty conveys the related Information logic. The Information unit encloses an invariant minimax impulse 1Nat which includes the Bit and free attracting Information. The attraction emerges from the cutting correlation connecting observing process impulses. The free Information of multiple interacting Bits self-organizes the Information process, creating the Information Observer.

**Sec 2** determines the impulse-cutting functions preserving the Markov process's additive and multiplicative properties. That requirement limits the admissible class controls by two real and two complex functions. Applying these control functions identifies both the cutting invariant impulse and the transitional impulse action within. The VP extreme, imposed on the observing process, proves that each three invariant cutting impulses during their time intervals enable the generation of a single invariant Information impulse. The information impulses' time interval condenses the triple of the previous impulses intervals and entropies. It also proves that action starting the Information impulse captures the Markov multiplicative entropy increments. The emerging Bit includes three parts:

First delivers a multiplicative action by capturing entropy of random process;

Second delivers the impulse step-down cut of the process entropy;

Third delivers the impulse step-up control bringing Information which is transferring to nearest impulse. That keeps the Information connection between the impulses and provides persistent continuation of the impulse sequence during the observing process. All three parts hold invariant Information measures within the impulse measure. Since each cutting impulse preserves its invariant information measure, each third of the sequential cutting impulses triples its information density. It implies the dependence on process time course both the IPF impulse components and final IPF impulse, limiting each impulse time interval. The final time interval evaluates the density of this impulse Information, which also limits the total IPF information. Finally, a finite maximal information density limits the IPF finite minimal physical time interval in an accessible time course. Each finite three time intervals within an invariant impulse allow finding the discrete correlation function in a following cutting increment. The results verify the estimated entropy contributions in all parts of the impulse and the following Information increments.

**Sec.3** analyses a merge of neighbor impulses, generating interactive jump on each impulse border. The interactive jump might be at locality of Schrödinger's bridge 'edge in cutting a sub-Markov process. The merge converges, causing action with following reaction, superimposing the cause and effect. It could cover unpredictable events within the merge. Because the merge squeezes the inter-action interval to a micro-minimum, the interaction of bordered impulse initiated process called a microprocess within the impulse emerging under the impulse' interacting jump interval.

Mathematically the jump increases the Markov drift (speed) up to infinity.

The jump brings the extreme discrete curved displacement rotating observing asymmetrical entropy increments starting the microprocess. The interacting equal asymmetric entropies increments run the microprocess entropy to entanglement with equal multiplicative probabilities.

The jump transforms the impulse time interval to space intervals which preserve information measure within the cutting impulse. The *entanglement starts before its space is formed and ends with beginning the space during reversible relative time interval of* $0.015625\pi$ part of the impulse invariant measure $\pi$. When the space interval is forming, the initial *multiplicative* movement changes to the *additive* movement. *The microprocess emerges from multiple interactions starting with probabilities* $p_{+a} = 0.3679, p_{-a} = 0.3679$, *inverse entropy* $S^*_{\mp a} = 2$. *With growing probabilities, the observing maximal entropy of an impulse converts to the equal Information Bit.*

**Sec 4** identifies the impulse invariant measure holding metric π which preserves the impulse curvatures. The curvature of rotating impulse encloses its time, space and probabilistic logic collected in observations. With growing Bayes probability, the logic approaches certainty while the cutting impulse of *the real time microprocess is building its Bit* advancing to a gap of reality. When the impulse captures energy, the logical Bit becomes Information on an edge of the reality gap, and the following interacting impulse encodes the Information Bit. The encoding *memorizes* the Bit which becomes *irreversible*.

**Sec.5** studies the emergence of transitional impulse inside of curved impulse. A rotating conjugate movement starts discrete time-space micro-intervals within a transitional impulse at rotating angle $\pm\pi/4$. That implies symmetrical mirror copies of the observation holding within the transitional impulse. Formation of a qubit and Bit starts at the entanglement which confines an entropy volume of the pair superposition in the transitional impulse at angle of rotation $\pi/2$. The microprocess within entire impulse is reversible until the impulse ending action cuts its entropy. The microprocess's transitive gap separates the entropy and appearance of the equivalent information.

**Sec.6.** analyses interacting curvatures of step-up and step-down actions. Each impulse step-down action has negative curvature corresponding attraction, the step-up reaction has positive curvature corresponding repulsion, the middle part of the impulse having negative curvature transfers the attraction between these parts. The impuses' interacting curvatures enable mutial attraction at condition Sec. 1.5.

The opposite curved interaction lovers the capturing potential energy, compared to other interactions for generating a Bit. The energy of the curving impulse dynamically and naturally encodes Bits in Information process. *The rotating thermodynamic process with minimal Landauer energy performs natural memorizing of each natural Bit.*

The integrated Bit, condensed in the IPF, increases the Bit Information density in finite impulse size, which conserves growing energy of equivalent interacting physical particles-objects.

Applying the Jarzynski Equality (JE) for measuring energy of the entropy-Information unit in both the statistical thermodynamic microprocess and the natural encoding thermodynamic macroprocess shows that the impulse minimax extreme principle (EP) satisfies the JE .

**Sec.7 i**dentifies invariant energy measure which each Bit encloses, starting Maxwell Demon.

We prove that the Bit creating from the impulse interactive observation reveals the structure of the Wheeler Bit, which participates in creation of the impulse Yes-No logic, encoding and memorizing the Bit. Such Bit-Participator is the primary Information Observer formed without any *a priori* physical law.

Extracting Hidden Information from the observing process, we found how Information Observer emerges with a certain logical Information Bit, and how the impulse interactive process encodes Bits, while the self-participating Bits generate the Information Observer.

That identifies the Information Observer as an extractor and holder of this Information.

Multiple units of the IPF functional measure connect and encode all fractions of the observed information process, performing *computation* of information.

**PART I. HIDDEN STOCHASTIC AND QUANTUM INFORMATION**

**1. Entropy functional on trajectories of Markov diffusion process**

Let have the $n$-dimensional controlled stochastic Ito differential equation:

$$d\tilde{x}_t = a(t,\tilde{x}_t,u_t)dt + \sigma(t,\tilde{x}_t)d\xi_t, \tilde{x}_s = \eta, t \in [s,T] = \Delta, s \in [0,T] \subset R^1_+, \qquad (1.1)$$

with the standard limitations [29,30] on drift function $a(t,\tilde{x}_t,u_t) = a^u(t,\tilde{x}_t)$, depending on a control $u_t$, diffusion $\sigma(t,\tilde{x}_t)$, and Wiener process $\xi_t = \xi(t,\omega)$, which are defined on a probability space of the elementary random events $\omega \in \Omega$ with the variables located in $R^n$; $\tilde{x}_t = \tilde{x}(t)$ is a diffusion process, as a solution of (1.1) under control $u_t$; $\Psi(s,t)$ is a $\sigma$-algebra created by the events $\{\tilde{x}(\tau) \in B\}$, and $P(s,\tilde{x},t,B)$ are transition probabilities on $s \le \tau \le t$; $P_{s,x} = P_{s,x}(A)$ are the corresponding conditional probability's distributions on an extended $\Psi(s,\infty)$; $E_{s,x}[\bullet]$ are the related mathematical expectations.

Suppose control function $u_t$ provides transformation of an initial process $\tilde{x}_t$, with transition probabilities $P(s,\tilde{x},t,B)$, to other diffusion process

$$\varsigma_t = \int_s^t \sigma(v,\zeta_v)d\zeta_v, \qquad (1.1a)$$

with transition probabilities

$$\tilde{P}(s,\varsigma_t,t,B) = \int_{\tilde{x}(t)\in B} \exp\{-\varphi_s^t(\omega)\}P_{s,x}(d\omega), \qquad (1.2)$$

where $\varphi_s^t = \varphi_s^t(\omega)$ is an additive functional of process $\tilde{x}_t = \tilde{x}(t)$ [26,29,30], measured regarding $\Psi(s,t)$ with probability 1, and $\varphi_s^t = \varphi_s^\tau + \varphi_\tau^t$, $E_{s,x}[\exp(-\varphi_s^t(\omega))] < \infty$.

The process $\varsigma_t$ is a transformed version of process $\tilde{x}_t$ whose transition satisfies (1.2).

At this transformation, the transitional probability's functions $\tilde{P}(s,\varsigma_t,t,B)$ (1.2) determine the corresponding extensive distributions $\tilde{P}_{s,x} = \tilde{P}_{s,x}(A)$ on $\Psi(s,\infty)$ with a density measure

$$p(\omega) = \frac{\tilde{P}_{s,x}}{P_{s,x}} = \exp\{-\varphi_s^t(\omega)\}. \qquad (1.3)$$

which according to Girsanov Theorem [29] satisfies the form [30]:

$$\varphi_s^T = 1/2\int_s^T a^u(t,\tilde{x}_t)^T (2b(t,\tilde{x}_t))^{-1} a^u(t,\tilde{x}_t)dt + \int_s^T (\sigma(t,\tilde{x}_t)^{-1} a^u(t,\tilde{x}_t)d\xi(t), \qquad (1.4)$$

$$2b(t,\tilde{x}) = \sigma(t,\tilde{x})\sigma^T(t,\tilde{x}) > 0 \qquad (1.4a)$$

for the considered controllable process with its upper limit $T$.

Using the definition of conditional entropy [7] of process $\tilde{x}_t$ regarding process $\varsigma_t$, we have

$$S(\tilde{x}_t / \varsigma_t) = E_{s,x}\{-\ln[p(\omega)]\}, \qquad (1.5)$$

where $E_{s,x}$ is a conditional mathematical expectation, taken along the process trajectories $\tilde{x}_t$ at a given $\tilde{x}_s$ (by an analogy with [39]) which hold transformations (1.3). From (1.3,1.4,1.5) we get

$$S(\tilde{x}_t / \varsigma_t) = E_{s,x}[\varphi_s^t(\omega)], \qquad (1.6)$$

where the additive functional, at its fixed upper limit $T$, has the form (1.4).

Since the transformed process $\varsigma_t$ (1.1a) has the same diffusion matrix but zero drift, we have

$$E_{s,x}[\int_s^T (\sigma(t,\tilde{x}_t)^{-1} a^u(t,\tilde{x}_t)d\xi(t)] = 0 \text{ and}$$

$$E_{s,x}[\varphi_s^T] = E_{s,x}[\bar{\varphi}_s^T], \bar{\varphi}_s^T = 1/2\int_s^T a^u(t,\tilde{x}_t)^T (2b(t,\tilde{x}_t))^{-1} a^u(t,\tilde{x}_t)dt. \qquad (1.6a)$$

We come to the entropy functional, expressed via parameters of the initial controllable stochastic equation (1.1) in the form:

$$S(\tilde{x}_t / \varsigma_t) = 1/2 E_{s,x}[\int_s^T a^u(t,\tilde{x}_t)^T (2b(t,\tilde{x}_t))^{-1} a^u(t,\tilde{x}_t)dt]. \qquad (1.7)$$

Conditional mathematical expectation on the process' trajectories (1.7) (with density measure (1.3) for the above processes) is invariant at non-degenerative transformations.

Measures $\tilde{P}_{s,x} = \tilde{P}_{s,x}(A)$, defined for diffusion process $\varsigma_t$ (1.1a) (with transitional probability (1.2)) holds the same dispersion $b(t,\tilde{x})$ as $\tilde{x}_t$ but zero drift, modeling standard perturbations in controllable systems, which is practically usable.

Formulas (1.2), (1.3), (1.6) and (1.7) are in [50,59] with the citations and references.

Entropy functional (EF) (1.5,1.6) is an *information indicator* of a distinction between the processes $\tilde{x}_t$ and $\varsigma_t$ by these processes' measures; it measures a *quantity of information* of process $\tilde{x}_t$ regarding process $\varsigma_t$. This quantity is zero for the process' equivalent measures, and is a positive for the nonequivalent measures.

## 2. The information evaluation of the process' cutoff operation by an impulse control

Control $u_t$ is defined on the space $KC(\Delta,U)$ of a piece-wise continuous function of $t \in \Delta$:

$$u_+ \overset{def}{=} \lim_{t \to \tau_k + o} u(t,\tilde{x}_{\tau_k}),\ u_- \overset{def}{=} \lim_{t \to \tau_k - o} u(t,\tilde{x}_{\tau_k}), \tag{2.1}$$

which is differentiable, excluding the set

$$\Delta^o = \Delta \setminus \{\tau_k\}_{k=1}^m, k = 1,...,m. \tag{2.1a}$$

and applied on diffusion process $\tilde{x}_t$ from moment $\tau_{k-o}$ to $\tau_k$, and then from moment $\tau_k$ to $\tau_{k+o}$, implementing the process' transformations

$$\tilde{x}_t(\tau_{k-o}) \to \tilde{x}_t^{\sigma}(\tau_k) \to \tilde{x}_t(\tau_{k+o}).$$

At a vicinity of moment $\tau_k$, between the jump of control $u_-$ and the jump of control $u_+$, we consider a control *impulse*:

$$\delta u_{\pm}(\tau_k) = u_-(\tau_{k-o}) + u_+(\tau_{k+o}). \tag{2.2}$$

The following *Proposition* evaluates the entropy functional contributions at such transformations.

Entropy functional (1.5, 1.6) at the switching moments $t = \tau_k$ of control (2.2) takes the values

$$\Delta S[\tilde{x}_t(\delta u_{\pm}(\tau_k))] = 1/2, \tag{2.3}$$

and at locality of $t = \tau_k$ : at $\tau_{k-o} \to \tau_k$ and $\tau_k \to \tau_{k+o}$, produced by each of the impulse control's step functions in (2.2), is estimated by

$$\Delta S[\tilde{x}_t(u_-(\tau_k))] = 1/4,\ u_- = u_-(\tau_k),\ \tau_{k-o} \to \tau_k \tag{2.3a}$$

and

$$\Delta S[\tilde{x}_t(u_+(\tau_k))] = 1/4,\ u_+ = u_+(\tau_k),\ \tau_k \to \tau_{k+o}. \tag{2.3b}$$

*Proof.* The jump of the control function $u_-$ in (2.1) from a moment $\tau_{k-o}$ to $\tau_k$, acting on the diffusion process, might cut off this process after moment $\tau_{k-o}$.

The cut off diffusion process has the same drift vector and the diffusion matrix as the initial diffusion process. The additive functional for this cut off has the form [29,30]:

$$\varphi_s^{t-} = \begin{cases} 0, t \le \tau_{k-o}; \\ \infty, t > \tau_k. \end{cases} \tag{2.4a}$$

The jump of the control function $u_+$ (2.1) from $\tau_k$ to $\tau_{k+o}$ might cut off the diffusion process *after* moment $\tau_k$ with the related additive functional

$$\varphi_s^{t+} = \begin{cases} \infty, t > \tau_k; \\ 0, t \le \tau_{k+o}. \end{cases}. \tag{2.4b}$$

For the control impulse (2.2), the additive functional at a vicinity of $t = \tau_k$ acquires the form of an *impulse function*

$$\varphi_s^{t-} + \varphi_s^{t+} = \delta\varphi_s^{\mp}, \tag{2.5}$$

which summarizes (2.4a) and (2.4b).

Entropy functional (1.5,1.6), following from (2.4a,b), takes the values

$$\Delta S[\tilde{x}_t(u_-(t \le \tau_{k-o}; t > \tau_k))] = E[\varphi_s^{t-}] = \begin{cases} 0, t \le \tau_{k-o} \\ \infty, t > \tau_k \end{cases}, \tag{2.6a}$$

$$\Delta S[\tilde{x}_t(u_+(t > \tau_{k-o}; t \le \tau_{k+o}))] = E[\varphi_s^{t+}] = \begin{cases} \infty, t > \tau_{k-o} \\ 0, t \le \tau_{k+o} \end{cases}, \tag{2.6b}$$

changing from 0 to $\infty$ and back from $\infty$ to 0 and acquiring an *absolute maximum* at $t > \tau_k$, between $\tau_{k-o}$ and $\tau_{k+o}$.

The multiplicative functionals [19,30] related to (2.4 a,b), are:

$$p_s^{t-} = \begin{cases} 0, t \le \tau_{k-o} \\ 1, t > \tau_k \end{cases}, \quad p_s^{t+} = \begin{cases} 1, t > \tau_k \\ 0, t \le \tau_{k+o} \end{cases}. \tag{2.7}$$

Impulse control (2.2) provides an impulse probability density in the form of multiplicative functional

$$\delta p_s^{\mp} = p_s^{t-} p_s^{t+}, \tag{2.8}$$

where $\delta p_s^{\mp}$ holds $\delta[\tau_k]$-function, which determines the process' transitional probabilities with $\tilde{P}_{s,x}(d\omega) = 0$ at $t \le \tau_{k-o}, t \le \tau_{k+o}$ and $\tilde{P}_{s,x}(d\omega) = P_{s,x}(d\omega)$ at $t > \tau_k$.

For the cutoff diffusion process, the transitional probability (at $t \le \tau_{k-o}$ and $t \le \tau_{k+o}$) turns to zero, and states $\tilde{x}(\tau_k - o), \tilde{x}(\tau_k + o)$ become independent, while their mutual time correlations *are dissolved*:

$$r_{\tau_k - o, \tau_k + o} = E[\tilde{x}(\tau_k - o), \tilde{x}(\tau_k + o)] \to 0. \qquad (2.9)$$

Entropy increment $\Delta S[\tilde{x}_t(\delta u_{\pm}(\tau_k))]$ for additive functional $\delta\varphi_s^{\mp}$ (2.5), which is produced within, or at a border of the control impulse (2.2), is defined by the equality

$$E[\varphi_s^{t-} + \varphi_s^{t+}] = E[\delta\varphi_s^{\mp}] = \int_{\tau_{k-o}}^{\tau_{k+o}} \delta\varphi_s^{\mp}(\omega) P_{\delta}(d\omega), \qquad (2.10)$$

where $P_{\delta}(d\omega)$ is a probability evaluation of the impulse $\delta\varphi_s^{\mp}$.

Taking integral of symmetric $\delta$-function $\delta\varphi_s^{\mp}$ between the above time intervals, we get on the border

$$E[\delta\varphi_s^{\mp}] = 1/2 P_{\delta}(\tau_k) \text{ at } \tau_k = \tau_{k-o}, \text{ or } \tau_k = \tau_{k+o}. \qquad (2.11)$$

The impulse, produced by deterministic controls (2.2) for each random process location and dimension, is random with probability at $\tau_k$-locality

$$P_{\delta c}(\tau_k) = 1, k = 1,...,m. \qquad (2.12)$$

This probability holds a jump-diffusion transition Markovian probability, which is conserved during the jump [57].

From (2.10)-(2.12) we get the entropy functional's increment, when impulse control (2.2) is applied (at $t = \tau_k$ for each $k$) in the form

$$\Delta S[\tilde{x}_t(\delta u_{\pm}(\tau_k))] = E[\delta\varphi_s^{\mp}] = 1/2, \qquad (2.13)$$

which proves (2.3).

Since that, each of the symmetrical information contributions (2.6a,b) at a vicinity of $t = \tau_k$ is measured by

$$\Delta S[\tilde{x}_t(u_-(t \le \tau_{k-o}; t > \tau_k))] = 1/4, \ u_- = u_-(\tau_k), \ \tau_{k-o} \to \tau_k; \qquad (2.13a)$$

$$\Delta S[\tilde{x}_t(u_+(t > \tau_{k-o}; t \le \tau_{k+o}))] = 1/4, \ u_+ = u_+(\tau_k), \ \tau_k \to \tau_{k+o}, \qquad (2.13b)$$

which proves (2.3a,b). •

Entropy functional (EF) (1.5), defined through Radon-Nikodym's probability density measure (1.3), holds all properties of the considered cutoff controllable process, where both $P_{s,x}$ and $\tilde{P}_{s,x}$ are defined. That includes abilities for measuring $\delta -$cutoff information and extracting hidden process information not measured by known information measures.

Hence, *information measures the cutoff interaction* which had bound and hidden by the interaction's uncertainty measure. According to the definition of entropy functional (1.5), it is measured in natural $\ln$, where each of its Nat equals $\log_2 e \cong 1.44 bits$.

Thus, measure (1.5, 1.6) is not using Shannon entropy measure.

From (2.6a, b), (2.9), (2.13) and (1.6, 1.3) follow *Corollaries:*

**A.** Step-wise control function $u_- = u_-(\tau_k)$, implementing transformation $\tilde{x}_t(\tau_{k-o}) \to \tilde{x}_t^{\sigma}(\tau_k)$, converts the entropy functional from its minimum at $t \le \tau_{k-o}$ to the maximum at $\tau_{k-o} \to \tau_k$;

**B.** Step-wise control function $u_+ = u_+(\tau_k)$, implementing transformation $\tilde{x}_t^{\sigma}(\tau_k) \to \tilde{x}_t(\tau_{k+o})$, converts the entropy functional from its maximum at $t > \tau_k$ to the minimum at $\tau_k \to \tau_{k+o}$;

**cr**imples control function $\delta u_{\tau_k}^{\mp}$, implementing transformations $\tilde{x}_t(\tau_{k-o}) \to \tilde{x}_t^{\sigma}(\tau_k) \to \tilde{x}_t(\tau_{k+o})$, switches the entropy functional from its minimum to maximum and back from maximum to minimum, while the absolute maximum of the entropy functional at a vicinity of $t = \tau_k$ allows the impulse control to deliver *maximal amount* of information (2.13) from these transformations, holding principle of extracting maxmin- minmax of the EF measure;

**C.** Dissolving the correlation between the process' cutoff points (2.9) leads to *losing the functional connections* at these discrete points, which evaluate the Feller kernel measure of the Markov diffusion process [9].

**D.** The relation of that measure to additive functional in form (1.3),(1.4) allows evaluating the *kernel's information* by the entropy functional (1.6).

**E.** The jump action (2.1) on Markov process, associated with "killing its drift", selects the Feller measure of the kernel, while the *functional cutoff* provides entropy-*information measure* of the Feller kernel, and *it* is *a source of a kernel information,* estimated by (2.13).

In a multi-dimensional diffusion process, each of the stepwise control, acting on the process' all dimensions, sequentially stops and starts the process, evaluating the multiple functional information.

The dissolved element of the functional's correlation matrix at these moments provides independence of the cutting off fractions, leading to orthogonality of the correlation matrix for these cutoff fractions.

**F.** Multi-dimensional delta-distribution is the minimax *optimal* to hold the dissolving *interacting correlations*, which approaches the Tracy-Widom' distribution for complex interactions [60 ]. •

Let us consider a sum of increments (2.13) under impulse control $\delta u(\tau_k)$, cutting process $x_t$ at moments $\tau_k, k=1,2,...,m,$ along the process' trajectory on intervals $s>\tau_1>t_1>\tau_2>t_2,...,t_{m-1}>\tau_m>t_m=T$.

Applying additive principle for the process' information functional, measured at the moments of dissolving the correlation, which provides maximal cut off information, we get sum

$$S_m=\sum_{k=1}^{m}\Delta S_k[\tilde{x}_t(\delta u(\tau_k))]=\Delta S_1[\tilde{x}_t(\delta u(\tau_1))]+\Delta S_2[\tilde{x}_t(\delta u(\tau_2))]|,...,+\Delta S_m[\tilde{x}_t(\delta u(\tau_m))]. \qquad (2.14)$$

Impulses $\delta u(\tau_k)$ implement the transitional transformations (1.2), initiating the Feller kernels along the process and extracting total kernel information for $n$ -dimensional process with $m$ cutoff.

This maximal sum measures the interstates information connections held by the process along the trajectories during its time $(T-s)$.

It measures information hidden by the process correlating states, which is not covered by traditional Shannon entropy measure.

This sum of *extracted* information approaches theoretical measure (1.6) at

$$S_m|_s^T \xrightarrow[m\to\infty]{} S[\tilde{x}_t/\varsigma_t]_s^T, \qquad (2.15)$$

if all local time intervals $t_1-s=o_1, t_2-t_1=o_2,...,t_m-t_{m-1}=o_m$, at $t_m=T$ satisfy condition

$$(T-s)=\lim_{m\to\infty}\sum_{t=s,m}^{t=T}o_m(t). \qquad (2.15a)$$

Realization of (2.15) requires applying the impulse controls at each instant $(\tilde{x},s),(\tilde{x},s+o(s))$ of the conditional mathematical expectation (1.5) along the process trajectories.

However for any *finite* number $m$ (of instant $(\tilde{x},s),(\tilde{x},s+o(s))$) the integral process information (1.5) cannot be exactly composed from the information, measured for the process' fractions.

Indeed, sum $S_{mo}|_s^T$ of additive fractions of (2.4) on the finite time intervals: $s,t_1;t_1+o_1,t_2;..,t_{m-1},t_{m-1}+o_m;t_m=T$ :

$$S_{mo}|_s^T=\Delta S_{1o}[\tilde{x}_t/\varsigma_t]|_s^{t_1}+\Delta S_{2o}[\tilde{x}_t/\varsigma_t]|_{t_1+o_1}^{t_2},...,+\Delta S_{mo}[\tilde{x}_t/\varsigma_t]|_{t_{m-1}+o_m}^{t_m} \qquad (2.16)$$

is less than $S[\tilde{x}_t/\varsigma_t]_s^T$, which is defined through the additive functional (1.4).

As a result, the additive principle for a process' information, measured by the EF, is *violated*:

$$S_m|_s^T<S[\tilde{x}_t/\varsigma_t]_s^T. \qquad (2.17)$$

If each $k$-cutoff "kills" its process dimension after moment $\tau_{k+o}$, then $m=n$, and (2.15) requires infinite process dimensions. A finite dimensional model approximates finite number of probing impulses checking the observing frequencies.

For any of these conditions, the EF measure, taken along the process trajectories during time $(T-s)$, limits maximum of total process information, extracting its hidden cutoff information (during the same time), and brings more information than Shannon traditional information measure for *multiple states* of the process.

Therefore, maximum of the process cutoff information, extracting its total integrated hidden information, approaches the EF information measure.

Since the maxmin impulse automatically minimize current entropy measure, its actual measurement is not necessary.

Sum (2.14) for multiple process' independent components considered during $(T-s)$, acquires the form of matrix trace:

$$S_{mo}|_s^T = Tr[\Delta S_k[\tilde{x}_t(\delta u(\tau_k))], k=1,.....n, m=n\ , \tag{2.18}$$

which relates (2.18) to Von Neumann entropy for quantum ensemble [42].

Relation (2.18) satisfies such impulse control that kills each dimension by its stopping time at the cutoff.

For $n \to \infty$, the Von Neumann entropy (2.18) equals to uncertain entropy functional (EF)(1.6), [53, 59], which is defined on irreversible random process.

*Information path functional* (IPF) defines distributed actions of multi-dimensional delta-function on entropy functional (1.7) through the additive functional for all dimensions:

$$I_{pm} = \delta_m\{S[\tilde{x}_t / \varsigma_t]|\} = 1/2E\{\int_s^T \delta_m[a(t,\tilde{x})^T(2b(t,\tilde{x}))^{-1}a(t,\tilde{x})dt)]\}, \tag{2.19}$$

which determines sum of information measures $\Delta I_k[\tilde{x}_t(\delta u(\tau_k))]$ along the path on cutting process intervals (2.15a).

In a limit it leads to

$$I_p = \lim_{m\to\infty}\sum_{k=1}^{m}\Delta I_k[\tilde{x}_t(\delta u(\tau_k))]. \tag{2.20}$$

Formal definition (2.19) allows the IPF representation by Furies integral leading to frequency analysis with Furies series.

The IPF is the sum of *extracted* information which approaches theoretical measure (2.16):

$$I_p = \lim_{m\to\infty} I_{mo} |_s^T = \lim_{m\to\infty} S_{mo} |_s^T \to S[\tilde{x}_t / \varsigma_t]_s^T, \quad (2.21)$$

if all finite time intervals, at $t_m = T$, satisfy condition (2.15a).

Since each cutting interval encloses invariant information measure, the limited $I_p$ limits the initially undefined (in (1.7)) upper time $T$ of the EF integral.

It also brings direct connection $I_p$ and $T-s$.

Therefore, at infinite sequence of the time intervals, this sequence has limit

$$\lim_{m\to\infty} o(t_m) \to 0, \quad (2.22)$$

and sum of such sequence is finite.

Realization (2.20),(2.21),(2.22) requires applying the impulse controls at each instant $(\tilde{x}, s), (\tilde{x}, s + o(s)), \ldots$ along the process trajectories of conditional math expectations.

However for any *finite* number $m$ of these instants, the *integral* process' information, composed from the information, measured for the process' fractions, is not complete.

Cutting the EF by impulse delta-function determines the increments of information for each impulse:

$$\Delta I[\tilde{x}_t] |_{t=\tau_k^{-o}}^{t=\tau_k^{+o}} = \begin{cases} 0, t < \tau_k^{-o} \\ 1/4Nat, t = \tau_k^{-o} \\ 1/4Nat, t = \tau_k^{+o} \\ 1/2Nat, t = \tau_k, \tau_k^{-o} < \tau_k < \tau_k^{+o} \end{cases} \quad (2.23)$$

with total

$$\sum_{t=\tau_k^{-o}}^{t=\tau_k^{+o}} \Delta I[\tilde{x}_t]_{\delta t} = 1Nat. \quad (2.23a)$$

The IPF along the cutting time correlations on trajectory $x_t$ in a limit determines eq.

$$I[\tilde{x}_t / \varsigma_t]_{x_t} = -1/8 \int_s^T Tr[(r_s \dot{r}_t^{-1}] dt = -1/8 Tr[\ln r(\mathrm{T}) / r(s)] \quad (2.24)$$

where $\dot{r}_t = 2b(t), r_s = E_{s,x}[x^2(t)]$.

Eqs. (1.7), (2.24) establish direct *connection entropy, information and the process time interval*.

A final finite integral information approaches that generated by the last impulse during the final finite $\tau_{k=n}$, which is Kronicker' impulse-discrete analog of Dirac' delta-function taking values 0 and 1 and preserving (2.23a).

Conditional probability satisfies Kolmogorov's 1-0 [71, p.116-117] law for function $f(x)|\xi$ of $\xi, x$ infinite sequence of independent random variables:

$$P_\delta(f(x)|\xi) = \begin{cases} 1, f(x)|\xi) \geq 0 \\ 0, f(x)|\xi) < 0 \end{cases} \quad (2.25)$$

This probability measure has applied for the impulse probes on an observable random process, which holds opposite Yes-No probabilities – as the unit of probability impulse step-function preserving the max-min.

*Comments*.

Number $M$ of equal probable possibilities determines Hartley's quantity of information $H = \ln M$ measured on Nats, which for the impulse $M = 2$ holds $H = \ln 2 Nat$.

The impulse information measured in bits is $I = 1/\ln 2 \ln M = \lg M bit = 1 bit$.

The correlation cutting by the impulse brings information $0.75 Nat$ from which $\delta S_u \cong 0.0568 Nat$ delivers the impulse controls. Since each cutoff brings invariant $1 Nat$ (2.23), the difference $(1-0.7)Nat \cong 0.3 Nat$ presents "free information" for each cutting impulse.

The IPF integral information evaluates maximal speed of enclosing information in the finite impulse time interval; the instant of the impulse time that cuts correlation's hidden information equals to $\mathrm{Bit} = \ln 2$.

All integrated information enfolds the Feller kernel whose time and energy evaluate results [11].

Minimal physical time interval limits the light time interval $\delta t_\tau \cong 1.33 \times 10^{-15} \sec$ defined by the light wavelength $\delta l_m \cong 4 \times 10^{-7} m$. That allows us estimate maximal information density for 1 bit:

$$I_{ko_k} \cong \ln 2 / 1.33 \times 10^{-15} \cong 5.2116 \times 10^{+15} Nat/s. \quad (2.26)$$

Or for each invariant impulse $1 Nat$, the maximal density estimates

$$I_{ko_{k1}} \cong 1/1.33 \times 10^{-15} \cong 7.5188 \times 10^{+15} Nat/s. \quad (2.26a)$$

Variety of the impulse $\downarrow\uparrow$ physical interactions unites the considered impulse information model, which both EF-IPF integrate. •

**3. An information form of Schrödinger's equation**

Applying the results of Secs.1-2, let us consider transformation of transition probability

$$P = P(s, \tilde{x}_t, t, B) \quad (3.1)$$

of diffusion process $\tilde{x}_t = \tilde{x}_t(t, x, \xi_t)$ (1.1) to transition probability function $\tilde{u}(s, x)$ with aid of additive functional $\varphi_s^t$:

$$E_{t,x}[\int\limits_{\tilde{x}(t)\in B} \exp\{-\varphi_s^t(\omega)\}] = \tilde{u}(t,x) \ . \qquad (3.2)$$

*Proposition 3.1.*

**1**. The evolution of function $\tilde{u} = \tilde{u}(t,x)$ satisfies the Kolmogorov differential equation for the probability:

$$\frac{\partial \tilde{u}}{\partial t} = a(t,x)\frac{\partial \tilde{u}}{\partial x} + b(t,x)\frac{\partial^2 \tilde{u}}{\partial x^2} - V(t,x)\tilde{u} \ , \ \bar{\varphi}_s^t = 1/2\int_s^t V dt \qquad (3.3)$$

with additive the functional (1. 4).

Equation (3.3) is basic equation in the following considerations.

**2**. The differential equation for evolution of entropy functional (1.5)

$$\tilde{S}(\tilde{x}_t / \varsigma_t) = E_{t,x}\{\varphi_s^t(\omega)\} \qquad (3.4)$$

satisfies the Kolmogorov's differential equation for math expectations of additive functional in form:

$$-\frac{\partial \tilde{S}}{\partial t} = (a^u)^T \frac{\partial \tilde{S}}{\partial x} + b\frac{\partial^2 \tilde{S}}{\partial x^2} + 1/2(a^u)^T (2b)^{-1} a^u \ . \qquad (3.5)$$

*Proof.* Since both probability $\tilde{u} = \tilde{u}(t,x)$ and entropy function $\tilde{S}(t,x)$ are defined on trajectories of diffusion process (1.1), their evolutions are connected by the same function $V(t,x)$ of additive functional (3.3, 1.4).

**3**. Complex conjugated information wave functions $(\tilde{Q},\tilde{Q}^*) = \bar{Q}$ are the solutions of a system of diffusion equations, which are equivalent to the dual *Schrödinger's Equations* [17] in the form

$$\frac{\partial \tilde{Q}}{\partial t} + b(t,x)\frac{\partial^2 \tilde{Q}}{\partial x^2} + a(t,x)\frac{\partial \tilde{Q}}{\partial x} + V_\psi(t,x)\tilde{Q} = 0 \,, \qquad (3.6a)$$

$$-\frac{\partial \tilde{Q}^*}{\partial t} + b(t,x)\frac{\partial^2 \tilde{Q}^*}{\partial x^2} - a(t,x)\frac{\partial \tilde{Q}^*}{\partial x^2} + V_\psi(t,x)\tilde{Q}^* = 0 \,, \qquad (3.6b)$$

where $V_\psi = -j\lambda\tilde{\psi}$ is a characteristic equivalent to function $V$ in (3.3) following from characteristic function of a random functional [30]:

$$\tilde{u}_\phi(t,x,\lambda) = E_{t,x}\exp[j\lambda\tilde{\Psi}], \tilde{\Psi} = \int_s^t \tilde{\psi}(\tau,\tilde{x}(\tau))d\tau \qquad (3.7)$$

which, at a fixed parameter $\lambda$ , $-\infty < \lambda < \infty$, satisfies Kolmogorov equation for math expectation:

$$\frac{\partial \tilde{u}_\phi}{\partial t} = a(t,x)\frac{\partial \tilde{u}_\phi}{\partial x} + b(t,x)\frac{\partial^2 \tilde{u}_\phi}{\partial x^2} - j\lambda\tilde{\psi}\tilde{u}_\phi \,, \qquad (3.7a)$$

where $\tilde{\psi}(t,x)$ is a real function and

$$\exp[j\lambda\tilde{\Psi}] = \bar{Q} = (\tilde{Q},\tilde{Q}^*) \qquad (3.7b)$$

is an information equivalent of a wave function in Quantum Mechanics. The joint eqs (3.7, 3.7a,b) *prove* (3.6a,b). This information form accumulates hidden information that integrates functional (3.4) during the evolving observations.

**4.** Probability density $p$ in (1.3) for the complex conjugated wave functions $(\tilde{Q}, \tilde{Q}^*)$, according to relation [16, 17,19, 20] satisfies equations

$$p = \tilde{Q} \times \tilde{Q}^* = |\bar{Q}|^2, |\bar{Q}|^2 = |\tilde{Q}|^2 + |\tilde{Q}^*|^2 + 2\,\mathrm{Re}(\tilde{Q} \times \tilde{Q}^*), \qquad (3.8)$$

where the absolute value

$$|\bar{Q}| = |\,\mathrm{e\,xp}[(j\lambda\tilde{\Psi})]\,| \qquad (3.9)$$

connects positive $p$ (in (3.8)) with positive additive functional $\bar{\varphi}_s^t$ (3.3) by equation

$$|\bar{Q}| = |\sqrt{p}\,| = |\sqrt{\exp(-\bar{\varphi}_s^t)}\,|. \qquad (3.9a)$$

The *proof* follows from joint eqs (3.8-3.9) and (3.3).

**5**. Entropy functional $\tilde{S} = \tilde{S}(\tilde{x}_t / \varsigma_t)$ on trajectories of conjugated wave functions with $\exp[j\lambda\tilde{\Psi}] = \bar{Q}$, expressed via function $V(t,x)$, holds

$$\tilde{S} = E[-\ln|\bar{Q}|^2] = -E\ln|\exp(j\lambda\tilde{\Psi})|^2 = E[\int_s^t V(\tau, x(\tau))d\tau] \quad . \qquad (3.10)$$

*Proof*. Taking logarithms from both sides (3.9a), we have:

$$\ln p = \ln|Q|^2] = \ln|\exp(j\lambda\tilde{\Psi})|^2 = -\bar{\varphi}_s^t \,. \qquad (3.11)$$

Using entropy functional, expressed via mathematical expectations of the wave function:

$$\tilde{S} = E[-\ln p] = -E[\ln|Q|^2] = -E\ln|\exp(j\lambda\tilde{\Psi})|^2, \qquad (3.11a)$$

and additive functional (1.4, 3.3) on the trajectories of Ito's eq.(1.1), we get relation

$$E[-\ln|\bar{Q}|^2] = E[\bar{\varphi}_s^t] \ . \qquad (3.11b)$$

From which follows (3.10) on the trajectories of the conjugated wave functions and diffusion process. •

**4. Schrödinger's bridge**

*Reversible* probabilities density (3.8), decomposed on a product of *forward and backward densities*, defines *Schrödinger's bridge* on Markovian path between forward and backward movement of the path states.

E. Schrödinger has considered the forward and backward densities as *information "waves"*.

For considering class of *reciprocal Markovian diffusion* [21, 22, 31], the entropy function for the Schrödinger's bridge follows from probability density $p$, satisfying (3.9), is connected to the wave function in form [21]:

$$\bar{Q} = \exp(R \pm jI), \tilde{Q} = \exp(R), \tilde{Q}^* = \exp(\pm jI), \qquad (3.12)$$

and by relation

$|\bar{Q}|^2 = p = \exp 2R$ . (3.13)

The transitional probability densities *between* states $(s,x)$ and $(t,y)$ on the reciprocal diffusion satisfy

$p = p(s,x;t,y) = p(s,x)p(t,y), p(s,x) = \exp(R - I), p(t,y) = \exp(R + I)$ . (3.13a)

*Proposition 3.2.*

1. Following relation (3.9), (3.12), (3.13), we get

$\tilde{S} = E(-\ln p) = E(-2\ln|\bar{Q}|), \tilde{S} = E[-2\ln(\exp|\bar{Q}|)] = 2E[-\ln(\exp(R \pm jI)] = 2E[(-R) \pm (-jI)]$(3.14)

where, we may designate a real part of $\tilde{S}$ as its real entropy:

$\tilde{S}_a = \mathrm{Re}(1/2\tilde{S}) = E(-R)$ (3.14a)

and imaginary part of $\tilde{S}$ as its imaginary entropy:

$\mathrm{Im}\,\tilde{S} = \tilde{S}_b = \pm E[-I] = E[\mp I] = \tilde{S}_b^{\pm}$ . (3.14b)

*Then*

$\tilde{S} = 2(\tilde{S}_a \pm j\tilde{S}_b)$ . (3.14c)

Subsequently, probability density (3.13a) at reversing its time course from $p(-t,y)$ back to $p(s,x)$ holds

$p^* = p(-t,y;s,x) = p(-t,y)p(s,x) = \exp(-R - I)\exp(R - I) = \exp(-2I)$ , (3.15)

or entropy:

$E[-\ln p^*] = \tilde{S}^* = -2\tilde{S}_b^{\pm} = 2\tilde{S}_a = \tilde{S}$ , (3.15a)

*satisfies the reversibility applied to Markov path between* states $(s,x)$ and $(t,y)$ of the *Schrödinger bridge:*

$p^* = p$ . • (3.16)

Then we come to *information evaluation of the bridge*

$\tilde{S}_a = 1/2\ln 2 = E[1/2\bar{\varphi}_s^t], \tilde{S}_b^{\pm} = -E[1/2\bar{\varphi}_s^t] = -1/2\ln 2$ , $\tilde{S} = -\ln 2, p = 1/2$ . (3.17)

*Proof*. From requirements (3.12) - (3.13a) we have (3.8) in form

$|\bar{Q}|^2 = (\exp R)^2 + |\exp(\pm jI|^2 + 2[(\exp R) \times (\exp \pm I) = \exp(2R)$ . (3.18)

The reversibility at the bridge implies mutual compensation the conjugated parts of the wave function, leading to the equality between the relations for the wave function (3.14) and its interactive components:

$\exp(2R) = 2(\exp R) \times \exp(\pm I)$ . (3.18a)

Logarithm of (3.18a), math expectations from both sides, and definition (3.14a,b) lead to:

$\tilde{S} = \tilde{S}_{ab} = -\ln 2 + \tilde{S}_a + \tilde{S}_b^{\pm} = -\ln 2$ , (3.18b)

where relation $\tilde{S}_a = -\tilde{S}_b^{\pm}$ follows from (3.15a), and $\tilde{S}_{ab}$ is entropy of the interactive components.

Satisfaction of (3.13), (3.18b) proves (3.17). We also get $R = |I| = 1/2\ln 2$ . •

**2**. In a *more general* case, when Markovian diffusion is not a reciprocal process, but the bridge is reversible, conditions (3.12-3.14) are not satisfied. Then we come to

$$E[\ln|\bar{Q}|^2] = 2E(\ln|\tilde{Q}|) + 2E[(\ln|\tilde{Q}^*|) + E[\ln(2\operatorname{Re}[|\tilde{Q}|\times|\tilde{Q}^*|]), \quad (3.19)$$

$$E[\ln(2\operatorname{Re}[|\tilde{Q}|\times|\tilde{Q}^*|]) = \ln 2 + E(\ln|\tilde{Q}|) + E[(\ln|\tilde{Q}^*|), \quad (3.19a)$$

where by analogy with (3.14a,b), we define the related entropies by :

$$E[-\ln|\bar{Q}|] = \tilde{S}, E(-\ln|\tilde{Q}|) = \tilde{S}_a, \pm E[(-\ln|\tilde{Q}^*|) = \tilde{S}_b^{\pm}. \quad (3.19b)$$

Then the entropy of the interactive components of the wave function (3.19a) holds:

$$-E[\ln(2\operatorname{Re}[|\tilde{Q}|\times|\tilde{Q}^*|]) = -\ln 2 - \operatorname{Re}[\ln|\tilde{Q}| + \ln|\tilde{Q}^*|] = \tilde{S}_{ab}, \quad (3.20)$$

keeping connection to total entropy $\tilde{S}$ , the real, and imaginary entropies at :

$$\tilde{S} = \tilde{S}_a + \tilde{S}_b^{\pm} + \tilde{S}_{ab}. \quad (3.20a)$$

At the bridge reversibility, entropies of conjugated parts of the wave function mutually compensate:

$$\tilde{S}_a = -\tilde{S}_b^{\pm}, \ -\operatorname{Re}\ln|\tilde{Q}| - \operatorname{Re}\ln|\tilde{Q}^*| = 0. \quad (3.21)$$

Then the interactive component satisfies

$$\tilde{S}_{ab} = \tilde{S} = -\ln 2 \quad (3.21a)$$

being equal to the bridge total entropy.

For information conjugated processes (3.6a,b), the interactive components hold the entropies (3.4) of *an action's functional* (satisfying (3.5)), which produces the bridge entropies at the interactions.

*Remarks*. At satisfaction of (3.9), (3.9b) with a normalized probability $p \le 1$ and relations

$$|\bar{Q}|^2 = p, -E\ln[|\bar{Q}|^2 = -E[\ln p] = \tilde{S}, \tilde{S} = E[\varphi_s^t], p = \exp(-\varphi_s^t), \varphi_s^t \ge 0, \tilde{S} \ge 0, p = \exp(-\tilde{S}) \le 1 \quad (3.21b)$$

the equality $p = \exp(-\tilde{S}) = \exp(\ln 2) = 2 > 1$ has no meaning.

While the connection to a hidden information

$$H_I = -\tilde{S}, \quad (3.21c)$$

at $p = \exp(-H_I) = \exp(-\ln 2) = 1/2$, imposes limitations on probabilities (3. 9a).•

The interactive part of the wave function *concentrates total Schrödinger bridge'* hidden information measure $\ln 2$ .

**3**. The process of cutting the entropy functional (Sec.1) has probability density $p^o = \exp(-0.5) \cong 0.7$ .

If this density is decomposed on a product of forward and backward reversible densities in a Markovian path defining Schrödinger's bridge, *then, the* previous relations (3.19a,b), (3.20), (3.21), at

$$\tilde{S} = -0.5 = \tilde{S}_{ab}, \text{ and } \tilde{S} = 0.5 = \tilde{S}_{ab} = \ln 2 + \tilde{S}_a + \tilde{S}_b^{\pm}, \quad (3.22)$$

bring

$$\tilde{S}_a = -\tilde{S}_b^{\pm} = -0.1. \bullet \quad (3.22a)$$

**4.** At any other given normalized probability density $p^o = p(t) = p^*(-t)$, satisfying (3.8), (3.9a) and (3.21), the condition of reversibility on a Brownian path as Schrödinger's bridge, *requires*

$$\tilde{S}_a = -\tilde{S}_b^{\pm} \tag{3.23}$$

at

$$p(t) = \exp-[\tilde{S}_a(t) + \tilde{S}_b(t) + 1/2\ln 2],\ p^*(-t) = \exp-[\tilde{S}_a(-t) + \tilde{S}_b(-t) + 1/2\ln 2]\,. \tag{3.23a}$$

5. The entropies, corresponding opposite time directions (forward $t$ and backward $-t$) of the functional:

$$\tilde{S}^+(t) = \tilde{S}_a(t) + \tilde{S}_b(t) + \ln 2 = \tilde{S}_{ab}(t), \tilde{S}^-(t) = \tilde{S}_a(-t) + \tilde{S}_b(-t) + \ln 2 = \tilde{S}_{ab}(-t) \tag{3.24}$$

on the reversible bridge:

$$\tilde{S}_{ab}(t) = \tilde{S}_{ab}(-t) \tag{3.24a}$$

satisfy equality

$$\tilde{S}_a(t) = \mp\tilde{S}_b(t) \tag{3.24b}$$

Here the relation

$$\tilde{S} = \tilde{S}^+(t) + \tilde{S}^-(-t) \tag{3.25}$$

is an information analog of the symmetrized process' functional action in [40] at

$$p(t) = \exp-(1/2\ln 2) \cong 0.7,\ p^*(-t) = \exp-(1/2\ln 2) \cong 0.7\,. \tag{3.25a}$$

**4.1. The evaluations of both quantum information of Schrödinger's path to the bridge and entanglement**

The quantum information of wave function in form of Schrödinger's Eqs. (3.6a,b) includes the wave superposition with the quantum probabilities (3.8),(3.9a) for both Markovian equivalents of reciprocal diffusion and a less restricted Markov diffusion process.

In both of these cases, condition (3.24b) leads to a Schrödinger's path to the bridge, specifically, with distinctive values of these probabilities and the entropies, concentrated in the bridge.

Quantum correlations arise at interaction of the superimposed components of a wave function, which produces these reversible probability densities on the path to the bridge.

Since quantum correlations entangle the superimposed components of a wave function, they bring *quantum entanglement*, which (at condition (3.24)) takes place on Schrödinger's path up to the bridge.

The distributed interactions, producing quantum entanglement, allow both its locality and non-locality.

Each localized or not localized components of wave function may interact by their local or non-local correlations, which connect them at entanglement that unites in a common unit. However natural interactions have a limited distance, defined in [37] by a "distance between a given state and the boundary of separable states with entangled states".

The distance is measured by the probabilities' trace distance between the nearest interacting probabilities [36, 37, 38], or that distance measures the minimal time intervals between the interactions [41].

After dissolving the interactive component of the wave function by "killing" the correlations at the moment $t = t^+$, we come to $\tilde{S}_{ab}(t^+) = 0$, and from (3.24) and (3.24ab) it follows

$$\tilde{S}_a(t^+) + \tilde{S}_b(t^+) = -\ln 2,$$

and

$$\tilde{S}(t^+) = -\ln 2 .$$

Killing the bridge by the control's cutoff (Sec.1) releases information, concentrated in the bridge:

$$\tilde{S}_{ab}^{2o} = H_{ab}^{2o} = 0.5 ,$$

from its total entropy

$$\tilde{S}_{ab}^{1o} = -\ln 2 .$$

Transformation from $\tilde{S}_{ab}^{1o}$ to $\tilde{S}_{ab}^{o}$ changes entropy from $-\ln 2$ to information $0.5$:

$$0.5 - \ln 2 \cong 0.2 Nats$$

which the impulse control spends for the bridge destruction.

Changing entropy from $\tilde{S}(t^+) = -\ln 2$ to information $H_I(t^+) = \ln 2$ requires no additional contributions:

$$H_I(t^+) - \tilde{S}(t^+) = 0 .$$

Entropy of the bridge $\tilde{S}_{ab}^{o}$ is equal to Von Neumann's entropy [42] and (2.18):

$\tilde{S} = Tr[E(-\ln p)]$ (at $p = \exp(-H_{ab}^{o} |)$ ),

while for both parts satisfy

$$\tilde{S}_{ab}^{1o} = -\ln 2 = -H_{ab}^{1o}$$

and

$$\tilde{S}_{ab}^{2o} = 0.5 = H_{ab}^{2o} .$$

The related probability densities take a diagonal forms in a multi-dimensional process, while its summary entropy is defined by math expectations of the probability densities over all process.

A total *bridge* information (entropy) in such process:

$$\tilde{S} = \tilde{S}_{ab}^{\Sigma o} = Tr[\tilde{S}_{ab}^{o}], \; \tilde{S}_{ab}^{o} = (\tilde{S}_{ab}^{1o}, \tilde{S}_{ab}^{2o}),$$

includes different unpredictable combinations of $\tilde{S}_{ab}^{1o}, \tilde{S}_{ab}^{2o}$, which a priori are unknown.

However, the frequency of appearing $H_{ab}^{2o}$ depends on the frequency of the cutoff information which measures one cutoff for each dimension of $n$-dimensional Markov process.

For so-called "Werner states" (with the entangled both *pure* state and pure *entangled* state), the relative entropy of entanglement, in general, is less than that for the entanglement with entropy $\tilde{S}_{ab}^{1o}$.

This means that, even at a slow rate of growing the entropy during formation of entanglement, an *irreversible* process of destruction could prevail, that leads to disentanglement at the end with releasing information. Such a formation is called Squashed-Compressed Dense Condensed entanglement [38].

Since an entanglement is a result of natural interaction of the wave functions, carrying some real and imaginary components, satisfaction of condition $\tilde{S}_a(t)+\tilde{S}_b(t)=0$ for both kinds of considered Markovian processes, corresponds to compensation of the interactive imaginary part by its real at the entanglement.

*It implies natural creation information as a phenomenon of interaction cutting the correlated entanglement.*

A collapse of the wave functions, ending of the bridge's formation, disentangles the interacting components, which is an equivalent of killing the entanglement and releasing the above entangled information.

Killing at moment $t=t^{+}$, or disentanglement (at this moment) requires changing the sign any of these entropy components.

For the particles' information units with their spins, correlated by a clock-wise and counter clock-wise directions, it needs changing anyone of these directions.

Since information $H_{ab}^{2o}$, transferred from such entanglement, is positive, it leads to the entanglement's instability and a possibility of a self-disentanglement.

Because the probability of such unstable entanglement is higher (0.7) than the probability (0.5) of a stable entanglement (which possesses entropy $\tilde{S}_{ab}^{1o}$), such *unstable entanglements might naturally exist more often than the stable entanglement.*

The self-disentanglement (a self-destruction) during a finite time, is known as effect of Entanglement Sudden Death, which was revealed for both local and non-local entanglements in the forms of a local Sudden Death (ESD) and Bell's non-locality sudden death (BNSD) [38]. This study demonstrates that both effects ESD and BNSD, having a finite time of destroying entanglement, are more probable, compared to forming the stable entanglement. The study suggests that these short living entanglements, which involve energy transfer and temperature effects, are more likely in Universe than long-lived entanglement.

In some cases, the self-destruction leads to the spontaneous emission from the entangled pair [43]**.**

It's shown that "decaying rate of an entangled atom is different from that in a product state, modifying the temporal emission distribution and life time of the atoms".

The review of the sadden death and the relationship between decoherence and disentanglement [38] concludes that the nature of the loss of entanglement consists of *lossing of state coherence.* Even though both the ESD and BNSD quantum effects behave very similar, the ESD is an extreme case in which "decoherence persists asymptotically, whereas the entanglement is entirely eliminated in finite time" [38].

The study found that entanglement, being measured by the states' coherence, decays at a different rate, compared to the coherence, measured by the reduction of off-diagonal density matrix elements.

And the time scale of disentanglement was always less or equal to the time scale of decoherence. The Sudden death evolution's dynamics affect the dynamics of evolution of information that *memorize* both ESD- BNSD disentanglements.

According to [38], the evolution decreases the initial probability of a potential entanglement in $\cong 8$ time for the probability $p^o = 0.5$ (3.17).

From that, the related ratio of the time interval of a sudden death $\Delta t_d$ to the time interval of the potential non-decaying entanglement $\Delta t_e$ evaluates

$$\Delta t_d / \Delta t_e = 1/8 \cong 0.125, \qquad (3.26)$$

which estimates also ratio of the related entangled information in forms of qubits.

Ratio (3.26) closes to relation these times in [35, 37], estimating with a less accuracy but in more wide range.

The evaluation of the time death along with information, kept in the entanglement, is important for estimation of the memorized quantity information and its time conservation in both forms of entanglements.

Quantum computation [46, 47, others], which uses quantum superposition and entanglement to perform operations on data, involves transfer an energy and the temperature effects. Specifically, in the quantum error correction protocols, these factors "degrade entanglement and coherence in addition to other sorts of phase and amplitude damping errors".

However, the entanglement and energy are neither mapped in a one-to-one fashion nor evolve at the same rate" [38].

Nevertheless reference [48] states that *in* a "tele-transportation, associated with entanglement, as an instantaneous non-local exchange of information, there is no involvement of energy or matter transfer".

This might be because the interconnected entangled particles only transfer of a pure information entropy through a distance by "a resonance through a quantum-tunneling effect" . This tunneling may exclude any forms of interactions.

Getting information, concentrated in the bridge, requires an *interaction with* the entangled information, which dissolves the correlations and the bridge.

Such a natural interaction could produce an impulse represented by asymmetrical delta-distribution. For example, when an electron or photon hits an entangled quantum superposition with a shared single state, the interaction releases these states with information bound in the bridge. Such a hitting impulse should carry information, compensating for entropy $\tilde{S}^o = \ln 2$, or information $\tilde{S}^o_o = -0.5$ accordingly.

If such an impulse is not a natural, performing, for example, a discrete quantum measurement, the measured device should carry the above quantities of information [49], and the information measurement must be taken exactly at the moment following completion of the entanglement's formation.

This strong requirement satisfies the information path functional whose information measure coincides with the measured time interval.

The EF-IPF measure statistical information for different Markov processes, and/or for other stochastic processes.

In such forms of measurement, artificial created impulses, or natural interactive impulses, their discrete controls functions, performing the measurements, apply at the moment of the transformation of a controllable process to the Brownian movement creating a kernel which concentrates the measurement.

Moreover, such control functions can automatically implement this transformation through a maximization of the measured entropy (Sec.2).

In this case, max entropy is $\tilde{S}^o = -\ln 2$ ( $0.7qNats \cong 1qbit$ ), if the measurement is taken at the moment of entanglement (which the measurement dissolves).

The control measurement should carry such quantum information, which is necessary to cut entropy $\tilde{S}^o$ bringing information unit $H_I^o = 0.5$ $Nats$ .

To get a total potential information 0.7 $Nats$ when cutting 0.5 $Nats$ , each of impulse stepwise controls should spend information 0.1 $Nats$ , which concurs with (3.22a,b).

Such a hitting impulse should carry information, compensating for entropy $\tilde{S}^o = \ln 2$ , or information $\tilde{S}_o^o = -0.5$ accordingly.

In such forms of measurement, discrete controls functions, performing the measurements, apply at the moment of the transformation of a controllable process to the Brownian movement creating a kernel which concentrates the measurement. These control functions could create artificial impulses or natural interactive impulses.

Moreover, such control functions can automatically implement this transformation through a maximization of the measured entropy (Sec.2). In this case, max entropy is $\tilde{S}^o = -\ln 2$ ( $0.7qNats \cong 1qbit$ ), if the measurement is taken at the moment of entanglement (which the measurement dissolves). The control measurement should carry such quantum information, which is necessary to cut entropy $\tilde{S}^o$ bringing information unit $H_I^o = 0.5$ $Nats$. To get total potential information 0.7 $Nats$ when cutting 0.5 $Nats$, each of the impulse stepwise controls should spend information 0.1 $Nats$, which concurs with (3.22a,b).

**5. The solution of variation problem for the entropy functional applying to information wave functions**

Applying the variation principle to the entropy functional, we consider an integral functional

$$S = \int_s^T L(t, x, \dot{x}) dt = S[x_t] \quad , \tag{5.1}$$

which minimizes the entropy functional (3.4) of the controlled process in form

$$\min_{u_t \in KC(\Delta, U)} \tilde{S}[\tilde{x}_t(u)] = S[x_t],\ Q \in KC(\Delta, R^n)\,. \tag{5.1a}$$

<u>*Proposition 5.1.*</u>

**1**. An *extremal solution* of variation problem (5.1a, 5.1) for the entropy functional (3.4), (1.7) brings the following equations of extremals for vector $x$ and conjugate vector $X$ accordingly at $(t, x) \in Q$,

$$\dot{x} = a^u \tag{5.2}$$

$$\dot{X} = -\partial P / \partial x - \partial V / \partial x\,, \tag{5.3}$$

where

$$P = (a^u)^T \frac{\partial S}{\partial x} + b^T \frac{\partial^2 S}{\partial x^2} \quad , \tag{5.4}$$

is function of action $S(t, x)$ on extremals (5.2,5.3); $V(t, x)$ is the integrant function (3.3) for the additive functional (1.4) in (3.2), which defines the probability function $\tilde{u} = \tilde{u}(t, x)$.

*Proof*. Using the Jacobi-Hamilton (JH) equations [51, 52] for function of action $S = S(t, x)$, defined on the extremals $x_t = x(t), (t, x) \in Q$ of functional (5.1), we have

$$-\frac{\partial S}{\partial t} = H, H = \dot{x}^T X - L, \tag{5.5}$$

where $X$ is a conjugate vector for $x$ and $H$ is a Hamiltonian for this functional.

(All derivations here and below have vector form).

From (5.1a) it follows

$$\frac{\partial S}{\partial t} = \frac{\partial \tilde{S}}{\partial t}, \frac{\partial \tilde{S}}{\partial x} = \frac{\partial S}{\partial x} \tag{5.5a}$$

where for the JH we have $\frac{\partial S}{\partial x} = X, -\frac{\partial S}{\partial t} = H$ .

This allows us to join eqs (5.5), (5.5a) and (3.5) in form

$$-\frac{\partial \tilde{S}}{\partial t} = (a^u)^T X + b\frac{\partial X}{\partial x} + 1/2a^u(2b)^{-1}a^u = -\frac{\partial S}{\partial t} = H, \quad (5.6)$$

where a dynamic Hamiltonian holds $H = V + P$, which includes function $V(t,x)$ and function of a potential

$$P(t,x) = (a^u)^T X + b^T \frac{\partial X}{\partial x}. \quad (5.7)$$

Applying to (5.6) Hamilton equations $\frac{\partial H}{\partial X} = \dot{x}$ and $\frac{\partial H}{\partial x} = -\dot{X}$, we get the extremals for vector $x$ and $X$ in forms (5.2) and (5.3) accordingly. •

**2**. A *minimal solution* of variation problem (5.1a, 5.1) for the entropy functional (3.4) brings the following equations of extremals for $x$ and $X$ accordingly:

$$\dot{x} = 2bX_o, \quad (5.8)$$

$$\dot{X}_o = -2H_o X_o, \quad (5.9)$$

satisfying condition

$$\min_{x(t)} P = P[x(\tau)] = 0. \quad (5.10)$$

Condition (5.10) is a dynamic constraint, which is imposed on the solutions (5.2), (5.3) at some set of the functional's field $Q \in KC(\Delta, R^n)$, where the following relations hold:

$$Q^o \subset Q,\ Q^o = R^n \times \Delta^o, \Delta^o = [0,\tau], \tau = \{\tau_k\}, k = 1,...,m \quad (5.11)$$

for process $x(t)_{t=\tau} = x(\tau)$.

Hamiltonian

$$H_o = -\frac{\partial S_o}{\partial t} \quad (5.12)$$

is defined for the function of action $S_o(t,x)$, which on the extremals (5.8,5.9) satisfies the condition

$$\min(-\partial \tilde{S}/\partial t) = -\partial \tilde{S}_o/\partial t. \quad (5.13)$$

Hamiltonian (5.6) and eq. (5.8) determine a second order differential equation of extremals:

$$\ddot{x} = \dot{x}[\dot{b}b^{-1} - 2H]. \quad (5.14)$$

*Proof*. Using (5.4) and (5.6), we find equation for Lagrangian in (5.1) in the form

$$L=-b\frac{\partial X}{\partial x}-1/2\dot{x}^T(2b)^{-1}\dot{x}\ . \qquad (5.15)$$

On extremals $x_t = x(t)$ (5.2, 5.3), both $a^u$ and $b$ (in 1.1, 1.6) are nonrandom.

After their substitution to (5.1) we get the integral functional $\tilde{S}$ on the extremals:

$$\tilde{S}[x(t)]=\int_s^T 1/2(a^u)^T(2b)^{-1}a^u dt\ , \qquad (5.15a)$$

which should satisfy the variation conditions (5.1a), or

$$\tilde{S}[x(t)]=S_o[x(t)], \qquad (5.15b)$$

where both integrals are determined on the same extremals.

From (5.15), (5.15a,b) it follows

$$L_o=1/2(a^u)^T(2b)^{-1}a^u\text{, or }\ L_o=\dot{x}^T(2b)^{-1}\dot{x}\ . \qquad (5.16)$$

Both expressions for Lagrangian (5.15) and (5.16) coincide on the extremals, where potential (5.7) satisfies condition (5.10) in the form

$$P_o=P[x(t)]=(a^u)^T(2b)^{-1}a^u+b^T\frac{\partial X_o}{\partial x}=0\,, \qquad (5.17)$$

for both Hamiltonian (5.12) and function of action $S_o(t,x)$ according to (5.13).

From (5.15b) it also follows

$$E\{\tilde{S}[x(t)]\}=\tilde{S}[x(t)]=S_o[x(t)]\,. \qquad (5.17a)$$

Applying Lagrangian (5.16) to Lagrange equation

$$\frac{\partial L_o}{\partial \dot{x}}=X_o\,, \qquad (5.17b)$$

we get equations for vector

$$X_o=(2b)^{-1}a^u \qquad (5.17c)$$

on extremals (5.8).

The JH solution for the EF determines both the entropy Lagrangian and Hamiltonian.

Lagrangian (5.16) satisfies the principle maximum [52] for functional (5.15), from which also follows (5.17a). Functional (5.1) reaches its minimum on extremals (5.8), while on the extremals (5.2), (5.3) this functional reaches some extreme values corresponding Hamiltonian (5.6).

This Hamiltonian, at satisfaction of (5.17), reaches its minimum:

$$\min H=\min[V+P]=1/2(a^u)^T(2b)^{-1}a^u=H_o\,, \qquad (5.18)$$

from which it follows

$$V=H_o \qquad (5.19a)$$

at $\min_{x(t)} P = P[x(\tau)] = 0$ . (5.19b)

Function $-\partial \tilde{S}(t,x)/\partial t = H$ in (5.6) on extremals (5.2,5.3) reaches a *maximum* when the constraint (5.10) is not imposed.

Both the minimum and maximum are conditional with respect to the constraint imposition.

The variation conditions (5.18), imposing constraint (5.10), selects Hamiltonian

$$H_o = -\frac{\partial S_o}{\partial t} = 1/2(a^u)^T(2b)^{-1}a^u \quad (5.20)$$

on the extremals (5.2,5.3) at discrete moments $(\tau_k)$ (5.11).

The variation principle identifies two Hamiltonians: $H$ satisfying (5.6) with function of action $S(t,x)$, and $H_o$ (5.20), whose function action $S_o(t,x)$ reaches absolute minimum at the moments $(\tau_k)$ (5.11) of imposing constraint $P_o = P_o[x(\tau)]$. After substituting (5.2) and (5.17b) in (5.16) and (5.20) both Lagrangian and Hamiltonian on the extremals take form:

$$L_o(x, X_o) = 1/2\dot{x}^T X_o = H_o . \quad (5.21)$$

Using $\dot{X}_o = -\partial H_o / \partial x$, we have $\dot{X}_o = -\partial H_o / \partial x = -1/2\dot{x}^T \partial X_o / \partial x$,

and from constraint (5.10), we get

$$\partial X_o / \partial x = -b^{-1}\dot{x}^T X_o \text{, and } \partial H_o / \partial x = 1/2\dot{x}^T b^{-1}\dot{x}^T X_o = 2H_o X_o , \quad (5.21a)$$

which after substituting (5.17b) leads to extremals (5.9).

Using eq. for the conjugate vector (5.3), we write constraint (5.10) in the form

$$\frac{\partial X_o}{\partial x} = -2X_o X_o^T , \quad (5.21b)$$

which follows from (5.7), (5.8) and (5.17c).

By differentiating (5.8) we get a second order differential eq on the extremals:

$\ddot{x} = 2b\dot{X}_o + 2\dot{b}X_o$, which after substituting (5.9) leads to

$$\ddot{x} = 2X_o[\dot{b} - 2bH], \quad (5.21c)$$

or to (5.14). •

At imposing constraint (5.17), we get the following relations

$$H(x(\tau)) = H_o , \tilde{S}[x(t)] = S_o[x(t)] , \quad (5.23)$$

$$S_o[x(t)] = E[\bar{\varphi}_s^t] = \bar{\varphi}_s^t[x(t)] , E\{\tilde{S}[x(t)]\} = \tilde{S}[x(t)], \quad (5.23a)$$

which are satisfied on extremals (5.8),(5.9) with additive functional $\bar{\varphi}_s^t[x(t)]$ (1.4, 3.3).

The variation principle leads to dynamic forms extremal trajectories (5.1), (5.23, 5.23a) of the EF. (The detailed proofs are in [41,58]).

Connection of the wave function and probability (Sec.3) with functions (3.13),( 3.12) and functionals (5.23, 5.23a) follows from relations:

$|\bar{Q}|^2 = \exp 2\tilde{Q} = p$ , *at* $\tilde{Q} = \mathrm{Re}|\exp j\lambda\tilde{\Psi}|$

For that $\tilde{Q}$ , we get

$$\tilde{Q} = -1/2\bar{\varphi}_s^t \ , \qquad (5.24)$$

where at real $\tilde{\Psi}$ , $\tilde{Q}$ in (5.24), real $p$ (in (3.9), (3.9b), and (3.13)) corresponds to real $\bar{\varphi}_s^t$. The real functional relations:

$$\exp\bar{Q} = -1/2\bar{\varphi}_s^t, |\bar{Q}|^2 = \exp 2\tilde{Q}, \tilde{Q} = \mathrm{Re}[\exp j\lambda\tilde{\Psi}], \text{ and } j\lambda\tilde{\Psi} = -1/2\bar{\varphi}_s^t, \quad (5.24a)$$

with real $\bar{\varphi}_s^t$, and real $\tilde{\Psi}$ require also real $j\lambda = c$ .

In particular at $c = 1, \lambda = -j$ , we have

$$\tilde{\Psi} = -1/2\bar{\varphi}_s^t, \ \psi = V_\psi = V . \qquad (5.25)$$

Using (5.23, 5.23a) we have probability functional on trajectories:

$$p = p_o = \exp(-\bar{\varphi}_s^t[x(t)]), \qquad (5.26)$$

and applying (5.25) we get equation for function of action, related to that in (5.22):

$$-\frac{\partial \tilde{S}_o}{\partial t} = V(t, x(t)) = H_o. \qquad (5.27)$$

*Proposition 5.2.*

Let us consider diffusion process $\tilde{x}(s,t)$ at a locality of states $x(\tau_k - o)$, $x(\tau_k)$ , $x(\tau_k + o)$ , formed by the impulse control's cutoff action (Sec. 2). The process is cutting off *after* each moment $t \le \tau_k - o$ *-at* $t > \tau_k$ and each moment $t \ge \tau_k + o$ *following* the cut-off, where $(\tau_k - o) < \tau_k < (\tau_k + o)$ .

Since the additive and multiplicative functionals (Sec.2) satisfy eqs (2.4a,b), (2.7) at these moments, the constraint (5.17, 5.10) acquires operator form $\tilde{\tilde{L}}$ in equation

$$-\frac{\partial \Delta\tilde{S}}{\partial s} = \tilde{\tilde{L}}\Delta\tilde{S}\,, \Delta\tilde{S}(s,t) = \begin{cases} 0, t \le \tau_k - o; \\ \infty, t > \tau_k; \end{cases} \qquad (5.28)$$

which at $\Delta\tilde{S}(s, t \le \tau_k - o) = 0$ satisfies equation

$$\tilde{\tilde{L}} = (a^u)^T \frac{\partial}{\partial x} + b\frac{\partial^2}{\partial x^2} = 0. \qquad (5.29)$$

The *proof* applies [30], where it is shown that $\Delta\tilde{S}(s,t \le \tau_k - o) = E_{s,x}[\varphi_s^{t-}]$ satisfies eq (5.28) with operator $\tilde{\tilde{L}}$ in (5.29), which is connected with operator $\tilde{L}$ of the initial Kolmogorov eq. (3.5) by relation

$$\tilde{\tilde{L}} = \tilde{L} - 1/2(a^u)^T(2b)^{-1}a^u . \quad (5.29a)$$

From these relations, at completion of (5.28), we get (5.29) and then

$$E_{s,x}[(a^u)^T \frac{\partial\Delta\tilde{S}}{\partial x} + b\frac{\partial^2\Delta\tilde{S}}{\partial x^2}] = 0, \quad (5.30)$$

where

$$\Delta\tilde{S}(s,t \le \tau_k - o) = E_{s,x}[\varphi_s^{t-}] = S_-$$

is the process' entropy functional, taken before the moment of cutting-off, when constraint (5.17) is still imposed. •

From Props. 5.1 and 5.2 it follows that impulse control's cutoff action implements the VP at the *locality* of these states in the form of *maxmin* and *minimax*, depending on the impulse's step-down and step-up actions accordingly (sec.2).

From [30] it also follows that solutions of (5.30) allow classifying the states $x(\tau) = \{x(\tau_k)\}, k = 1,...,m$, considered to be the *boundary* points of a diffusion process at $\lim_{t\to\tau} \tilde{x}(t) = x(\tau)$.

A boundary point $x_\tau = x(\tau)$ *attracts* only if the function

$$R(x) = \exp\{-\int_{x_o}^{x} a^u(y)b^{-1}(y)dy\}, \quad (5.31)$$

defining the general solutions of (5.30), is integrable at a locality of $x = x_\tau$, satisfying the condition

$$|\int_{x_o}^{x_\tau} R(x)dx| < \infty . \quad (5.32)$$

A boundary point *repels* if (5.31) does not have the limited solutions at this locality**;** it means that eq. (5.31) is not integrable in this locality. •

The boundary attracting dynamic states carry *hidden dynamic* connections between the process' cut-off correlating states.

**6. The operator form of the information wave function equation**

Eq (3.3), written in the operator form:

$$-\frac{\partial \tilde{u}}{\partial t} = \tilde{L}\tilde{u} - V\tilde{u}, \qquad (6.1)$$

where

$$\tilde{L} = a(t,x)\frac{\partial}{\partial x} + b(t,x)\frac{\partial^2}{\partial x^2} \qquad (6.1a)$$

on extremals (5.2), (6.1) holds

$$-\frac{\partial \tilde{u}_o}{\partial t} = \tilde{L}\tilde{u}_o - V\tilde{u}_o. \qquad (6.2)$$

On extremals (5.8,5.9), at condition (5.19a), with imposing constraint (5.17), this equation satisfies

$$\tilde{L}\tilde{u}_o = 0, \qquad (6.2a)$$

which brings (6.2) to form

$$\frac{\partial \tilde{u}_o}{\partial t} = H_o\tilde{u}_o. \qquad (6.3)$$

Using (6.1), eq (5.6) acquires operator form:

$$-\frac{\partial \tilde{S}}{\partial t} = \tilde{L}\tilde{S} + V, \qquad (6.4)$$

which on the extremals, satisfying constraint $\tilde{L}\tilde{S}_o = 0$, leads to

$$-\frac{\partial \tilde{S}_o}{\partial t} = H_o. \qquad (6.5)$$

In the Kolmogorov equation for function $\tilde{u}_\phi(t,x,\lambda)$ (3.7a), real function $\tilde{\psi}(t,x)$, at $\tilde{\psi}(t,x) = V$ and condition (5.19a), is connected to Hamiltonian (6.5) in form:

$$\tilde{\psi}(t,x) = H_o. \qquad (6.6)$$

Then *equation for wave function $\tilde{u}_\phi$ (3.7a) with information Hamiltonian* $H_o$ holds Schrödinger's information form:

$$\frac{\partial \tilde{u}_\phi}{\partial t} = a(t,x)\frac{\partial \tilde{u}_\phi}{\partial x} + b(t,x)\frac{\partial^2 \tilde{u}_\phi}{\partial x^2} - j\lambda H_o\tilde{u}_\phi. \qquad (6.7)$$

On the extremals (5.8,5.9), where

$$\tilde{L}\tilde{u}_\phi = 0$$

Eq.(6.7) acquires the form

$$\frac{\partial \tilde{u}_{\phi o}}{\partial t} = -j\lambda H_o\tilde{u}_{\phi o}. \qquad (6.8)$$

Wave function $\tilde{u}_{\phi}$ *is defined along a trajectory for which* function action $S(t,x)$ satisfies eq (5.6), while wave function $\tilde{u}_{\phi o}$ is defined along extremals (5.8,5.9) with function of action $S_o(t,x)$ minimizing the entropy functional (5.15a).

Let's consider Eqs. (6.7, 6.8) at a fixed $\lambda^o$ having physical meaning of maximal frequency $[\nu_{\max}] = \sec^{-1}$ for energy spectrum for information wave $\tilde{u}_{\psi o}(t,x,\lambda^o)$: $\nu_{\max} \cong 2.82k\Theta / h$ where $h$ is Plank constant and $\Theta$ is absolute temperature.

The related information frequency $\lambda^o$ (in $[\lambda^o] = Nats/\sec$) is equal to

$$\lambda^o = \nu_{\max} / Nats = 2.82 / hNats = \hat{h}^{-1}, \qquad (6.9)$$

where $\hat{h}$ is an information equivalent of Plank constant. At a room temperature, this equivalent evaluates

$$\hat{h} \cong 0.5643 \bullet 10^{-15}. \qquad (6.9a)$$

<u>*Proposition 6.1.*</u>

Eqs (6.8) for $\tilde{u}_{\psi o}(t,x,\lambda^o)$ at $\lambda = \lambda^o = \hat{h}^{-1}$ acquires form

$$\frac{\partial \tilde{u}_{\phi o}}{\partial t} = -j\hat{h}^{-1} H_o \tilde{u}_{\phi o} \ . \qquad (6.10)$$

Imposing constraint (5.10) at each of extremal's $x_t = x(t)$ instant $t = \tau_k$ determines $H_o(\tau_k)$.

The imposing constraint leads to following *conditions* for *the entropy forces, Hamiltonian, function action, and wave function* of this dynamic model:

**1**. For the model's conjugated vector of entropy forces $X_o = (X_{oi}(t), X_{ok}(t)), i,k = 1,...n$, the condition holds

$$X_{oi}(\tau_k) X_{ok}(\tau_k) = (X_{oi}(\tau_k))^2 = (X_{ok}(\tau_k))^2 \ , \qquad (6.11)$$

while its complex conjugated components: $X_{oi} = \operatorname{Re} X_{oi} + j \operatorname{Im} X_{oi}, X_{ok} = \operatorname{Re} X_{ok} - j \operatorname{Im} X_{oi}$,

$$\operatorname{Re} X_{oi} = \operatorname{Re} X_{ok}, \operatorname{Im} X_{oi} = \operatorname{Im} X_{ok} \ , \qquad (6.11a)$$

at the moment $t = \tau_k$ of applying the constraint take real values

$$X_{oi}(\tau_k) = \operatorname{Re} X_{oi}(\tau_k), X_{ok}(\tau_k) = \operatorname{Re} X_{ok}(\tau_k) \ . \qquad (6.11b)$$

**2.** The model's additive components of the entropy Hamiltonians $H_o = (H_{oi}(t), H_{ok}(t))$, defined on complex conjugated extremals $x_i(t)$ and $x_k(t)$, coincide at $t = \tau_k$:

$$| H_{oi}(\tau_k) | = | H_{ok}(\tau_k) | , \qquad (6.12)$$

where $H_{oi}(\tau_k) = \operatorname{Re} H_o(\tau)$ is real and $H_{ok}(\tau_k) = \operatorname{Im} H_o(\tau)$ is imaginary component of $H_o(\tau)$.

**3**. For the components of function of action $S_o = (S_{oi}(t), S_{ok}(t)), i, k = 1,...n$ the condition is

$$S_{oi}(\tau_k) S_{ok}(\tau_k) = S_{oi}^{ab}(\tau_k) = S_{ok}(\tau_k) S_{oi}(\tau_k),, \qquad (6.13)$$

where $S_{oi} = \operatorname{Re} S_{oi} + j \operatorname{Im} S_{oi}$, $S_{ok} = \operatorname{Re} S_{ok} - j \operatorname{Im} S_{ok}$, $S_{oi}^{ab}(\tau_k)$ satisfies (3.20).

**4**. For complex conjugated wave functions $\tilde{u}_{\phi oi}(x_i(t)), \tilde{u}_{\phi ok}(x_k(t))$ the condition is

$$\tilde{u}_{\phi oi}(x_i(\tau_k)) \tilde{u}_{\phi ok}((x_k(\tau_k)) = [\tilde{u}_{\phi oi}(x_i(\tau_k)]^2 = [\tilde{u}_{\phi ok}(x_k(\tau_k)]^2 = \tilde{u}_{\phi oi}(x_i(\tau_k) \tilde{u}_{\phi oi}(x_i(\tau_k)) \,. \qquad (6.14)$$

*Proofs.*

**1**. Since the constraint for model's conjugated vector $X_o = (X_{oi}, X_{ok}), i, k = 1,...n$ has form (5.22a):

$$\partial X_{oi} / \partial x_k = -2 X_{oi} X_{ok} = -2 X_{ok} X_{oi} = \partial X_{ok} / \partial x_i \,, \qquad (6.15)$$

its complex conjugated components (6.11a):

$$X_{oi} = \operatorname{Re} X_{oi} + j \operatorname{Im} X_{oi}, X_{ok} = \operatorname{Re} X_{ok} - j \operatorname{Im} X_{ok},$$

$$\operatorname{Re} X_{oi} = \operatorname{Re} X_{ok}, \operatorname{Im} X_{oi} = \operatorname{Im} X_{ok} \,, \qquad (6.16)$$

at moment $\tau_k$ of applying the constraint take real values (6.11b), and acquire the form (6.11).

**2.** Since both Hamiltonians $H_{oi}(t), H_{ok}(t)$ are invariants and additive on complex conjugated extremals $x_i(t)$ and $x_k(t)$, they coincide at $t = \tau_k$ ,which leads to (6. 12).

Therefore, both integrals of these Hamiltonians, taken during the same time, also coincide, leading to

$$| S_{oi}(\tau_k) | = | S_{ok}(\tau_k) | \,, \qquad (6.17)$$

which satisfy the condition (6.13)

**3**. The constraint (6.8), applied to (6.7) acquires the form

$$a(t,x) \frac{\partial \tilde{u}_\phi}{\partial x} = -b(t,x) \frac{\partial^2 \tilde{u}_\phi}{\partial x^2} \,, \qquad (6.18)$$

where wave function $\tilde{u}_\phi(x_t)$ is defined on the same extremals $x_t$ as function $S = S(x_t)$.

*Condition (6.18) corresponds superposition of regular and diffusion components of the wave function.*

Since that, at moment $t = \tau_k$ of imposing the constraint, the relations for the derivations of wave function $\tilde{u}_{\phi o}(x_t)$ hold the form analogous to (6.11):

$$\frac{\partial \tilde{u}_{\phi oi}}{\partial x_i}(\tau_k)\frac{\partial \tilde{u}_{\phi ok}}{\partial x_k}(\tau_k)=[\frac{\partial \tilde{u}_{\phi oi}}{\partial x_i}(\tau_k)]^2=[\frac{\partial \tilde{u}_{\phi ok}}{\partial x_k}(\tau_k)]^2=\frac{\partial \tilde{u}_{\phi ok}}{\partial x_k}(\tau_k)\frac{\partial \tilde{u}_{\phi oi}}{\partial x_i}(\tau_k). \quad (6.19a)$$

Writing functions

$$\tilde{u}_{\phi oi}=\int_{\tau_k}\frac{\partial \tilde{u}_{\phi oi}}{\partial x_i}\partial x_i(\tau),\tilde{u}_{\phi ok}=\int_{\tau_k}\frac{\partial \tilde{u}_{\phi oi}}{\partial x_k}\partial x_k(\tau) \quad (6.19b)$$

through the derivations, satisfying (6.16c) and (6.16d) accordingly, we get ,

$$\tilde{u}_{\phi oi}=\int_{\tau_k}\frac{\partial \tilde{u}_{\phi oi}}{\partial x_i}\dot{x}_i\partial(\tau),\tilde{u}_{\phi ok}=\int_{\tau_k}\frac{\partial \tilde{u}_{\phi ok}}{\partial x_k}\dot{x}_k\partial(\tau), \quad (6.20)$$

where

$$\frac{\partial \tilde{u}_{\phi oi}}{\partial x_i}\dot{x}_i=H_{\phi oi},\frac{\partial \tilde{u}_{\phi ok}}{\partial x_k}\dot{x}_k=H_{\phi ok} \quad (6.20a)$$

are the complex conjugated Hamiltonians for $\tilde{u}_{\phi oi}(x_i(t)),\tilde{u}_{\phi ok}(x_k(t))$, defined on the same extremals with the functions of actions $S_{oi}=S_{oi}(x_i(\tau_k)),S_{ok}=S_{ok}(x_k(\tau_k)$.

Following (6.12), Hamiltonians (6.20a) coincide at moment $t=\tau_k$ imposing the constraint, which leads to equality

$$\tilde{u}_{\phi oi}(x_i(\tau_k))=\tilde{u}_{\phi ok}(x_k(\tau_k)), \quad (6.21)$$

and relations (6.14).

Satisfaction (6.14) corresponds *entangling* wave functions $\tilde{u}_{\phi oi}(x_i(t)),\tilde{u}_{\phi ok}(x_k(t))$ at imposing the constraint.

Thus, Eq.(6.10) holds the conditions of superposition and entanglement (at the moment $\tau=\tau_k$ when (5.10) is satisfied).•

The entanglement, at imposing the constraint, follows from the conditions of *minimizing the entropy functional* on extremals (5.8), (5.9) and (5.14).

The minimal extremals (as solutions of (5.12)), with the related functions of actions, probabilities (3.1), and wave functions $\tilde{u}_\phi$, start at the moment of imposing the constraint. That holds a minimal Markov path.

A minimal path from Markov probability $\tilde{P}(\tau,\tilde{x}_\tau)$ to $\tilde{P}(t,\tilde{x}_t),t>\tau$ along the minimal Markov diffusion process is a *Schrödinger's* process that holds a mixture of Brownian bridges [14].

Following *Schrödinger's* reversal natural law: the bridge from probability $\tilde{P}(\tau,\tilde{x}_\tau)$ to $\tilde{P}(t,\tilde{x}_t),t>\tau$ is just the reversal of *Schrödinger's* bridge from probability $\tilde{P}(t,\tilde{x}_t)$ to $\tilde{P}(\tau,\tilde{x}_\tau)$, allows us to express the bridge's probability density (3.8), (3.13a), (3.16) via the above quantum conjugated wave functions in the form

$$p = \tilde{u}_{\psi o}(t,x)\tilde{u}_{\psi o}^{*}(t,x), \qquad (6.22)$$

which relates to that in [31].

The *Schrödinger's* process (Secs. 3,4) with the entanglement, at conditions (3.24), minimizes entropy functional (5.1a) via minimization of math expectation of additive functional (1.7, 3.4), while probability density (6.22) is also expressed by additive functional (5.26).

The minimum of entropy functional (1.5) [53] leads to condition (3.24), from which follows the Schrödinger's bridge condition (3.23).

Since the jump action on Markov process (sec.2), leading to killing its drift, selects Feller's measure of the kernel on the transition to Brownian diffusion, this cutoff of the EF provides *information measure of the Feller transitional kernel.*

The same cutoff actions on the Schrödinger Brownian's bridges, measured by the additive functional of the distributed Markov diffusion process, will select the Feller transitional kernel's *information from an entire current Markov movement.*

These results relate to other significant publications in this field [32],[19], [21] ,[10], [12], [13].

## 7. Analysis of information processes with the concurrent Markov diffusion and the equivalent quantum information process. The information of Feller's kernel and Schrödinger's bridge

The transformed Markov diffusion process builds the quantum dynamic process as its information equivalent, while both processes have the equivalent probability's densities, measured by the same entropy functional on the trajectories of these processes, and start simultaneously under the same cutoff controls.

The impulse control cuts the fraction of the Markov process, which encloses the operator, transforming Markov process (with a finite drift) to Brownian motion (with zero drift and the diffusion equivalent for both of them).

Since such transformation holds a Feller's kernel operator, the quantity of information, automatically selected from the Markov process by the cutoff control and measured by entropy functional during the cutoff, evaluates the kernel information (sec.2).

The Markovian equivalent of quantum process evaluates the Schrödinger's bridge information (4.1a, b), which coincides with Feller's kernel information measure of the entropy functional before cutting a potential Brownian –kernel motion on interval $\Delta_1$.

Thus, Feller's kernel on the Markov process *or* Schrödinger's bridge on the related quantum process encloses the same entropy $\tilde{S}^{1o}_{ab} = |\ln 2|$ for a stable bridge *or* a stable kernel. This quantity is released by the cutoff action of the impulse control, which spends information $\tilde{H}^{1o}_{ab} = \ln 2$ on this transformation by killing the Markov Feller's kernel, or the quantum Schrödinger's bridge.
We assume that the cutoff of the process' correlation provides such information from the external process as an outside source, for example, by some control device which measures this information.
The entropy, taken from the Markov diffusion by the cutoff control impulse, accounts only 0.5/0.7=71.43% of both Schrödinger's bridge and the total kernel entropy $|\ln 2| = 1 bits$.
Since the kernel covers an irreversible transformation of the Markov diffusion to Brownian motion, its time interval $\Delta_1$ holds a unit of an *irreversible time* interval.
Time interval $\delta_{imp}$ of the cutting control is less than the total kernel's time interval $\Delta_1$ in the same ratio $\delta_{imp} / \Delta_1 = 0.5 / \ln 2$, when $\delta_{imp}$ evaluates only a part of that irreversible unit.
Interval $\delta_{imp}$ estimates *a minimum of a maximal time interval* $\Delta_1$***, or a maximum of its minimal information amount since*** such a cutoff implements the minimax variation principle.
Thus, cutoff interval $\delta_{imp}$ evaluates a minimal irreversible time course of the Markov diffusion, which is a model of more broad-spectrum irreversible processes [55, 56].
For the Brownian path to Quantum Bridge with its entangled wave functions, time interval $\Delta_1$ evaluates its reversible time interval.
Cutting the kernel deletes Markov's irreversibility, while cutting the bridge deletes quantum reversibility. Interval between the nearest impulses $\Delta_{1x}$ evaluates a *minimal time-delay*, as an *optimal waiting time* between delivering new information.
In $m$-dimensional part of the process, at other current instants $\tau^1 = (\tau_1^1, \tau_1^2, \tau_1^3, ..., \tau_1^m)$, another related control emerges, and the situation is repeating along a time course the processes.
According to eq. (6.9), $\lambda^o$ has a physical meaning of maximal frequency for energy spectrum, which measures information invariant of Plank constant (6.9a). Applying $\hat{h}$ we estimate a maximum of minimal delay $\Delta_{1mx} = \max \Delta_{1x}$. This interval estimates ratio of the $\Delta_{1x}$ entropy measure $\ln 2$ to a *minimal* real eigenvalue $\lambda_\tau^{om} = 2\alpha_{\min}$ (information speed in Nat/sec), generated at moment $t = \tau_k$, when the conjugated information functions of actions superimpose.
Thus, interval $\Delta_{1mx}$ preserves invariant quantity of information

$\mathbf{a}_o = \lambda_\tau^{om} \Delta_{1mx} = \ln 2$. (7.1)

Then we get

$\Delta_{1mx} = \ln 2 / \hat{h} \cong \ln 2 \times 1.772107035 \bullet 10^{15} \sec$, (7.2)

which estimates a maximal difference in the time between non-local entanglements.

The related *space distance* $L_1$ can be estimated using speed of light $c = 300 \times 10^3 km/\sec$:

$L_{1mx} = \ln 2 \times 1.772107035 \bullet 10^{15} \times 300 \bullet 10^3 \cong 3.72141735 \bullet 10^{20} km$. (7.3)

Since the impulse, cutting the Markov diffusion, produces minimal hidden information $0.5 Nats$, the impulse's stepwise controls, carrying information $0.2 Nats$, require an external source for getting that information.

If this information delivers an interactive impulse of Markov diffusion process, it should carry $0.7 Nat \cong \ln 2 Nats = 1 bit$, which is maxmin information, required for covering the kernel.

The minimal interval $\Delta_{1mn} = \delta_{imp} = 0.5 / 2\alpha_{\max}$, at $\alpha_{\max} \cong 0.9 Nats/\sec$ [41], brings

$\delta_{imp} \cong 0.5/1.8 = 0.277 \sec$.

This means, each $0.277 \sec$ the interactive Markov process is able to produce entropy $0.5$ Nats, which estimates a *unit of instance* generating an elementary *hidden* information.

However, the current time interval, as the interval of imposing the constraint, depends on the value of both invariant (6.9) and the process initial eigenvalues $\alpha_{io}$, whose spectrum is changing.

The impulse time interval between the step-down and step-up controls $\delta_{imp}$ are varied by the IPF application.

The diapason of such variations, at changing the initial eigenvalue for the model dimensions from $n_o = 2$ to $n = 22$, is: from $\Delta t_o = 1.0 \sec$ to $\Delta t_n = 0.0015 \sec$.

The information dynamics *originate* from a Bernstein-Markov process [22], at the step-up transformation from the Brownian movement to the Markov process, which provides a "transition kernel" [40] with a normalized Feynman-Kac measure.

The process, studied as a sub-markov process with Markov path, loops [44], is connected with Schrodinger bridge [31] as unique Markov process in the class of reciprocal processes.

This is specific "Euclidean quantum mechanics" *process,* whose dynamics hold probabilistic description of process' analytics [40].

The transition kernel of this process has a reversal density probability corresponding (3.23a), (6.22), which is represented by a product of the conjugated wave functions and related anti-symmetric information action functional (3.24b), (3.25a).

Such a process satisfies the *least action principle* [40] as an extension of the variation principle (sec.5), which we implement through the step-up control, imposing the dynamic constraint (5.10).

The least action principle holds the form of minimum functional (5.1a) (which in [45] corresponds the minimum of the time's forward action functional). It has also the form of an absolute minimum of a sum of the anti-symmetric action functionals, which satisfies Eq.(6.13) at each ending moment of the extremal movement.

Thus, the Bernstein-Markov process here is described in terms of information dynamics, whose impulse control's action, intervening in the Brownian movement, injects the quantity information (4.7), in the transition kernel during the above transformation.

Transforming the hidden information in Bernstein-Markov process gives the start to Informational Macrodynamics [41], which implement the variation principle (sec.5) on the trajectories of information path functional [59].

The considered dynamic model of the Markov process generates its concurrent interactive time interval $\delta_{imp}$, which determines the limited life time (7.3).

The eqs (secs. 3-6) describe the emerging kernel and bridge during each of the cutoff intervals.

The cutting off random process originates an *information observer*, whose impulse, cutting the minimax, acts as both a producer and consumer of information.

The observer interacts with an external observing stochastic process at close locality of the cutting edges of each impulse, which provides an interactive "window".

The information observer emerges through the impulse cutting information *concurrently* with transforming Markovian-Brownian diffusion and generating the quantum information dynamics that initiate a Schrödinger's bridge and entanglement within the impulse' cutting edges.

## REFERENCES TO PART I

[1]. C.E. Shannon, W. Weaver, *The Mathematical Theory of Communication*, Illinois Press, Urbana, 1949.

[2]. E.T. Jaynes, *Information Theory and Statistical Mechanics in Statistical Physics*, Benjamin, New York, 1963.

[3]. A.N. Kolmogorov, Logical basis for information theory and probability theory, *IEEE Trans. Inform. Theory,* **14** (5), 662–664, 1968.

[4]. S. Kullback, *Information Theory and Statistics*, Wiley and Sons, New York, 1959.

[5]. R. L.Stratonovich, *Theory of information*, Sov. Radio, Moscow, 1975.

[6]. W. Feller, The general diffusion operator and positivity preserving semi-groups in one dimension. *Ann Math.* **60**, 417-436, 1954.

[7]. W. Feller, On boundaries and lateral conditions for the Kolmogorov differential equations. *Ann. Math.* **65**, 527–570, 1957.

[8]. V.S. Lerner, The entropy functional measure on trajectories of a Markov diffusion process and its estimation by the process' cut off applying the impulse control, arXiv: 1204.5513.

[9]. V.S. Lerner, An observer's information dynamics: Acquisition of information and the origin of the cognitive dynamics, *Journal Information Sciences*, 184, 111-139, 2012.

[10]. N. Privaulta, J-C. Zambrini, Markovian bridges and reversible diffusion processes with jumps, *Ann. Inst. H. Poincaré, Probab. Statist.*, 40, 599–633, 2004.

[11]. M. Fukushima, P. He and J. Ying, Time changes of symmetric diffusions and Feller measures, *An. Probability*, **32**(4), 3138–3166, 2004.

[12]. J. Barral, N. Fournier, S. Jaffard and S. Seuret. A pure jump Markov process with a random singularity spectrum, *The Annals of Probability*, **38**(5), 1924–194, 2010.

[13]. P. He, Time change, jumping measure and Feller measure, *Osaka J. Math*., **4**4, 459–470, 2007.

[14]. H. P. McKean, Jr. and H. Tanaka, Additive functionals of the Brownian path, *Mem. Coll. Sei. Univ. Kyoto Ser. A Math*. **33**, 479-50, 1961.

[15]. E.Schrödinger, Uber die Umkehrung der Naturgesetze, *Sitzungsberichte der Preuss Akad*. Wissen, Berlin, Phys. Math. Klasse, 144-153, 1931.

[16]. M. Nagasawa, Transformations of diffusions and *Schrödinger* processes, *Prob. Th. Rel. Fields* 82, 109-136, 1989.

[17]. M. Nagasawa, *Schrödinger Equation and Diffusion Theory*, Birkhauser, Basel, 1993.

[18]. J. C. Zambrini, Variational processes and stochastic versions of mechanics, *J. Math. Phys*.,**27**, 2307, 1986.

[19]. N. Ikeda, S. Watanabe, *Stochastic Differential Equations and Diffusion Process*, College Press, Univ. of Beijing, 1998.

[20]. R.Z.Abi, *Schrödinger Diffusion Processes*, Birkhauser, Basel, 1996.

[21]. B.C.Levy and A.J.Krener, Dynamics and Kinematics and of reciprocal diffusions, *J.Math.Phys.*, **37**(2), 769-803, 1993.

[22]. A.B. Cruzeiro, Liming Wu and J.C. Zambrini. Bernstein processes associated with a Markov process, *Proceedings of the Third International Workshop on Stochastic Analysis and Mathematical Physics*: ANESTOC'98, Ed. R. Rebolledo, Trends in Mathematics, Birkhauser.

[23]. B.Z. Jamison, The Markov process of Schrödinger, *Z. Wahrsch. Verw. Gebiete* **32**, 323-331,1975.

[24]. P. Garbaczewski and R. Olkiewicz, Feynman-Kac kernels in Markovian representations of the Schrödinger interpolating dynamics, *J. Math. Phys.*,**37**, 732 , 1996.

[25]. A. Andrisani and N. C. Petroni, Markov processes and generalized Schrödinger equations, *J. Math. Phys.*, **52**, 113509, 2011.

[26]. E.B.Dynkin Additive functional of a Wiener process determined by stochastic integrals, *Teoria. Veroyat. i Primenenia*, **5** ,441-451, 1960.

[27]. A.D. Ventsel, Non-negative additive functionals of Markov processes, *Dokl. Akademi Nayk CCCP*, **137**, 17-20, 1961.

[28]. R.M Blumenthal, R.K. Getoor, Additive functionals of Markov processes in duality,*Trans. Am. Math.Soc.*, **112**( 1), 131-163,1964.

[29]. E.B.Dynkin,*Theory of Markov Processes*, Pergamon Press, New York, 1960.

[30]. Y.V Prokhorov, Y.A Rozanov, *Theory Probabilities*, Nauka, Moscow, 1973.

[31]. M. Pavon, Quantum Schrödinger bridges, arXiv: 0306052.

[32]. H. Follmer and N. Gantert, Entropy minimization and Schrodinger processes in infinite dimensions, *Annals of Probability,* **25**(2), 901-926,1997.

[33]. J. M. Borwein, A.S. Lewis and R.Z.Nussbaum, Entropy minimization, DAD problems, and doubly stochastic kernels. *J. Funct. Anal.* **123**, 264-307, 1994.

[34]. C.Leonard, Minimization of entropy functional, arXiv: 0710.1462.

[35]. T.Yu and J. H. Eberly, Finite-time disentanglement via spontaneous emission, *Phys Rev Lett*., **93** (14), 2004.

[36]. T. Yu and J. H. Eberly Negative Entanglement Measure, and What It Implies, arXiv: 0703083.

[37]. M.B. Plenio, S.Virmani, An introduction to Entanglement Measures, arXiv*:* 0504163.

[38]. K. Ann, G. Jaeger, Finite-time destruction of entanglement and non-locality by environmental influences, *arXiv*/ quant-ph/0903.0009, 2009.

[39]. M. Kac, *Probability and Related Topics in Physical Sciences*, Boulder, Colorado, 1957.

[40]. S.Albeverio, K. Yasue, and J. C. Zambrini, Euclidean quantum mechanics: analytical approach, *Ann.Inst H. Poincare*, Sec A, **50** (3), 259-308, 1989.

[41]. V.S. Lerner, *Information Path Functional and Informational Macrodynamics,* Nova Sc.Publ., New York, 2010.

[42]. John Von Neumann, *Mathematical Foundations of Quantum Mechanics*, Princeton University Press, 1996.

[43]. P.Sancho,l.Plaja, Entanglement of unstable atoms: modifications of the emission properties, *J. Phys. B: At. Mol. Opt. Phys.* **42**, 165008 , 2009.

[44]. Yves Le Jan, Markov paths, loops and fields, *Lecture Notes in Mathematics,* vol. 2026, Springer, Heidelberg, 2011.

[45]. R.P.Feynman, *The Feynman Lectures on Computation,* Eds A. J. G. Hey and R. W. Allen, Addison-Wesley, 1996.

[46]. C.H.Bennett, E.Bernstein, G.Brassard, U. Vazirani, The strengths and weaknesses of quantum computation, *SIAM Journal on Computing,* **26**(5): 1510–1523, 1997.

[47]. R.P. Poplavskii, Thermodynamical models of information processing, *Uspekhi Acad. Nauk*, 115 (3): 465–501, 1975.

[48]. P. Manzelli, Entanglement Theory, *The General Science Journal*, 9, 2007.

[49]. E.P. Wigner, The problem of measurement. *Am. J. Phys*., **31**, 6-15, 1963.

[50]. V.S. Lerner, *Variation principle in Informational Macrodynamics*, Kluwer Acad. Publs., Boston, 2003.

[51]. I.M. Gelfand, S.V.Fomin, *Calculus of Variations* , Prentice Hall, New York,1962.

[52]. V.M. Alekseev, V.M.Tichomirov, S.V. Fomin, *Optimal Control*, Nauka, Moscow, 1979.

[53]. V.S.Lerner, Solution to the variation problem for information path functional of a controlled random process functional, *Journal of Mathematical Analysis and Applications*, **334**,441-466, 2007.

[54].A.A.Yushkevich, Continuous-time Markov decision deterministic process with intervention, *Stochastics*, **9**,235-274,1983.

[55]. R.L. Dobrushin, Yu.M. Sukhov, I. Fritts, A.N. Kolmogorov-founder of the theory of reversible Markov processes, *Uspekhi Mat. Nauk.*, **43** (6)(264), 167–188, 1988.

[56]. L. Henderson and V. Vedral, Information, Relative Entropy of Entanglement, and Irreversibility, *Phys. Rev. Letts*, 84(10),2000.

[57]. F.B.Handson, *Applied Stochastic Processes and control for jump-diffusions: modeling, Analysis and computation*, Univ. of Illinois, Chicago, 2006.

[58]. V.S.Lerner, Dynamic approximation of a random information functional, *J. Mathematical Analysis and Applications*, 327(1), 494-514, 2007.

[59].V.S. Lerner, The information path functional approach for a dynamic modeling of a controllable stochastic process, *arXiv*: 1201.0035, 2011.

## PART II. APPLICATION TO INFORMATION OBSERVER

### 1. PRINCIPLES OF OBSERVATION AND OBSERVER

A physical approach to the observer, developed the Copenhagen's interpretation of quantum mechanics [1-5], requires an act of observation, as a physical carrier of the observer knowledge. But this observer's role does not describe the formalism of quantum mechanics. As N.Bohr believed, *probability itself a fundamental nature of reality*.

J.A. Wheeler and R. Feynman developed the time-symmetric direct interparticle action theory [6] referring to all phenomena captured by classical electrodynamics [7] that "the curves of action and reaction cross".(All these references below apply only to the introduction).

According to D. Bohm ontological interpretation of quantum physics [8]: physical processes are determined by information, which "is a difference of form that makes a difference of content, i.e., meaning". Bohm believed that "meaning unfolds into intention, intention into actions"; and "intention generally arises out of a previous perception of meaning or significance of a certain total situation". "That observer entails mental processes".

The quantum approach of J. C. Eccles [9] "is to find a way for the 'self' to control its brain."

J.A. Wheeler introduces physical theory [10-14] of information-theoretic origin of an observer. Wheeler hypothesized that the Bit participates in the origin of all physical processes. Summarizing his physical theory [9-14] of an Observer-Participator, he introduced the doctrine "It from Bit". In his memoir [13], J.A.Wheeler divided his life into three themes or periods which reflect the historical development of modern physics. The first period he called "everything is particles." In the second period, "everything is fields." And in the third period, "everything is Information." These three stages represent an increasing generality of worldview."

But Wheeler's theory does not explain *how* the Bit self-creates.

Previously, many physicists [1-14], including Einstein [15], Penrose [16], and others, defined the Observer as having a separate and *physical* origin.

The problem of probability in quantum mechanics, writes Weinberg [17], "is that in quantum mechanics, the way that wave functions change with time is governed by an equation, the Schrödinger equation, *that does not involve probabilities*. It is just as deterministic as Newton's equations of motion and gravitation... So if we regard the whole process of measurement as being governed by the equations of quantum mechanics, and these equations are perfectly deterministic, how do probabilities get into quantum mechanics?"

D. Tong [18] argues that Quantum Fields are the real building blocks of the universe. The origin of physical particles is the natural probabilities of the vacuum. The comparative review of Wheeler theory and contemporary physics [19] shows that "Everything is From Field". Due to the quantum origins, "The elementary act of observer-participatorship transcends the category of time (delayed-choice double slit)" [19].

Still problem consists in unification classical and quantum physics.

But since Information originates in quantum processes, its study should focus not on the physics of the observing process's interacting particles, but on its Information-theoretical essence.

That leads to possibility of such unification using *Information formalism.*

A.N. Kolmogorov [20] established Probability Theory as the foundation of Information Theory and logic. Kolmogorov defined *random* simply as "the absence of periodicity" [21, p. 664].

C.E. Shannon's *Mathematical Theory of Communication* [22] measures relative entropy, which applies to the random states of an Information process. Kullback-Leibler's divergence [23], also known as relative entropy, measures the relative Information connections between the states of an observed process. The probabilistic origin of Information is well established [21-23, others], along with its unit, the Bit.
There are many studies of Information mechanisms employing various physical phenomena to account for intelligence. E. T. Jaynes [24] applied Bayesian probabilities to propose a plausible reasoning mechanism [25] whose rules of deductive logic connects maximum Bayes Information (entropy) to human mental activities, as a subjective observer.
The Observer is an *interactant*, present in all phenomena.
Understanding all these starting with definition and role information developments in our Information Age have become a critical task for scientific researchers, technological and economic institutions**.**
Wikipedia defines Information though its universal action: "Information is any entity or form that provides the answer to a question of some kind or resolves uncertainty"[38].
The cited references, along with many others, studying information mechanisms in intelligence, explain these through various physical phenomena, whose specifics are still mostly unknown.
Science knows that interactions have built structure of the Universe as its fundamental phenomena.
There have been many studies these interactions specifics; however, no one approach has unified the study of all their common information origins, regularities, and differentiation.
The first approach unifying these studies was published in [27-29], and extended results were published in [30-37]. This unified approach focuses on observations as interactions producing the Observer itself.
The Information Observer emerges by observing a random interactive process.
The essence is the probabilistic tracing of interacting events, which an Information path measures.
During the observation, the uncertainty of the random interactive process is converted into certainty. Thus, certainty is a source of Information.

Any single certain interaction is a “Yes-No” action which identifies a Bit, the elementary unit of Information. Multiple observations generate the Bit-moving dynamics, or Informational dynamics.
Bits organize themselves in triplets, which logically self-organize and assemble an Informational network. In the process of assembling the network, the triplets merge and interact with each other. Triplet interactions are memorized and become nodes of the Informational network. Then, the nodes themselves organize logically. A sequence of the logically organized triplet nodes defines a code of the network. This code integrates and carries all prior observations in the emerging Information Observer.
The Information Observer emerges from probabilistic observation without any pre-existing physical law.
The observation includes physical processes interacting with energies of different qualities. Quality energy evaluates the level of its order (disorder) or symmetry (asymmetry), which measures minimal entropy ln2 equivalent to Bit [27]. Each process high quality compensates for entropy of lesser quality. That erases the symmetrical process’ reversible logic equivalent to ln2, bringing the asymmetrical information logic. The physical erasure of the entropy by observation creates a certainty. Transferring entropy during interaction unifies multiple physical observations.

The well-known Shannon approach defines entropy as probability measures of the uncertainty of the observation. If the entropy of the observation decreases, uncertainty disappears, instead appearing as an equal certainty. Revealing certainty from uncertainty is the scientific path which determines the facts of reality.
Expressed as a mathematical formalism, the Bit evolves from the abstract probability of the observation as an elementary observer itself. Every step of this approach is substantiated here through a unified formalism of mathematics and logic.
Shannon’s Communications Theory [22] shows the following:

1. Shannon H-entropy measures the set of probabilities of symbols of a message as a source of signals.
2. Maximum H is the most uncertain situation.
3. Minimum H-entropy measures the maximal probability.
4. Channel capacity measures its entropy.
5. H is maximized when H is equal to the Channel capacity entropy.
6. Encoding the source message in Bit equalizes the entropy of channel capacity with the entropy of the source (thereby maximizing the equal uncertainty). According to Landauer [26], encoding requires the expenditure of energy quantified by this maximal entropy. (Therefore, the energy of encoding erases the maximal entropy-uncertainty to zero, reaching maximal probability.)
7. "H=0 if and only when we are certain of the outcome does H vanish."
8. H measures the amount of Information bits encoding the entropy source of message.

These principles of Shannon's theory of communications agree with the main principle of our approach: Entropy, as measure of uncertainty, erases energy, converting it to the equal Information, measuring certainty.

The principles of our information approach, describing below, have published since 1972 [27-37].

The approach's main contribution is to extend these principles to any observing random process whose entropy and Information integrate the related path functionals. The integral encoding structures an observer of this information.

Shannon wrote [22]: "A physical system, or a mathematical model of a system which produces a sequence of symbols governed by a set of probabilities, is known as a stochastic process... Conversely, any stochastic process which produces a discrete sequence of symbols chosen from a finite set may be considered a discrete source... Stochastic processes of the type described above are known mathematically as discrete Markoff processes and have been extensively studied in the literature."

In our formal model of the observation, the impulse observation runs axiomatic probabilities of a random field, linking Kolmogorov 0-1 law and Markov process probabilities [20]. The field connects sets of possible and actual events with their probabilities. The field's energy covers actual events. This triad specifies the observation.

*The Kolmogorov 0-1 probabilities' act of observation generates the Markov process within the field.*

The Markov process models the arising observer's process, collecting observations which change its measure of probabilities similar to the sequence of *a priori-a posteriori* Bayes probabilities. These objective probabilities, being an immanent part of the process, virtually observe and measure the Markov correlation connecting states-events, discretely changing the entropy of correlation which generates probabilistic impulses.

Each such impulse virtually cuts the observing entropy-uncertainty hidden in the cutting correlation. The cutting entropy decreases the initial Markov entropy, and increases the entropy of the cutting impulse.

Such multiple interactions minimize the uncertainty of the Markov process and maximize the entropy of each subsequent observing impulse. That runs the minimax principle for each observing impulse along the Markov interactive impulses.

When the observing probability approaches 1, the impulse cutting entropy converts to Information. Overcoming the entropy-information gap, where the energies of different qualities interact, brings the information Bit. Such Bit is an information model of a physical unit.

During the observation, the merging impulse curves and rotates the interactive yes-no conjugated entropies of the microprocess. The entropy entanglement starts within the impulse time interval before its space forms, and ends at the beginning of the space during the reversible relative time interval $0.15625\pi$ part of impulse entropy measure $\pi$.

The opposite curvature, enclosing the entropy of the interacting impulses, lowers the potential energy of an external process performing function of logical Maxwell Demon. Delivering Landauer's energy [26] memorizes the physical Bit. Minimal energy creates the curvature of the Bit's geometry. The curvature brings ability of binding. That creates the Bit's Free Information, enabling Information attraction and binding at the natural encoding.

Sequential interactive cuts along the process integrate the cutoff Hidden Information in the Information macroprocess, with a irreversible time course. Each memorized Information binds the reversible microprocess within an impulse with the irreversible Information macroprocess along the multi-dimensional process. The impulse observation consecutively converts entropy to Information in the emerging Information observer, conveying Information causality, certain logic, and complexity. The curving interaction is the main information mechanism connecting the emerging information structures of Observer analogous to gravitation in physics.

A triplet, as the elementary unit of bound bits, interacting with other triplets, connects them in a macroprocess.

The physics of the information macroprocess describe irreversible thermodynamics of interacting particles.

Multiple interacting Bits self-organize the Information process, encoding Information causality, logic, and complexity. The trajectory of the observation process carries the wave function both probabilistic and certain, which self-builds the Information macrounits-triplets.

Macrounits logically self-organize Information Networks (IN), encoding the units in geometrical structures which enclose the triplet code.

Multiple INs bind their ending triplets, enclosing the Observer's Information cognition and intelligence.

The Observer cognition assembles common units through multiple attractions and resonances as it forms the IN triplet hierarchy. The maximal number of accepted triplet levels in multiple INs measures the Observer maximum comparative Information intelligence.

The intelligent Observer recognizes and encodes digital images in message transmission. Being self-reflective, this enables it to understand the meaning of the message. The cognitive logic self-controls the process encoding the intelligence in a double helix coding structure.

Integrating the process entropy in the Entropy Functional and the Bits in the Information Path Integral's measures formalizes the variation problem in the minimax law, determining all regularities of the processes. Solving the problem mathematically describes the micro-macro processes, the IN, and the invariant conditions of the Observer's self-organization and self-replication. Information becomes equivalent of Observer time.

The values of quality energy and information, transferring during interactions, identify anatomy of information units: from qubits, bits, Free information, triplet, IN, and ending triplet's binding multiple INs. That determines different physical structures of physical units, starting from elementary structure of the particles in Standard Model to various macro units: molecules, multiple physical-chemical forms, leaving cells, organisms, and humans.

Thus, the probability field of observing impulses enables generating various information–physical units satisfying extreme mathematical law which dictates the allowable combinations of the invariant units. The fundamental constant and emerging constraints provide the values of specific properties to each allowable unit's combination.

Multiple physical triples units adjoin the IN hierarchical structure whose free information produces new units at higher level node and encodes triple code logic. Each unit unique position in IN hierarchy defines location of each code logical structure. The IN node hierarchical levels classify both quality of assembled information and energies establishing their connection.

Thus, the probability field of observing impulses enables generating various information–physical units satisfying extreme mathematical law, which dictates the allowable combinations of these invariant units. The fundamental constant and emerging constraints provide the values of specific properties to each allowable unit's combination.

These functional regularities create a united Information mechanism whose integral logic self-operates, transforming interacting uncertainties to physical reality-matter. This enables the exchange of human Information and the design of Artificial Intelligences.

Both Information and Information processes emerge as phenomena of natural interactions. Each specific field triad generates an Information process, creating its Observer. The Information equations described here finalize the main results, validate them numerically, and present Information models of many interactive physical processes.

The approach focuses on formal Information mechanisms in an Observer, without reference to the specific physical processes which originate these mechanisms in the Observations. The Information formalism describes a self-building information machine which creates both Humans and Nature.

*Comments1.* A bridge connecting of physical results [6, 7] and [39] with our approach.

Many years passed since Schrödinger introduced his equation of Quantum Mechanics as a new physical microscopic theory of interacting particles. However, up to now, the scientific origin of the connection between Quantum Mechanics and classical physics has not been established That must include linking a wave function to a probabilistic field, and connecting Quantum Mechanics to Quantum Information Theory.

Resuts [6,7] have shown that "theory of direct inter-particle interaction, associated with a particle acting upon itself, derives from the motion of a system of charged particles under the influence of electromagnetic forces."

However, the inter-particle interaction in the electrodynamics Maxwell field deals with problem that action and symmetric (adjunct) reaction should merge. Satisfaction of this requirement allows the connection of Maxwell's equations with the equations for atomic particles using the variation principle for total energy as the equivalent of a conservation law for such adjunct interactions. Solution of the obtained equation leads to a *discrete* action crossing reaction.

Study [39] obtains Schrödinger's equation in Quantum Mechanics from Maxwell equations. The equations for energy, momentum, frequency and wavelength of the electromagnetic wave in the atom are derived using the model of atom by analogy with the transmission line. The balance of electromagnetic energy in the atom satisfies the structural constant for the atom $so =$ 8.277 56. This constant connects to the physical structure constant $1/h_{\alpha}^{o*} \cong 137.036$ (the updated value) by relation $so = (1/2h_{\alpha}^{o*})^{1/2}$.

Results (Sec.2.2.3) identify a *bridge between minimal uncertainty and a certainty* measured by the entropy invariant $S_{\mp a}^{*} = 2h_{\alpha}^{o}$ which enables creation of an initial Information macrounit—a triplet with probability $p_{\pm a} = \exp(-2h_{\alpha}^{o}) = 0.98555075021 \rightarrow 1$ approximating the certainty.

This is the *bridge between micro-and macroprocesses* emerging along the path of observing the impulse interactions from maximal uncertainty to Information certainty [40].

The invariant connects this microprocess, which arises at the merge of interactive action and reaction (Sec. 2.2), with the motion of the interactive adjunct charged particle in a Maxwell field.

This proves the requirement for the *discrete action merging reaction*, which leads to impulse interaction rising the microprocess. Since the merging microprocess emanates for random field, it indicates that equations of the electromagnetic wave in the atom also originate in random field.
Moreover, the Schrödinger equation, describing the microprocess, emerges from the *initial random impulses* of the merging actions and reactions, while both references [6, 7] and [39] have studied the *deterministic* processes.
The invariant constant also binds the emerging micro-macroprocess with Maxwell equations extended to an equation of the interacting atom particles. In addition, the extended model of the atoms, covering three of the four fundamental interactions (electro-magnetic, weak and strong interactions), allows the *Information description*, which confirms "It from bit". The merging impulses 1-0 and 1-0 also explain *creation of qubit* $|0\rangle$ and $|1\rangle$ *in the emerging microprocess* during the entanglement. (Sec.5.4.1.). •

Study [41] has shown "that gravity, just as electromagnetism in Wheeler-Feynman's time symmetric electrodynamics, also be an "adjunct field" instead of an independent entity".
In [40] we calculate a weak information forces' analogy with the gravitational force. •

Reference [42] has revealed that "the entangling space-time works just like a quantum error-correcting code, protecting information in jittery qubits to store it not in individual qubits, but in patterns of entanglement among many", starting with a triple.

Sec. 2 details the mechanism with emerging a space interval *during reversible time interval at the entanglement.* •

Here we focus on the entropy integral measure and it application to the Observer.

## 2. THE EVALUATION OF IMPULSES IN THE INTERACTIVE OBSERVING PROCESS

### 2.1. Class of Discrete Step-Down and Step-Up Functions in Impulse

Let us find a class of step-down $u_-^t = u_-(\tau_k^{-o})$ and step-up $u_+^t = u_+(\tau_k^{+o})$ functions acting on the cutting discrete interval $o(\tau_k) = \tau_k^{+o} - \tau_k^{-o}$, which will preserve the Markov diffusion process' additive and multiplicative functions within each impulse of the process.

*Lemma* 2.1.

1. Opposite discrete functions $u_-^t$ and $u_+^t$ in form

$$u_-(\tau_k^{-o}) = \downarrow_{\tau_k^{-o}} \bar{u}_-, u_+(\tau_k^{+o}) = \uparrow_{\tau_k^{+o}} \bar{u}_+ \quad (1.1)$$

satisfy conditions of additivity:

$$[u_+^t - u_-^t] = U_a \text{ (a)}$$

or

$$[u_+^t + u_-^t] = U_a \text{(b)} \quad (1.1A)$$

and and multiplicativity:

$$[u_+^t - u_-^t] \times [u_+^t + u_-^t] = U_m \quad (1.1B)$$

at

$$\mathrm{U}_a = U_m = \mathrm{U}_{am} = c^2 > 0 \quad (1.1C)$$

**If** instance-jump $\downarrow_{\tau_k^{-o}}$ has plain (time) interval $\bar{u}_- > 0$, and instance jump $\uparrow_{\tau_k^{+o}}$ has high (space) interval $\bar{u}_+ > 0$, where their real values for relation (1.1A)(a) satisfy

$$\bar{u}_- = 0.5, \ \bar{u}_+ = 1, \bar{u}_+ = 2\bar{u}_-, , \tag{1.2a}$$

and for relation (1.1A) (b) the considered intervals' real values hold

$$\bar{u}_-^o = \bar{u}_+^o = 2.. \tag{1.2b}$$

2. Complex functions

$$u_t(u_\pm^{t1}, u_\pm^{t2}), u_\pm^{t1} = [u_+ = (j-1), u_- = (j+1)], j = \sqrt{-1} \tag{1.2c}$$

satisfy conditions (1.1aA), (1.1B) in forms

$$u_+ - u_- = (j-1) - (j+1) = -2, u_+ \times u_- = (j-1) \times (j+1) = (j^2 - 1) = -2,$$

which however do not preserve positive value (1.1C).

Therefore, it holds $c^2 < 0$ at imaginary opposite complex functions

$$u_t(-u_\pm^{t1}) = u_t(u_\pm^{t2}), u_\pm^{t2} = [u_+ = (j+1), u_- = (j-1)] \tag{1.2d}$$

satisfying (1.1bA)-(1.C).

2a. At the equal absolute values $|u_+^t| = |u_-^t|$, the imaginary functions

$$u_+^t = j\sqrt{2}, u_-^t = -j\sqrt{2} \tag{1.2d1}$$

satisfy only multiplicative part $U_m = u_+^t \times u_-^t = -2$ when the impulse additive measure holds $\mathrm{U}_a = 0.$

*Proofs* are straight forward.

Assuming both opposite functions apply on borders of impulse interval $o(\tau_k) = (\tau_k^{+o}, \tau_k^{-o})$ in forms

$$u_-^{t1} = u_-(\tau_k^{-o}), u_+^{t1} = u_+(\tau_k^{-o}),$$

and

$$u_-^{t2} = u_-(\tau_k^{+o}), u_+^{t2} = u_+(\tau_k^{+o}) \tag{1.2e}$$

at

$$u_-^{t1} u_+^{t1} = c^2(\tau_k^{-o}), u_-^{t2} u_+^{t2} = c^2(\tau_k^{+o}), t = \tau_k^{+o}, \tag{1.2f}$$

It follows that only by the end of this time interval at $t = \tau_k^{+o}$ both Markov properties (1.1A, B) satisfy, while at beginning $t = \tau_k^{-o}$, the starting process satisfies only (1.1A). •

*Corollary* 2.1

1. Conditions 1.1A-1.1C imply that $c^2(\tau_k^{-o}), c^2(\tau_k^{+o})$ are discrete functions of actions (1.2f) switching on interval $\Delta_\tau = \tau_k^{+o} - \tau_k^{-o}$.

Let us construct impulse discrete function on interval $\Delta_\tau$ in form

$$\delta^o u_{t=\tau_k} = [u_-(\tau_k^{-o}) - u_+(\tau_k^{+o})] / (\tau_k^{+o} - \tau_k^{-o}) \ . \tag{1.3}$$

Substitution in (1.1) relations (1.2b) for $\bar{u}_+^o = \bar{u}_+ = 2$ brings

$$u_{-}(\tau_k^{-o}) = -1_{\tau_k^{-o}} 0.5, u_{+}(\tau_k^{+o}) = +1_{\tau_k^{+o}} 2\,. \qquad (1.3a)$$

Substitution (1.3a) in (1.3) implies equality

$$\delta^o u_{t=\tau_k} = [u_{+}(\tau_k^{+o}) - u_{-}(\tau_k^{-o})] / (\tau_k^{+o} - \tau_k^{-o}) > 0\,. \qquad (1.3a)$$

which satisfies positivity of $c^2 > 0$

2. Discrete function on interval $\Delta = (s_k^{+o}, \tau_k^{-o})$ preceding interval $\Delta_\tau$

$$u_{+}(s_k^{+o}) = +1_{s_k^{+o}} \bar{u}_{+}, u_{-}(\tau_k^{o}) = -1_{\tau k}^{+o} \bar{u}_{-} \qquad (1.3b)$$

are multiplicative:

$$(u_{-}(\tau_k^{-o}) - u_{+}(s_k^{+o})) \times (u_{-}(\tau_k^{-o}) - u_{+}(s_k^{+o})) = [u_{-}(\tau_k^{-o}) - u_{+}(s_k^{+o})]^2\,.$$

2a. Discrete functions (1.2e) in form

$$\bar{u}_{+} = j\bar{u}, \bar{u}_{-} = -j\bar{u}, \bar{u} \neq 0 \qquad (1.3c)$$

satisfy only condition (1.1A) which for functions (1.3b) holds

$$[u_{-}(\tau_k^{-o}) - u_{+}(s_k^{+o})]^2 = -(j\bar{u})^2 [-1_{\tau_k^{-o}} - 1_{s_k^{+o}}]^2 > 0. \qquad (1.3d)$$

**2. 2. Discrete impulse action on the Entropy Functional**

Let us find discrete analog of the EF integral increments under impulse *discrete* function (1.3a) with step-down function $u_{-}(\tau_k^{-o}) = -1_{\tau_k^{-o}} \bar{u}_{-}$ and step-up $u_{+}(\tau_k^{+o}) = +1_{\tau_k^{+o}} \bar{u}_{+}$ function:

$$\delta[u_{-}(\tau_k^{-o}), u_{+}(\tau_k^{+o})] = (\downarrow 1_{\tau_k^{-o}} - \uparrow 1_{\tau_k^{+o}}) \bar{u}_k. \qquad (1.3e)$$

*Proposition* 2.1

1. Applying (1.3e) to the EF integral in form of discrete delta-function [24] leads to

$$\Delta S[\tilde{x}_t / \varsigma_t]\big|_{t=\tau_k^{-o}}^{t=\tau_k^{+o}} = \begin{cases} 0, t < \tau_k^{-o} \\ 1/4 u_{-}(\tau_k^{-o}) o(\tau_k^{-o}) / \tau_k^{-o}, t = \tau_k^{-o}, 1/4 \downarrow 1_{\tau_k^{-o}} \bar{u}_{ko} \\ 1/2 (u_{-}(\tau_k^{-o}) - u_{+}(\tau_k^{+o})) o(\tau_k) / (\tau_k^{+o} - \tau_k^{-o}), t = \tau_k, \tau_k^{-o} < \tau_k < \tau_k^{+o}, 1/2 (\downarrow 1_{\tau_k^{-o}} - \uparrow 1_{\tau_k^{+o}}) \bar{u}_{km} \\ 1/4 u_{+}(\tau_k^{+o}) o(\tau_k^{+o}) / \tau_k^{+o}, t = \tau_k^{+o}, 1/4 \uparrow 1_{\tau_k^{+o}} \bar{u}_{k1} \end{cases} ,(1.4)$$

where the entropy units on the impulse left border $\bar{u}_{ko}$, middle part $\bar{u}_{km}$, and right border $\bar{u}_{k1}$ determine the inner impulse time intervals:

$$\bar{u}_{ko} = \bar{u}_{-} \times o(\tau_k^{-o}) / \tau_k^{-o}, \bar{u}_{km} = (\bar{u}_{+} - \bar{u}_{-}) \times o(\tau_k) / (\tau_k^{+o} - \tau_k^{-o}), \bar{u}_{k1} = \bar{u}_{+} \times o(\tau_k^{+o}) / \tau_k^{+o},$$
$$\bar{u}_{km} = 1/2(\bar{u}_{+} - \bar{u}_{-}) = 0.75, o(\tau_k) = \tau_k^{+o} - \tau_k^{-o}, o(\tau_k^{-o}) / \tau_k^{-o} = 0.5, o(\tau_k^{+o}) / \tau_k^{+o} = 0.1875, \qquad (1.5)$$

and $|\bar{u}_{-} \times \bar{u}_{+}| = |1/2 \times 2| = |\bar{u}_k| = |1|_k$ is multiplicative measure of that impulse.

Measuring the middle impulse interval $\bar{u}_{sm}$ in (1.4) by single impulse entropy unit $\bar{u}_k = |1|_k$ defines finite size of the parameters $\bar{u}_{ko}, \bar{u}_{km}, \bar{u}_{k1}$ in (1.5), which estimate value on the unit border through $\bar{u}_{km}$ :

$$\bar{u}_{ko} = 0.25 = 1/3 \bar{u}_{km}, \bar{u}_{k1} = 2 \times 0.1875 = 0.375 = 0.5 \bar{u}_{km}\,. \bullet \qquad (1.6)$$

*Proofs* follow from Proposition 1.3 below.

Let us introduce an impulse entropy unit $\bar{u}_s = |1|_s$ at moments $(s_k^{-o}, s_k^o, s_k^{+o})$ of its left border, middle, and right border accordingly, acting prior to impulse $\bar{u}_k = |1|_k$.

Then we will find increments of entropy $\Delta S[\tilde{x}_t / \varsigma_t]|_{s_k^+}^{\tau_k^{-o}}$ on the impulse $\bar{u}_k$ border at its interval $\Delta_{\tau s+} = \delta_{sk\pm} = (s_k^{+o} - \delta_k^{\tau-})$ that borders impulse $\bar{u}_s$, and interval $\Delta_{\tau s-} = \delta_{sk\mp} = (\delta_k^{\tau-} - \delta_k^{\tau+})$ that borders impulse $\bar{u}_k$ under the impulse $\bar{u}_s = |1|_s$ border functions $u_+(s_k^{+o})$, and the impulse $\bar{u}_k = |1|_k$ border functions $u_-(\delta_k^{\tau-})$ and $u_+(\delta_k^{\tau+})$.

That brings following discrete functions on these borders:

$$\delta^o u_{\tau=(s_k^{+o}-\delta_k^{\tau-})} = (u_+(s_k^{+o}) - u_-(\delta_k^{\tau-}))(s_k^{+o} - \delta_k^{\tau-})^{-1} = \uparrow 1_{s_k^{+o}} \bar{u} - \downarrow 1_{\delta_k^{\tau-}} \bar{u} = [\uparrow 1_{s_k^{+o}} - \downarrow 1_{\delta_k^{\tau-}}]\bar{u}, \tag{1.7}$$

$$\delta^o u_{\tau=(\delta_k^{\tau-}-\delta_k^{\tau+})} = (u_-(\delta_k^{\tau-}) - u_+(\delta_k^{\tau+}))(\delta_k^{\tau-} - \delta_k^{\tau+})^{-1} = \downarrow 1_{\delta_k^{\tau-}} \bar{u} - \uparrow 1_{\delta_k^{\tau+}} \bar{u} = [\downarrow 1_{\delta_k^{\tau-}} - \uparrow 1_{\delta_k^{\tau+}}]\bar{u}, \tag{1.8}$$

$$\delta^o u_{\tau=(\delta_k^{\tau+}-\tau_k^{-o})} = (u_+(\delta_k^{\tau+}) - u_-(\tau_k^{-o}))(\delta_k^{\tau+} - \tau_k^{-o})^{-1} = \uparrow 1_{\delta_k^{\tau+}} \bar{u} - \downarrow 1_{\tau_k^{-o}} \bar{u} = [\uparrow 1_{\delta_k^{\tau+}} - \downarrow 1_{\tau_k^{-o}}]\bar{u}. \tag{1.9}$$

Here equal unit $\bar{u} > 0$ evaluates each impulse interval, which according to the optimal principle is an invariant.

Since the EF is additive functional, applying functions (1.7)-(1.9) leads to additive discrete sum of the entropy increments:

$$\Delta S[\tilde{x}_t / \varsigma_t]|_{s_k^+}^{\tau_k^{-o}} = \Delta S[\tilde{x}_t / \varsigma_t]|_{s_k^+}^{\delta_k^{\tau-}} + \Delta S[\tilde{x}_t / \varsigma_t]|_{\delta_k^{\tau-}}^{\delta_k^{\tau+}} + \Delta S[\tilde{x}_t / \varsigma_t]|_{\delta_k^{\tau+}}^{\tau_k^{-o}} \tag{1.10}$$

collected along time interval

$$\Delta_{\tau sk\pm} = s_k^{+o} - \delta_k^{\tau-} + \delta_k^{\tau-} - \delta_k^{\tau+} + \delta_k^{\tau+} - \tau_k^{-o} = s_k^{+o} - \tau_k^{-o} = \Delta_{\tau s}. \tag{1.10a}$$

*Proposition* 2.2

A. The increments of the entropy functional (1.1.10), collected on intervals (1.10a), under functions (1.7)-(1.9), bring the entropy contribution:

$$\Delta S[\tilde{x}_t / \varsigma_t]|_{s_k^+}^{\delta_k^{\tau-}} = 1/2(u_+(s_k^{+o}) - u_-(\delta_k^{\tau-}))o(s_k^{+o} - \delta_k^{\tau-}))(s_k^{+o} - \delta_k^{\tau-})^{-1} = 1/2[\uparrow 1_{s_k^{+o}} - \downarrow 1_{\delta_k^{\tau-}}]\bar{u}_{ks} \tag{1.11}$$

on interval

$$\bar{u}_{ks} = \bar{u}(o(s_k^{+o} - \delta_k^{\tau-})(s_k^{+o} - \delta_k^{\tau-})^{-1}, \tag{1.11a}$$

the entropy contribution

$$\Delta S[\tilde{x}_t / \varsigma_t]|_{\delta_k^{\tau-}}^{\delta_k^{\tau+}} = 1/2(u_-(\delta_k^{\tau-}) - u_+(\delta_k^{\tau+}))o(\delta_k^{\tau-} - \delta_k^{\tau+}))(\delta_k^{\tau-} - \delta_k^{\tau+})^{-1} = 1/2[\downarrow 1_{\delta_k^{\tau-}} - \uparrow 1_{\delta_k^{\tau+}}]\bar{u}_{k\delta s}, \tag{1.12}$$

on interval

$$\bar{u}_{k\delta s} = \bar{u} \times (o(\delta_k^{\tau-} - \delta_k^{\tau+}))(\delta_k^{\tau-} - \delta_k^{\tau+})^{-1}, \tag{1.12a}$$

and the entropy contribution

$$\Delta S[\tilde{x}_t / \varsigma_t]|_{\delta_k^{\tau+}}^{\tau_k^{-o}} = 1/2(u_+(\delta_k^{\tau+}) - u_-(\tau_k^{-o}))o(\delta_k^{\tau+} - \tau_k^{-o})(\delta_k^{\tau+} - \tau_k^{-o})^{-1} = 1/2[\uparrow 1_{\delta_k^{\tau+}} - \downarrow 1_{\tau_k^{-o}}]\bar{u}_{k\delta}, \tag{1.13}$$

under function

$$[\uparrow 1_{\delta_k^{\tau+}} - \downarrow 1_{\tau_k^{-o}}]\bar{u}_{k\delta} = [\uparrow 1_{\delta_k^{\tau+}} + \uparrow 1_{|\tau_{k|}^{-o}}]\bar{u}_{k\delta}. \quad (1.13a)$$

Here on the impulse invariant interval $\bar{u}$ , each impulse acquires the entropy measure:

$$\bar{u}_{k\delta} = \bar{u}\times(o(\delta_k^{\tau+} - \tau_k^{-o}))(\delta_k^{\tau+} - \tau_k^{-o})^{-1} = \bar{u}\times o(\delta_k^{\tau+})(\delta_k^{\tau+} - \tau_k^{-o})^{-1} + \bar{u}\times o(\tau_k^{-o})(\tau_k^{-o})^{-1}(\delta_k^{\tau+} - \tau_k^{-o})^{-1}\tau_k^{-o}$$

(1.14)

Relation (1.14) allows representing the impulse interval entropy measure in form

$$\bar{u}_{k\delta} = \bar{u}_{k\delta o} + \bar{u}_{k\delta 1} \quad (1.14a)$$

with its components

$$\bar{u}_{k\delta o} = \bar{u}\times(o(\delta_k^{\tau+}))(\delta_k^{\tau+} - \tau_k^{-o})^{-1}), \ \bar{u}_{k\delta 1} = \bar{u}_{ko1}\times\bar{u}_{ko2}, \quad (1.14b)$$

$$\bar{u}_{ko1} = \bar{u}\times(o(\tau_k^{-o}))(\tau_k^{-o})^{-1}, \ \bar{u}_{ko2} = \bar{u}^{-1}\times\tau_k^{-o}(\delta_k^{\tau+} - \tau_k^{-o})^{-1}. \quad (1.14c)$$

B. Intervals $\bar{u}_{ko1}$ and $\bar{u}_{ko2}$ are multiplicative parts of impulse $\bar{u}_{k\delta 1}$ step-up function, which connect to $\bar{u}_{k\delta o}$ by relations

$$\bar{u}_{k\delta o} = \bar{u}_{k\delta 1} = 1/2\bar{u}_{k\delta}, \ \bar{u}_{k\delta 1} = \bar{u}_{ko1}. \quad (1.15)$$

Here the invariant impulse $|\bar{u}_{k\delta}| = |1|_s$ (1.14a) holds step-up and step-down functions acting on time interval $\delta_k^{\tau+} = 2\tau_k^{-o}$ , which measure

$$\bar{u}_{k\delta} = \bar{u}_{ks} \text{ at} |\bar{u}_{k\delta 1}| = 1/2\bar{u}_{sm}. \quad (1.15a)$$

Function $\bar{u}_{ko}$ and $\bar{u}_{k1}$ act on the above relative time intervals accordingly:

$$o(\tau_k^{-o})(\tau_k^{-o})^{-1} = 0.5, \ o(\tau_k^{+o})/\tau_k^{+o}) = 0.1875. \quad (1.15b)$$

The impulse $\bar{u}_{k\delta}$ multiplicative step-up and step-down functions apply on two equal time intervals:

$$\tau_k^{-o} = \delta_k^{\tau+}/2 \quad (1.16a)$$

and

$$(\delta_k^{\tau+} - \tau_k^{-o}) = \delta_k^{\tau+}/2. \quad (1.16b)$$

On first interval (1.16b), its step-up part $[\uparrow 1_{\delta_k^{\tau+}}]$ with entropy measure $\bar{u}_{-}$ captures entropy increment

$$\Delta S[\tilde{x}_t/\varsigma_t]|_{\delta_k^{\tau+}}^{\tau_k^{-o}} = 1/2[\uparrow 1_{\delta_k^{\tau+}}]\bar{u}_{-} = 1/8[\uparrow 1_{\delta_k^{\tau+}}]. \quad (1.16)$$

On second interval (1.16a), its step-down multiplicative part in (1.14b) at $\bar{u}_{ko2} = \bar{u}^{-1}$ transfers entropy (1.16) to the impulse starting action $[\downarrow 1_{\tau_k^{-o}}]$ which cuts is within impulse entropy measure (1.4) at

$$\bar{u}_{ko1} = 1/2\bar{u}_{ko}$$

where $\bar{u}_{k\delta 1}$ in (1.14b) multiplies

$$\bar{u}[\uparrow 1_{\delta_k^{\tau+}}]\delta_k^{\tau+}/2\times\bar{u}^{-1}[\downarrow 1_{\tau_k^{-o}}]\tau_k^{-o}. \quad (1.16c)$$

Both equal time intervals in (1.16a), (1.16b) are on the impulse border where opposite inverse entropy increments and orthogonal.

C. The applied *extreme* solution (Proposition 1.2), decreasing time intervals (1.3.1.5b), brings minimal increment (1.10a) and

(a)-persistence continuation a sequence of the process impulses;

(b)- balance condition for the entropy contributions;

(c)-each impulse invariant unit $\bar{u}_k = |1|_k$, supplied by entropy unit $\bar{u}_s = |1|_s$, *triples* Information *increasing information density* in each following Information unit.•

*Proofs.*

The additive sum of entropy increments under invariant impulses (1.7-1.9) satisfies balance condition:

$$\Delta S[\tilde{x}_t / \varsigma_t]|_{s_k^+}^{\tau_k^{-o}} = \Delta S[\tilde{x}_t / \varsigma_t]|_{s_k^+}^{\delta_k^{\tau-}} + \Delta S[\tilde{x}_t / \varsigma_t]|_{\delta_k^{\tau-}}^{\delta_k^{\tau+}} + \Delta S[\tilde{x}_t / \varsigma_t]|_{\delta_k^{\tau+}}^{\tau_k^{-o}} =$$

$$1/2[\uparrow 1_{s_k^{+o}} - \downarrow 1_{\delta_k^{\tau-}}]\bar{u}_{ks} + 1/2[\uparrow 1_{\delta_k^{\tau-}} - \uparrow 1_{\delta_k^{\tau+}}]\bar{u}_{k\delta s} + 1/2\uparrow 1_{\delta_k^{\tau+}}\bar{u}_{k\delta o} - 1/2\downarrow 1_{|\tau|_k^{-o}}\bar{u}_{k\delta 1} = 0, \quad (1.17)$$

where action $1/2\downarrow 1_{|\tau|_k^{-o}}\bar{u}_{k\delta 1} \rightarrow 1/4\downarrow 1_{|\tau|_k^{-o}}\bar{u}_{ko}$ transfers entropy increment $\Delta S[\tilde{x}_t / \varsigma_t](\tau_k^{-o}) = 1/4\downarrow 1_{|\tau|_k^{-o}}\bar{u}_{ko}$ on discrete locality $|\tau|_k^{-o}$ by step-down action $\downarrow 1_{|\tau|_k^{-o}}\bar{u}_{k\delta 1}$.

Fulfillment the relations

$$[\uparrow 1_{s_k^{+o}}\bar{u}_{ks} - \downarrow 1_{\delta_k^{\tau-}}\bar{u}_{ks} + \uparrow 1_{\delta_k^{\tau-}}\bar{u}_{k\delta s} - \uparrow 1_{\delta_k^{\tau+}}\bar{u}_{k\delta s} + \uparrow 1_{\delta_k^{\tau+}}\bar{u}_{k\delta o} - \downarrow 1_{|\tau|_k^{-o}}\bar{u}_{k\delta 1}] = 0$$

$$[\uparrow 1_{s_k^{+o}}\bar{u}_{ks} + \uparrow 1_{\delta_k^{\tau-}}[\bar{u}_{k\delta s} - \bar{u}_{ks}] + \uparrow 1_{\delta_k^{\tau+}}[\bar{u}_{k\delta o} - \bar{u}_{k\delta s}] = \downarrow 1_{|\tau|_k^{-o}}\bar{u}_{k\delta 1}, \downarrow 1_{|\tau|_k^{-o}}\bar{u}_{k\delta 1} = -1/2\downarrow 1_{|\tau|_k^{-o}}\bar{u}_{ko}$$

leads to sum of the impulse intervals:

$$\bar{u}_{ks} - \bar{u}_{ks} + \bar{u}_{k\delta s} + \bar{u}_{k\delta s} - \bar{u}_{k\delta s} + \bar{u}_{k\delta o} - \bar{u}_{k\delta 1} = 0,$$

and to relation

$$\bar{u}_{k\delta 1} = -1/2\bar{u}_{ko},$$

or to

$$\bar{u}_{k\delta o} = \bar{u}_{k\delta 1}. \quad (1.17a)$$

Impulse $[\uparrow 1_{\delta_k^{\tau+}}\bar{u}_{k\delta o} - \downarrow 1_{|\tau|_k^{-o}}\bar{u}_{k\delta 1}] = [\uparrow 1_{\delta_k^{\tau+}} + \uparrow 1_{|\tau|_k^{-o}}]\bar{u}_{k\delta}$ contains intervals

$$\bar{u}_{k\delta} = \bar{u}_{k\delta o} + \bar{u}_{k\delta 1},$$

where from relations (1.9) , (1.13a) it follows $\bar{u}_{k\delta} = \bar{u}$ , and (1.17a) leads to

$$\bar{u}_{k\delta o} = \bar{u}_{k\delta 1} = 1/2\bar{u}. \quad (1.17b)$$

Interval

$$\bar{u}_{k\delta} = \bar{u} \times [(o(\delta_k^{\tau+}))(\delta_k^{\tau+} - \tau_k^{-o})^{-1}) + (o(\tau_k^{-o}))(\tau_k^{-o})^{-1}(\delta_k^{\tau+} - \tau_k^{-o})^{-1}\tau_k^{-o}] \quad (1.17c)$$

consists of $\bar{u}_{k\delta}$ components:

$$\bar{u}_{k\delta o} = \bar{u} \times (o(\delta_k^{\tau+}))(\delta_k^{\tau+} - \tau_k^{-o})^{-1})$$

and

$$\bar{u}_{k\delta 1} = \bar{u}_{ko1} \times \bar{u}_{ko2} / \bar{u},, \quad (1.17d)$$

where

$$\bar{u}_{ko1} = \bar{u} \times (o(\tau_k^{-o}))(\tau_k^{-o})^{-1}, \ \bar{u}_{ko2} = \bar{u}^{-1} \times \tau_k^{-o}(\delta_k^{\tau+} - \tau_k^{-o})^{-1}.$$

Intervals $\bar{u}_{ko1}$ and [$\bar{u}_{ko2} / \bar{u}$ ] are multiplicative parts of impulse interval $\bar{u}_{k\delta 1}$ covered by the impulse starting interval $|\tau|_k^{-o}$.

From (1.17b) and relations (1.17d) it follows

$$\bar{u}_{k\delta o} = \bar{u} \times (o(\delta_k^{\tau+}))(\delta_k^{\tau+} - \tau_k^{-o})^{-1}) = 1/2\bar{u},$$

$$(o(\delta_k^{\tau+}))(\delta_k^{\tau+} - \tau_k^{-o})^{-1}) = 1/2, \quad (1.18)$$

and

$$\bar{u}_{ko1} = \bar{u} \times (o(\tau_k^{-o}))(\tau_k^{-o})^{-1} = 1/2\bar{u}. \quad (1.18a)$$

That leads to

$$(o(\tau_k^{-o}))(\tau_k^{-o})^{-1} = 1/2 \quad (1.18b)$$

and from (1.18) to relations

$$(\delta_k^{\tau+} - \tau_k^{-o})^{-1}) = (\tau_k^{-o})^{-1}, \delta_k^{\tau+} - \tau_k^{-o} = \tau_k^{-o}, \ \tau_k^{-o}(\delta_k^{\tau+} - \tau_k^{-o})^{-1} = 1. \quad (1.18c)$$

Then to

$$\tau_k^{-o} = 1/2\delta_k^{\tau+}. \quad (1.18d)$$

From (1.18c) it follows

$$\bar{u}_{ko2} = \bar{u}^{-1}.. \quad (1.18f)$$

Applying the sequence of Eqs. (1.7-1.9), at equal invariant $\bar{u}$ , leads to

$$\bar{u} = u_+(s_k^{+o}) - u_-(\tau_k^{-o}), \quad (1.19)$$

$$\bar{u} = u_+(\delta_k^{\tau+}) - u_-(\tau_k^{-o}) \quad (1.19a)$$

at

$$u_+(\delta_k^{\tau+}) - u_-(\tau_k^{-o}) = \bar{u}_{kb}.$$

That brings invariant $|\bar{u}_s| = |1|_s$ to both impulses (1.19) and (1.19a).

Relation

$$u_+(s_k^{+o}) + u_-(\tau_k^{-o}) = 2[u_-(\delta_k^{\tau-}) + (u_+(\delta_k^{\tau+})] = 0$$

following from the sequence of Equations (1.7-1.9) leads to

$$u_-(\delta_k^{\tau-}) = -u_+(\delta_k^{\tau+}), \quad (1.19b)$$

or to reversing and mutually neutralizing these actions on related moments $\delta_k^{\tau-} \cong \delta_k^{\tau+}$.

Impulse interval $\bar{u}_{k\delta}$, with $\bar{u}_{k\delta o}$ and $\bar{u}_{k\delta 1}$, starts interval of applying step-down action $o(\tau_k^{-o})(\tau_k^{-o})^{-1} = 0.5$ in (1.4) at

$$\bar{u}_{k\delta 1} = \bar{u}_{k\delta o} = 1/2\bar{u}_{k\delta}.$$

Invariant impulse $|\bar{u}_s| = |1|_s$, consists of two step-actions $[\uparrow 1_{\delta_k^{\tau+}} \downarrow 1_{|\tau|_k^{-o}}]\bar{u}_{k\delta}$, which measures intervals

$$\bar{u}_{kb} = \bar{u}_{sm} = \bar{u}_s$$

at

$\bar{u}_{k\delta 1} = 1/2\bar{u}_{sm}$. . (1.19c)

At conditions (1.18c, d), limiting time-jump in (1.13a), step-actions of impulse $\bar{u}_{k\delta}$ apply on two equal time intervals following from (1.19c).

On the first interval

$$(\delta_k^{\tau+} - \tau_k^{-o}) = \delta_k^{\tau+}/2$$

step-up part of $\bar{u}_{k\delta}$ -action $[\uparrow 1_{\delta_k^{\tau+}}]$ captures entropy increment

$$\Delta S[\tilde{x}_t / \varsigma_t]|_{\delta_k^{\tau+}}^{\tau_k^{-o}} = 1/2[\uparrow 1_{\delta_k^{\tau+}}]\bar{u}_- = 1/8[\uparrow 1_{\delta_k^{\tau+}}]. \quad (1.20)$$

On the second interval $\tau_k^{-o} = \delta_k^{\tau+}/2$, the captured entropy (1.20) through the step-down multiplicative part (1.17c, d) delivers to the cutting action $\bar{u}_{ko} = \bar{u}_- \times o(\tau_k^{-o})(\tau_k^{-o})^{-1}$ the equal contributions

$$\Delta S[\tilde{x}_t / \varsigma_t]|_{\delta_k^{\tau+}}^{\tau_k^{-o}} = 1/4[\downarrow 1_{\tau_k^{-o}}]\bar{u}_- = 1/8[\downarrow 1_{\tau_k^{-o}}]. \quad (1.20a)$$

The control action $[\downarrow 1_{\tau_k^{-o}}]$ at $\bar{u}_- = 0.5$ cuts the entropy of correlation in impulse (1.4) at $\bar{u}_{ko1} = 1/2\bar{u}_{ko}$ . •

*Comment* 2.1

Action $[\uparrow 1_{\delta_k^{\tau+}}]$ cuts the captured entropy from impulse $\bar{u}_s = |1|_s$, while multiplicative step-down part (1.17b) transforms the captured entropy to the cutting action in (1.4) at $\bar{u}_{ko2} = \bar{u}^{-1}$ . •

At the end of $k$ impulse, control action $\bar{u}_+$ transforms entropy (1.20) on interval $\bar{u}_{kio} = \bar{u}_- \times (o(\tau_k^{+o}) / \tau_k^{+o})$ to Information

$$\Delta I[\tilde{x}_t / \varsigma_t]|_{\delta_{k+}^{\tau+}}^{\tau_k^{+o}} = 1/4[\uparrow 1_{\tau_k^{-o}}]\bar{u}_{kio}\bar{u}_+ = 1/4 \times (-2\bar{u}_{kio})[\uparrow 1_{\tau_k^{-o}}] \quad (1.21)$$

and supplies it to $k+1$ impulse. (If between these impulses, the entropy increments on the process trajectory are absent, cut).

That leads to a balance equation for the information contributions to $k$ -impulse:

$$\Delta I[\tilde{x}_t / \varsigma_t]|_{\delta_k^{\tau+}}^{\tau_k^{-o}} + \Delta I[\tilde{x}_t / \varsigma_t]|_{\tau_k^{-o}}^{\tau_k} + \Delta I[\tilde{x}_t / \varsigma_t]|_{\tau_k}^{\tau_k^{+o}} = \Delta I[\tilde{x}_t / \varsigma_t]|_{\delta_{k+}^{\tau+}}^{\tau_k^{+o}}, \quad (1.21a)$$

where interval $\bar{u}_{ko}$ holds Information contribution $\Delta I[\tilde{x}_t / \varsigma_t]|_{\tau_k}^{\tau_k^{+o}} = 1/4\bar{u}_{km}$ satisfied (1.4) at $\bar{u}_+ = -2$, which measures $\bar{u}_{km} = 0.75$ (1.5).

That brings relations

$$0.125 + 0.75 + \bar{u}_{kio} = -2\bar{u}_{kio}, 0.125 + 0.75 + 3\bar{u}_{ko} = 0, \bar{u}_{k1} = 3\bar{u}_{ko} = 0.375 = \bar{u}_- \times o(\tau_k^{+0}) / \tau_k^{+0})$$

(1.22)

and

$$o(\tau_k^{+o}) / \tau_k^{+o} = 0.1875, \quad (1.22a)$$

$$\bar{u}_{ko} + \bar{u}_{km} + \bar{u}_{k1} = 1.25 = 5/3\bar{u}_{km}. \quad (1.22b)$$

From these and (1.21a) it follows

$$\Delta I[\tilde{x}_t / \varsigma_t]|_{\tau_k}^{\tau_k^{+o}} = 3\Delta I[\tilde{x}_t / \varsigma_t]|_{\delta_k^{\tau+}}^{\tau_k^{-o}} . \qquad (1.23)$$

Ratio $\bar{u}_{k1} / 2\bar{u}_{kio} = 3/2$ at $2\bar{u}_{kio} = 0.25$ evaluates part of $k$ impulse Information transferred to $k+1$ impulse.

Relations (1.17b,d), (1.18b,d,f), (1.19c), and (1.22a) *prove* Proposition 1.3 parts A-B.•

Since $\bar{u}_{-} = 0.5$ is cutting interval of impulse $\bar{u}_k$, it allows evaluate the additive sum of the discrete cutoff entropy contributions (1.4) during the entire impulse $(\downarrow 1_{\tau_k^{-o}} - \uparrow 1_{\tau_k^{+o}}) = \delta_k$ using $\bar{u}_{-} = \bar{u}_k$:

$$\Delta S[\tilde{x}_t / \varsigma_t]|_{\tau_k^{-o}}^{\tau_k^{+o}} = 1/4\bar{u}_k / 2 + 1/2\bar{u}_k + 1/4 \times 3/2\bar{u}_k = \bar{u}_k . \qquad (1.24)$$

That determines the impulse cutoff Information measure

$$\Delta S[\tilde{x}_t / \varsigma_t]_{\delta_k} = \Delta I[\tilde{x}_t / \varsigma_t]_{\delta_k} = (\downarrow 1_{\tau_k^{-o}} - \uparrow 1_{\tau_k^{+o}})\bar{u}_k = |1|\bar{u}_k , \bar{u}_k = |1|_k \; Nat \qquad (1.24a)$$

equals to $\cong 1.44$ Bit, which the cutting entropy functional from random process generates.

That single impulse unit $\bar{u}_k = |1|_k$ measures the following relative Information intervals:

$$\bar{u}_{ko} = 1/3\bar{u}_{km} , \; \bar{u}_{km} = 1 , \; \bar{u}_{kio} = 1/3\bar{u}_{km} = \bar{u}_{ko} , \qquad (1.24b)$$

and relative time

$$\tau_k^{+o} / \tau_k^{-o} = 3 \qquad (1.24c).$$

From relations

$$\bar{u}_{ko1} = 1/2\bar{u}_{sm} \text{ and } \bar{u}_{ko1} = 1/2\bar{u}_{ko} = 1/6\bar{u}_{km}$$

it follows

$$\bar{u}_{km} = 3\bar{u}_{sm} . \qquad (1.25)$$

That shows that impulse unit $\bar{u}_k = |1|_k$ triples Information supplied by entropy unit $\bar{u}_s = |1|_s$, or interval $\bar{u}_k$ compresses three intervals $\bar{u}_s$.

At satisfaction of the extremal principle, each impulse holds invariant interval size $|\bar{u}_k| = |1|_k$ proportional to the middle impulse interval $o(\tau)$ with Information $\bar{u}_{km}$ which measures $o(\tau)$, and vice versa, time $o(\tau)$ measures this Information.

Condition of decreasing $t - s_k^{+o} = o(t) \to 0$ with growing $t \to T$ squeezes sequence $s_k^{+o} \to \tau_{m-1}^{+o}, k = 1,2....m$ and leads to persistence continuation of the impulse sequence with transforming of the previous impulse entropy to Information of the following impulse: $\bar{u}_s = |1|_s \to \bar{u}_k = |1|_k$.

The sequence of growing and compressed Information increases at

$$\bar{u}_{k+1} = |3\bar{u}_k| = |1|_{k+1} . \qquad (1.25a)$$

The persistence continuation of the impulse sequence links intervals between sequential impulses ($\bar{u}_{ks}$ ,$\bar{u}_{k\delta s}, \bar{u}_{k\delta o}$) whose imaginary (virtual) function $[\uparrow 1_{s_k^{+o}} - \downarrow 1_{\delta_k^{\tau-}} + \uparrow 1_{\delta_k^{\tau+}}]u$ prognosis entropies increments (1.11), (1.12), (1.10).

Information contributions at each cutting interval $\delta_{k-1}, \delta_k, k, k+1,...,m$: $\Delta I[\tilde{x}_t / \varsigma_t]_{\delta_{k-1}}$, $\Delta I[\tilde{x}_t / \varsigma_t]_{\delta_k}$,.... determine the time distance interval $\tau_k^{-o} - \tau_{k-1}^{+o} = o_s(\tau_k)$ when each entropy increment

$$\Delta S[\tilde{x}_t / \varsigma_t]|_{\tau_{k-1}^{+o}}^{t \to \tau_k^{-o}} = 1/2uo(\tau_k) = \bar{u}_s \times o_s(\tau_k)$$

supplies each $\Delta I[\tilde{x}_t / \varsigma_t]_{\delta_k}$ satisfying

$$\bar{u} \times o(\tau_k) = \Delta I[\tilde{x}_t / \varsigma_t]_{\delta_k} \text{ at } \bar{u} \times o(\tau_k) = \bar{u}_k(\tau_k^{+o} - \tau_k^{-o}).$$

Hence, impulse interval

$$\bar{u}_k = \Delta I[\tilde{x}_t / \varsigma_t]_{\delta_k} / (\tau_k^{+o} - \tau_k^{-o}) \quad (1.26)$$

measures density of Information at each $\delta_k = \tau_k^{+o} - \tau_k^{-o}$. Squeezing of these intervals sequentially increases in each following Bit.

Relations (1.25a, b), (1.26) *prove* part C of Proposition 1.3.•

Such a Bit includes three parts:

- the first delivers multiplicative action (1.16c) by capturing entropy of random process;
- the second delivers the impulse step-down cut of the process entropy;
- the third is Information, which delivers the impulse step-up control and then transfers it to the nearest impulse.

That keeps the information connection between the impulses and provides persistence continuation of the impulse sequence during the process time $T$.

*Corollaries* 2.1

A. The additive sum of discrete functions (1.4) during the impulse intervals determines the impulse Information measure (Bit), generated from the cutting entropy functional of random process.

The step-down function generates $1/8 + 0.75 = 0.875 Nat$ from which it spends $1/8$ $Nat$ for cutting correlation while getting $0.75$ $Nat$ from the cut. Step-up function holds $1/8$ $Nat$ while 0.675 $Nat$ it gets from cutting $0.75$ $Nat$, from which $0.5 Nat$ it transfers to next impulse leaving $0.125$ $Nat$ within $k$ impulse.

The impulse has $1/8 + 0.75 + 1/8 = 1 Nat$ of total $1.25 Nat$ from which $1/8$ $Nat$ is the captured entropy increment from a previous impulse. The impulse actually generates $0.75 Nat \cong 1 Bit$, while the step-up action, using $1/8 Nat$, transfers $2/8 Nat$ Information to next $k$ impulse, capturing $1/8 Nat$ from the entropy impulse between $k$ and $k+1$ Information impulses (on interval $o_s(\tau_k)$).

B. From total maximum $0.875$ Nat, the impulse cuts minimum of that maximum $0.75$ $Nat$ implementing minimax principle, which validates variation condition (1.1.7) and results (1.17).

By transferring overall $0.375 Nat$ to next $k+1$ impulse, that $k$ impulse supplies it with its maximum of $1/3 \times 0.75 Nat$ from the cutting Information, thereafter implementing principle maximum of minimal cut.

C. Thus, each cutting Bit is active *Information unit* delivering Information from previous impulse and supplying Information to following impulse.

It includes: the cutting step-down control's Information delivered through capturing the external entropy of the random process; the cutoff Information, which the above control cuts from the random process; the Information delivered by the impulse step-up control, which, being transferred to the nearest impulse, keeps the Information connection between the impulses that provides persistence continuation of the impulse sequence.

D. The amount of Information that each second Bit of the cutoff sequence condenses grows in three times, which sequentially increases the Bit information density. At invariant increments of impulse (1.4), every $\bar{u}_k$ compresses three previous intervals $\bar{u}_{k-1}$ thereafter sequentially increase both density of interval $\bar{u}_k$ and density of these increments for each $k+1$ impulse. •

## 3. The Emerging Microprocess

As the Bayes *a posteriori* probabilities grow along observations, neighbor impulses may merge, generating interactive jumps on the border of each impulse.

The merge meets causing action with reaction, superimposing cause and effect and their probabilities.

It could cover unpredictable events within the merge.

Mathematically the jump increases Markov drift (speed) up to infinity (Section 1.1.3.2).

A starting jumping action $\uparrow$ interacting with opposite $\downarrow$ action of the bordered impulses initiates the impulse inner process $\tilde{x}_{otk} = \tilde{x}(t \in o(\tau_k)))$ called a microprocess.

(Because the merge squeezes the interaction interval to a micro-minimum).

*Comments* 2.2

In a sub-Markov process [1, 2], potential kernel negative curvature exposes Markov drifts convergence, which could lead to the merge. •

### 3. 1. The Conjugated Entropy Increments in the Microprocess

The microprocess is developing under step-function $u_{\pm}^{t1}$, $u_{\pm}^{t2}$ within the bordered impulse with the step-function $u_t(u_-^t, u_+^t) = c^2(t \in o(\tau_k))$ on a fixed impulse interval $o(\tau_k)$ within the discrete impulse (1. 4).

The impulse step-down $u_-^t = u_-(\tau_k^{-o})$ and step-up $u_+^t = u_+(\tau_k^{+o})$ functions, acting on the discrete interval $o(\tau_k) = \tau_k^{+o} - \tau_k^{-o}$ satisfying (1.1A-1.1C) and (1.2a-1.2d), generates the EF increments:

$$\Delta S_- = \Delta S_-[u_-^t], \Delta S_+ = \Delta S_+[u_+^t], \qquad (2.1)$$

which preserve the additive and multiplicative properties within the Markov process.

(But these merging actions may not simultaneously possess both these Markov properties).

Here, step function $u_{\pm}^{t1}$ (1.1c) is the analog of $\bar{u}_{k\delta 1}$ in (1.16c) at locality $\delta_k^{\tau +}/2$ of the beginning of impulse moment $\tau_k^{-o}$.

Opposite functions $u_{\pm}^{t1}(t^*)$ of jumps $\uparrow\downarrow$, starting at beginning of the process with relative time

$$t^* = [\mp\pi/2 \times \delta t^{\pm *}/\mathrm{o}(\tau_k)], \delta t^{\pm *} \in (\delta t_{ok}^{\pm} \to 1/2\,\mathrm{o}(\tau_k)), \qquad (2.2)$$

hold directions of opposite impulses

$$u_{\pm}^{t1} = [u_+ = \uparrow_{\mathrm{t}_o^{*-}} (j-1), u_- = \downarrow_{\mathrm{t}_o^{*+}} (j+1)] \qquad (2.3)$$

on interval $\delta_o[\mathrm{t}_o^{*-}, \mathrm{t}_o^{*+}] = \delta t^* < o(\tau)$ at a locality of the impulse initial time $\tau_k^{-o}$.

Controls (2.3), holding $u = c^2 < 0$, brings imaginable $u$ and minimal time interval

$$o = (\delta_k^{\tau +}/2)^2 = (\tau_k^{-o})^2. \qquad (2.3a)$$

The microprocess increments at interval $o$ do not possess Markov properties (1.C).

The jumps (3.3) initiate relative differential increments of entropy:

$$\frac{\delta S}{S} / \delta t^* = u^{t1}_{\pm}, \ [u_+ = \uparrow_{t_o^{*-}} (j-1), u_- = \downarrow_{t_o^{*+}} (j+1)], \qquad (2.4)$$

which in a limit leads to differential Equations:

$$\dot{S}_+(t^*) = (j-1)S_+(t^*), \dot{S}_-(t^*) = (j+1)S_-(t^*) . \qquad (2.5)$$

The applied (2.3) with symbol $j$ of orthogonality to the microprocess entropy increments rotates them.

Solutions of (2.5) describe the microprocess with opposite conjugated entropies functions on relative time $t^*$:

$$S_+(t^*) = [exp(-t^*)(\mathrm{Cos}(t^*) - jSin(t^*))]\,|^{1/2o(\tau_k)}_{t_o^{*-}}, S_-(t^*) = [exp(t^*)(\mathrm{Cos}(t^*) + jSin(t^*))]\,|^{1/2o(\tau_k)}_{t_o^{*+}}$$

(2.6)

with initial conditions $S_+(t_o^{*-})$, $S_-(t_o^{*+})$ at moment

$$t_o^{*+} = t_o^{*-} = [\mp\pi / 2\delta t_{ok}^{\pm}] . \qquad (2.6a)$$

The wide of step-function $u^{t1}_{\pm} : \delta t_o^{\pm} / \mathrm{o}(\tau_k) = 0.2 + 0.005 = 0.205$ relative to interval $\mathrm{o}(\tau_k)$ and the impulse beginning interval $\tau_k^{-o} / \mathrm{o}(\tau_k) = 0.25$ relative to that interval determine the relative moment $\delta t_{ok}^{\pm} = \delta t_o^{\pm} / \tau_k^{-o} = \pm 0.82$ of starting this function.

From that, numerical solutions of (3.6) by the moment of time $\delta t_{ok}^{\pm} = \pm 0.82$ follow:

$$S_+(t_o^+) = [exp(-\pi / 2\times 0.82)(\mathrm{Cos}(\pi / 2\times 0.82)) - jSin(\pi / 2\times -0.82))] \approx 0.2758\times 1,$$
$$S_-(t_o^-) = [exp(\pi / 2\times -0.82)(\mathrm{Cos}(-\pi / 2\times -0.82) + jSin(-\pi / 2\times -0.82))] \approx 0.2758\times 1 \qquad (2.7)$$

The numerical solutions by the moments of time

$$t^{*-} = -\pi / 2\times 1/2\,\mathrm{o}(\tau_k) / \mathrm{o}(\tau_k) = -\pi / 4 \qquad (2.8a)$$

and

$$t^{*+} = \pi / 2\times 1/2\,\mathrm{o}(\tau_k) / \mathrm{o}(\tau_k) = \pi / 4 \qquad (2.8)$$

are

$$S_+(t^{*-}) = S_+(t_o^{*-})\times \exp(-\pi / 4)[\mathrm{Cos}(\pi / 4) - jSin(\pi / 4)],$$
$$S_-(t^{*+}) = S_-(t_o^{*+})\times \exp(-\pi / 4)[\mathrm{Cos}(-\pi / 4) + jSin(-\pi / 4)] = S_-(t_o^{*+})\times \exp(-\pi / 4)[\mathrm{Cos}(-\pi / 4) - jSin(\pi / 4)] .$$

(2.9)

These vector-functions at opposite moments (2.6a) hold opposite signs of their angles $\mp\pi / 4$ with values:

$$S_+(t^{*-}) \cong 0.2758\times 0.455 \cong +0.125, S_-(t^{*+}) \cong 0.2758\times 0.455 \cong -0.125 . \qquad (2.10)$$

Function $u^{t2}_{\pm}$ (1.2d), starting these opposite increments, turns them on angle $\varphi_-^2 - \varphi_+^2 = \pi / 2$ that equalizes the increments and starts entangling both equal increments with their angles within interval $t = \tau_k \mp 0$:

$$S_-^2(t = \tau_k + 0) = \delta S_-^1(t = \tau_k - 0)\times \downarrow_{\tau_k+0} \pi / 2 = S_-^1(t = \tau_k - 0)\times \exp(\pi / 2\times t^{*+}_{\tau_k+0})[\mathrm{Cos}(\pi / 2\times t^{*+}_{\tau_k+0}) + \mathrm{jSin}(\pi / 2\times t^{*+}_{\tau_k+0}),$$
$$S_+^2(t = \tau_k + 0) = \delta S_+^1(t = \tau_k - 0)\times \uparrow_{\tau_k+0} \pi / 2 = S_-^1(t = \tau_k - 0)\times \exp(-\pi / 2\times t^{*-}_{\tau_k+0})[\mathrm{Cos}(-\pi / 2\times t^{*-}_{\tau_k+0}) + \mathrm{jSin}(-\pi / 2\times t^{*-}_{\tau_k+0})$$

(2.11) at moments

$t^{*\pm}_{\tau_k+0}=[\mp\pi\times 2\delta t^{\pm}_{1k}],\delta t^{\pm}_{1k}=\delta t^{\pm}_{1}/1/2\tau_k\cong 0.4375,\delta t^{\pm}_{1}=\pm(0.5-\delta t^{k1}_{\pm}),\delta t^{k1}_{\pm}=\tau^{-o}_k/\tau_k+\delta t^{ko}_{\pm}/\tau_k,$

$\delta t^{k1}_{\pm}=0.25+0.03125=0.2895$

where $\delta t^{ko}_{\pm}/\tau_k\cong 32^{-1}$ evaluates dissimilarities between functions $u^{t2}_{\pm}=[u_+=(j+1),u_-=(j-1)]$ switching from moment $t=\tau_k-0$ to moment $t=\tau_k$.

The resulting values at $t=\tau_k+0$ are

$S^2_-(t=\tau_k+0)=0.125\exp(\pi/2\times 0.4375)\times 1\cong 0.25,S^2_+(t=\tau_k+0)=0.125\exp(\pi/2\times 0.4375)\times 1\cong 0.25$ (2.13) which, being in the same direction, are summing at that locality:

$$S^o_{\mp}=2S^2_{\mp}[(\delta t^{ko}_{\pm}/\tau_k)]\cong\mp 0.5. \quad (2.14)$$

The entanglement, starting with entropy (2.13), continues to entropy (2.14) up to cutting all entangled entropy increments.

Thus, the entanglement starts at angle $(\pi/2)\times 0.4375<\pi/4$ takes relative time interval of the impulse $\delta t^{ko}_{\pm}/\tau_k\cong 0.03125$ to ends on angle $\pi/2$.

Since only at angle $\pi/2$ the space interval within impulse begins, it means that *the entanglement starts before the space is formed and ends with beginning the space.*

Here $\tau_k=1/2o(\tau),o(\tau)=1Nat$ and

$$\delta t^{ko}_{\pm}=0.03125\times 1/2o(\tau)=0.015625o(\tau)=\varepsilon_{ok}. \quad (2.14a)$$

*Comments* 2.3

A potential path during creation of both entanglement and space could be a wormhole -a shortcut in space-time predicted by General Relativity. But real *space curvature does not exist at this time*. It may emerge only after entanglement at the moment of forming a Bit at the end of the impulse. Hence, space curvature may form at the *end of a microprocess (analog of a quantum process)* when the Bit, as the elementary unit of a macroprocess, emerges.

*Since the entanglement has no space measure, the entangled states can be everywhere in a space.* •

The $t=\tau_k\mp 0$ locality evaluates the $0_k$ -vicinity of action of inverse opposite functions (2.9), whose signs imply the signs of increments in (2.14) and in the following formulas.

The subsequent step-up function changes increment (2.14) according to Equations

$$S_{\mp}(\tau^{+o}_k)=S^o_{\mp}(\delta t^{ko}_{\pm}/\tau_k)\times\exp(t^{*+}_{\tau^{+o}_k}),t^{*+}_{\tau^{+o}_k}=[\pi/2\delta t^{*o}_k],\delta t^{*o}_k\in(\delta t^{*o}_{1k}\to\tau^{+o}_k/\tau_k), \quad (2.15)$$

at

$$\delta t^{*o}_{1k}=\delta t^{\pm}_{1k}/1/2\tau,\delta t^{\pm}_{1k}=\pm(0.5-\delta t^{k1}_{\pm}),$$
$$\delta t^{k1}_{\pm}=\delta t^{ko}_{\pm}/\tau_k+\tau^{+o}_k/\tau_k=0.25+0.03125=0.2895, \quad (2.15a)$$
$$\delta t^{\pm}_{1k}=\delta t^{\pm}_{1}/1/2\tau_k\cong 0.4375$$

with resulting value

$$S_{\mp}(\tau^{+o}_k)=\mp 0.5\exp(\pi/2\times 0.4375)=\mp 0.5\times(\cong 2)\cong\mp 1,, \quad (2.16)$$

which measures total entropy of the impulse

$$\bar{u}_k=|1|_k=1Nat. \quad (2.17)$$

Trajectories (2.10-2.16) describe anti-symmetric conjugated dynamics of the microprocess within the impulse, which is reversible, generating entangled entropy increments (2.16) up to the cutting moment.

*Comments* 2.4

From relation (3.4) and Jacobi-Hamiltonian variation equation $\partial S / \partial t = -\tilde{H}$ it follows that the microprocess Hamiltonian gets the form

$$\tilde{H}(t^*) = -u_{\pm}^{t2} S(t^*). . \quad (2.17a)$$

That Equation admits the conjugated Hamiltonian with both real and imaginary parts:

$$\tilde{H}(t^*) = -[(j+1)S + (j-1)S] = -[(\dot{S}_+(t^*) / S_+(t^*) + \dot{S}_-(t^*) / S_-(t^*)]S(t^*). \quad (2.17b)$$

At the entanglement, the conjugated entropies are

$$S_+(t^*+) = S_-(t^*+), S = S_+(t^*+) + S_-(t^*+) = 2S_+(t^*+) = 2S_-(t^*+)$$

and the Hamiltonian is

$$\tilde{H}(t^*+) = -[(\dot{S}_+(t^*) / 2 + \dot{S}_-(t^*) / 2] = -\dot{S}(t^*+). \bullet \quad (2.17c).$$

Cutting this entangled joint entropy at moment $\tau_k^+ \cong 0_k + \tau_k^{o+}$ converts it to the equal Information contribution

$$S_{\mp}^o[\tau_k^+] = \Delta I[\tau_k^+] \cong 1.44 \,\text{bit} \quad (2.18)$$

which each impulse $\bar{u}_k$ produces.

An interacting impulse with the impulse microprocess delivers entropy on the $0_k$-vicinity of the cutting moment:

$$S_c^*(\tau_k^+) = \exp 0_k = 1. \quad (2.19)$$

Each current impulse step-up action $[\uparrow_{\tau_k^{+o}} \bar{u}_+^o]$ (in (3.6)) generates an Information Bit from the microprocess reversible entropy.

Thus, the jumping actions provide the minimal discrete displacement (2.3a, 2.2), which rotates the opposite entropy increments. The interactive jump generates a pair of random interactive actions on the bordered impulses, which are equally probable, reversible within the probabilities of multiple random interactive actions.

The curving shift initiates a microprocess within the bordered impulse running the superposition and entanglement of the conjugates entropy fractions during time interval starting with the jump. The entanglement starts before the space of the shift is formed and ends with beginning the space shift, being small part of impulse reversible time interval..

### 3.2. The Rotating Conjugated Dynamics of the Microprocess

*Starting step functions* $u_{\pm}^{t1}$ initiates increments of the entropies on interval $o(\tau_k - 0)$ by moment $t = \tau_k - 0$:

$$\delta S_+[u_+^{t1}] = \delta S_+^1(t = \tau_k - 0)) = \delta S_+^1(t = \tau_k^{-o}) \uparrow_{\tau_k^{+o}} (j-1), \delta S_-[u_-^{t1}] = \delta S_-^1(t = \tau_k - 0) = \delta S_-^1(t = \tau_k^{-o}) \downarrow_{\tau_k^{-o}} (j+1). \quad (2.20)$$

Step functions $u_{\pm}^{t2}$ (1.2d) starting at $t = \tau_k - 0$ contribute the entropy increments on interv

$$\delta S_+[u_+^{t2}] = \partial S_+^2(t = \tau_k) = \partial S_+^2(t = \tau_k - 0)) \uparrow_{\tau_k} (j+1), \delta S_-[u_-^{t2}] = \partial S_-^2(t = \tau_k) = \partial S_-^2(t = \tau_k - 0)) \downarrow_{\tau_k} (-j+1). \quad (2.21)$$

Complex function $u_+^{t1}$ turns on the multiplication of functions $\delta S_+^1(t = \tau_k^{-o})$ on angle $\varphi_+^1 = -\pi / 4$, and function $u_-^{t1}$ turns on the multiplication function $\partial S_-^1(t = \tau_k^{-o})$ on angle $\varphi_-^1 = \pi / 4$ by moment $t = \tau_k - 0$ .This brings the entropy increments

$$\delta S^1_+(t=\tau_k-0))=\delta S^1_+(t=\tau_k^{-o})\times\uparrow_{\tau_k^{+o}}-\pi/4, \delta S^1_-(t=\tau_k-0))=\delta S^1_-(t=\tau_k^{-o})\times\downarrow_{\tau_k^{+o}}\pi/4 \quad (2.22)$$

Analogously, step functions $u^{t2}_\pm$, starting at $t=\tau_k-0$, turn entropy increments (2.22) on angles $\varphi^2_-=\pi/4$ by moment $t=\tau_k$ and on angle $\varphi^2_+=-\pi/4$ the entropy increments by moment $t=\tau_k$:

$$\delta S^2_-(t=\tau_k)=\delta S^2_-(t=\tau_k-0)\times\downarrow_{\tau_k}\pi/4, \delta S^2_+(t=\tau_k)=\delta S^2_+(t=\tau_k-0)\times\uparrow_{\tau_k}-\pi/4 \ . \quad (2.23)$$

The difference of angles between the functions in (2.22): $\varphi^1_+-\varphi^1_-=-\pi/2$ is overcoming on time interval $o(\tau_k-0)=\tau_k^{-o}+1/2o(\tau_k)$.

After that, control $u^{t2}_\pm$, starting with opposite increments (2.23), turns them on angle $\varphi^2_--\varphi^2_+=\pi/2$ equalizing entropy increments (2.23).

That launches *entanglement* of entropy increments and their angles *within* interval $o(\tau_k)$ (on a middle of the impulse) at $t=\tau_k$:

$$\delta S^2_+(t=\tau_k)=\delta S^2_-(t=\tau_k)=\delta S^2_\mp . \quad (2.24)$$

Control $\bar{u}_-=0.5$, turning the time-located vector-function at the impulse beginning:

$$u^t_-=u_-(\tau_k^{-o}):\xleftarrow{\tau_k^{-o},\bar{u}_-=0.5,}\delta\varphi_1=0 \quad (2.25)$$

on angle

$$\delta\varphi_1=\varphi^1_+-\varphi^1_-=\pi/2,$$

transforms it to space vector $u_+(\tau_k-0)=\uparrow_{\tau_k-o}\bar{u}_+=1$ during a jump from moment $t=\tau_k^{-o}$ to moment $t=\tau_k-0$ on interval $o(\tau_k-0)$ in (2.22).

Then, vector-function $\downarrow_{\tau_k}\bar{u}^o_-=2$, starting on time $t=\tau_k-0$ with space interval $\bar{u}^o_-=2$, jumps to vector-function $\uparrow_{\tau_{k+0}}\bar{u}^o_+=2$ forming on time interval $o(\tau_k+0)=1/2o(\tau_k)+\tau_k^+$ the additive space-time impulse

$$u_\mp=[\downarrow_{\tau_k+0}\bar{u}^o_-]+[\uparrow_{\tau_k^{+o}}\bar{u}^o_+]. \quad (2.26)$$

The first part of (2.26) equalizes increments (2.24) within *space-time* interval $\bar{u}_-\times 1/2o(\tau_k)$ and then joins them on $\bar{u}_-\times o(\tau_k+0)$, which finalizes the entanglement.

The last part of impulse (2.26) cuts-kills the entangled increments on interval $\bar{u}_+\times\tau_k^+$ at ending moment $\tau_k^+$.

Section 2.3.4 details the time-space relation and their measures.

Relations (2.1-2.26) lead to following specifics of the microprocess.

3.1a. Step-functions $u^{t1}_\pm$ initiate microprocess $\tilde{x}_{otk1}=\tilde{x}(t\in o(\tau_k-0))$ on beginning of the impulse discrete interval $o(\tau_k-0)$ with only additive increments (2.2). Opposite step functions $u^{t2}_\pm$ continue microprocess $\tilde{x}_{otk2}=\tilde{x}(t\in o(\tau_k+0))$ within interval $o(\tau_k+0)$ with both additive and multiplicative increments (2.3) preserving the process Markov properties.

3.1b. Space-time impulse (2.16) within interval $o(\tau_k + 0)$ processes entanglement of increments (2.25) of microprocess $\tilde{x}_{otk2} = \tilde{x}(t \in o(\tau_k + 0))$ summing these increments on $o(\tau_k)$ locality of $t = \tau_k$:

$$S^o_{\mp} = 2\delta S^2_{\mp}[(o(\tau_k)] . \qquad (2.27)$$

Then it kills entropies (2.27) at ending moment $\tau_k^{o+} \to \tau_k^+$:

$$S^o_{\mp}[\tau_k^+] = 0 . \qquad (2.27a)$$

The microprocess, producing entropy increment (2.27) within the impulse interval, is reversible before killing which converts the increments in equal Information contribution

$$S^o_{\mp}[\tau_k^+] \Rightarrow \Delta I[\tau_k^+] . \qquad (2.27b)$$

The Information, emerging at the ending impulse time interval, finalizes the injection of energy with step-up control $[\uparrow_{\tau_k^{+o}} \bar{u}^o_+]$, which starts at a transitional impulse. The energy injection can be a result of the impulse's middle interaction with environment.

From the impulse ending moment starts an irreversible Information process of multiple Bits.

3.1c. Transferring the initial time-located vector to equivalent space-vector $\uparrow_{\tau_k - o} \bar{u}_+$ transforms the transition impulse, starting within a jump of time $\tau_k^{-o}$ on interval of $\bar{u}_- = 0.5$ up to creating space interval $\bar{u}_+ = 1$.

The opposite space vector $\downarrow_{\tau_k} \bar{u}^o_- = 2$, acting on relative time interval $1/2o(\tau_k)/(\tau_k^{+o} - \tau_k^{-o}) = 0.5$, forms space-time function $\downarrow_{\tau_k} \bar{u}^1_- : \bar{u}^1_- = 2 \times 0.5 = 1$, which, as inverse equivalent of opposite function $\uparrow_{\tau_k - o} \bar{u}_+$, neutralizes it to zero. Both time duration of $\bar{u}_- = 0.5$ and $\bar{u}_+ = 1$ concentrate these functions in transition interval $\tau_k - (\tau_k - 0) = 0_k$. After that, within the entire impulse, only step-down functions $[\downarrow_{\tau_k^{-o}} \bar{u}_-]$ on time interval $\bar{u}_- = 0.5$ and step-up function $[\uparrow_{\tau_k^{+o}} \bar{u}^1_+]$on space-time interval $\bar{u}^1_+ = \bar{u}_+ \times \tau_k^+ = 2_{\tau_k^+}$ are left. That determines size of the discrete $1-0$ impulse by multiplicative measure $U_m = |0.5 \times 2| = |1|_k = \bar{u}_k$ generating an Information Bit.

Therefore, functions $u_+(\tau_k - 0) = \uparrow_{\tau_k - o} \bar{u}_+$ and $u_-(\tau_k) \downarrow_{\tau_k} \bar{u}^o_-$ are transitional during formation of that impulse and creation time-space microprocess $\tilde{x}_{otk} = \tilde{x}(t \in 1/2o(\tau_k), h_k \in 2_{\tau_k^+})$ with final entropy increment (2.27) and a virtual logic. The microprocess transits from the entropy increment at $\tau_k$-locality (2.27) to actual Information (2.27b).

### 3.3. Probability Functions of the Microprocess

Amplitudes of the process probability functions at $S^*_{\mp}(\tau_k^{+o}) = |S^*_+| = |S^*_-| = 1$ are equal and independent:

$$p_{+a} = 0.3679, p_{-a} = 0.3679 . \qquad (2.28)$$

That leads to

$$p_{+a} p_{-a} = p^2_{\pm a} = 0.1353, S^*_{\mp a} = -\ln p^2_{a\pm} = 2 ,$$

or at

$S^*_{\mp a} = 2$, to

$p_{a\pm} = \exp(-2) = 0.1353$. (2.28a)

Where

$S^*_{\mp a} = S^*_{\mp}(\tau_k^{+o}+) + S^*_c(\tau_k^{+}+)$

includes the interactive components at $\tau_k^{+o}+$ following $k$ impulse.

Functions $u_+ = (j-1), u_- = (j+1)$, satisfying (1.IA), fulfill the additive property at the impulse starting interval $o[t_o^{\mp}]$, running the anti-symmetric entropy fractions.

Opposite functions $u_+ = (1+j), u_- = (1-j)$, satisfying (1.IB) by the end of impulse at $\uparrow_{\tau_{k+}^{+o}} \bar{u}_{\pm}$, entangle these entropy fractions within space interval $\bar{u}_{\pm} = \pm 2$ of impulse' $|1/2\times 2| = |\bar{u}_k| = |1|_k$.

The entangling fractions hold the equal impulse probabilities (2.28), which indicate appearance of entangled anti-symmetric fractions simultaneously with starting space interval.

Interacting probability amplitudes $p_{+a}, p_{-a}$ of $p_{\pm a}$ satisfy multiplicative relation $p_{\pm a} = \sqrt{p_{+a} p_{-a}}$.

However, the sum of non-interacting probabilities: $p_+ + p_- = \exp(-S^*_+) + \exp(-S^*_-) = p_{\pm} \neq \mathrm{p}_{a\pm}$ does not comply with it.

The summary probability $p_{\pm am} = 0.7358$ of the non-interacting entropies increments is unequal to probability $p_{\pm a}$ of interacting entropies.

The interacting probabilities in transitional impulse $[\uparrow 1_{\tau|_k^-} \downarrow 1_{\tau|_k^+}]\bar{u}_k$ on $\tau_k$-locality violate their additive property, but preserve additive of the entropy increments.

The impulse microprocess on the ending interval preserves both additive and multiplicative properties only for the entropy increments.

The basic relations for the impulse's entropy and probability are equivalent for quantum mechanics (QM) probability amplitudes relations.

However, the impulse cutting probabilities $p_+, p_-$ are the probability of random events in the hidden correlations, while probability amplitudes $p_{+a}, p_{-a}$ are attributes of the microprocess starting within the cutting impulse. That distinguishes the considered microprocess from the related QM equations, considered for physical particles.

The entropy of multiple impulses integrates the macroprocess along the observing random distributions.

With minimal impulse entropy ½ Nat starting a Virtual Observer, each following impulse' initial entropy $S_{\pm}(t_o) = 0.25Nat$ self-generates entropy $S^*_{\mp a} = 0.5Nat$.

Thus, the Virtual Observer's time-space microprocess starts with probability $p_{a\pm} = \exp(-0.5) = 0.6015$.

Probability $p_{a\pm} = 0.1353$ is relational to the impulse initial conditions, which evaluates appearance of time-space actual impulse (satisfying (2.26)) that *decreases* its initial entropy on $S^*_{\mp a} = 2$ Nat.

The impulse's invariant measure, satisfying the minimax, preserves $p_{a\pm}$ along the time-space microprocess for multiple time-space impulses.

Reaching the probability of appearance, the time-space impulse needs $m_p = 0.6015 / 0.1353 \cong 4.4457 \approx 5$ multiplications of invariant $p_{a\pm} = 0.1353$, which predicts the *a priori* probability of the impulse's reactive action.

The space interval, beginning the displacement shift, starts within the interval of entanglement (2.15a) having probability

$$P_{\Delta}^{*}(\delta t_{\pm}^{k1}) = \exp(-|S_{\mp}^{*}(\delta t_{\pm}^{k1})|, P_{\Delta}^{*}(\delta t_{\pm}^{k1}) = 0.821214 \qquad (2.28b)$$

at

$$\delta t_{\pm}^{k1} = 0.2895, S_{\mp}^{*}(\delta t_{\pm}^{k1}) = \mp 0.125\exp(\pi / 2 \times \delta t_{\pm}^{k1}) = \mp 0.1969415, \qquad (2.28c)$$

and continues during the shift, extending to the space part of the impulse multiplicative measure after the displacement ends. Hence, each reversible microprocess within the impulse generates invariant increment of entropy, which sequentially minimizes the starting uncertainty of the observation.

Assigning the entropy minimal uncertainty measure $h_{\alpha}^{o} = 1/137$, the physical structural parameter of energy, which includes the Plank constant's equivalent of energy, leads to relation:

$$S_{\mp a}^{*} = 2h_{\alpha}^{o}, p_{\pm a} = \exp(-2h_{\alpha}^{o}) = 0.98555075021 \to 1. \qquad (2.29)$$

This evaluates the probability of a real impulse's physical strength of the coupling independently chosen entropy fractions.

The initially orthogonal non-interacting entropy increments $S_{+a}^{*} = h_{\alpha}^{o}$ , $S_{-a}^{*} = h_{\alpha}^{o}$ at mutual interactive actions, satisfy multiplicative relation

$$S_{\mp a}^{*} = (h_{\alpha}^{o})^2[\mathrm{Cos}^2(\bar{u}t) + Sin^2(\bar{u}t)]|_{t_o^{\mp}}^{t=1/2\tau} = (h_{\alpha}^{o})^2 = inv \qquad (2.29a)$$

which at $S_{\mp a}^{*} = (h_{\alpha}^{o})^2 \to 0$ approaches $p_{\pm a}^{*} = \exp[-(h_{\alpha}^{o})^2] \to 1$.

The impulse interaction adjoins the initial orthogonal geometrical sum of entropy fractions in linear sum $2h_{\alpha}^{o}$.

Starting physical coupling with double structural $h_{\alpha}^{o}$ creates initial Information triple with probability (2.29).

*The microprocess initiates the merge that starts with the jumping actions' multiplication on the bordered impulse time according to (1.16b) succeeding displacement (2.3a) during the merge. Both follow from the EF extreme.*

*The multiplication violates the Markov property (1.1B) leading to a complex control (1.2c), which starts the microprocess within the displacement and rotates the initial conjugated entropy increments.*
*The microprocess (2.3.3) emerges from multiple interactions starting with probabilities (2.38), inverse entropy $S_{\mp a}^{*} = 2$, and injection of the related random energy. With growing probabilities up to 1, this energy increases rising the equivalent entropy, which is leading to the equal Information Bit.*
*The energy aspect is in Sec.2.6, where the Jarzynski Equality (JE) [3], applied to the evolving microprocess, measures thermodynamic energy connecting the JE with this process's Information measure. Thus, this developing microprocess presents a Stochastic Quantum process with evolving thermodynamics and a path to Information Macrodynamics* [4].

Examples.

Let us find which of the entropy functional expression meets requirements (1.1A,B) within discrete intervals $\Delta_t = (t - s) \to o(t)$, particularly on $\Delta_k = (\tau_k^{-o} - s_k^{+o}) \to o(\tau_k^{-o})$ under opposite functions $u_+, u_-$:

$$u_+(s_k^{+o}) = +1_{s_k^{+o}}\bar{u} = \bar{u}(s_k^{+o}), u_- = -u_+(s_k^{+o}) = -1_{s_k^{+o}}\bar{u} \,. \qquad (2.30)$$

Following relations (1.11), we get entropy increments

$$S_+[\tilde{x}_t / \varsigma_t]|_{s_k^+}^{t\to\tau_k^{-o}} = -1/2[u_+(s_k^{+o})](\tau_k^{-o} - s_k^{+o})^{-1}(s_k^{+o})^2 = -1/2[u_+(s_k^{+o})(s_k^{+o})^2 / s_k^{+o}(3-1)] = -1/4[u_+(s_k^{+o})s_k^{+o}] \quad (2.30a)$$

$$S_-[\tilde{x}_t / \varsigma_t]|_{s_k^+}^{t\to\tau_k^{-o}} = 1/2[u_-(s_k^{+o})](\tau_k^{-o} - s_k^{+o})^{-1}(s_k^{+o})^2 = 1/2[u_-(s_k^{+o})(s_k^{+o})^2 / s_k^{+o}(3-1)] = 1/4[u_-(s_k^{+o})s_k^{+o}], \quad (2.30b)$$

which satisfy

$$S_+[\tilde{x}_t / \varsigma_t]|_{s_k^+}^{t\to\tau_k^{-o}} = -S_-[\tilde{x}_t / \varsigma_t]|_{s_k^+}^{t\to\tau_k^{-o}}, \qquad (2.31a)$$

$$S_+[\tilde{x}_t / \varsigma_t]|_{s_k^+}^{t\to\tau_k^{-o}} - S_-[\tilde{x}_t / \varsigma_t]|_{s_k^+}^{t\to\tau_k^{-o}} = -1/2[\bar{u}(s_k^{+o})s_k^{+o}] = \Delta S[\tilde{x}_t / \varsigma_t]|_{s_k^+}^{t\to\tau_k^{-o}} \,. \qquad (2.31b)$$

Relations

$$4S_+[\tilde{x}_t / \varsigma_t]|_{s_k^+}^{t\to\tau_k^{-o}} / s_k^{+o} = -\bar{u}(s_k^{+o})(s_k^{+o}) = -2\times 1_{s_k^{+o}},$$

satisfy conditions

$$4S_+[\tilde{x}_t / \varsigma_t]|_{s_k^+}^{t\to\tau_k^{-o}} / s_k^{+o} \times 4S_-[\tilde{x}_t / \varsigma_t]|_{s_k^+}^{t\to\tau_k^{-o}} / s_k^{+o} = -\bar{u}(s_k^{+o}) \times \bar{u}(s_k^{+o}) = -(2\times 1_{s_k^{+o}}) \times (2\times 1_{s_k^{+o}}) = -(2\times 1_{s_k^{+o}})^2, \quad (2.32a)$$

$$4S_+[\tilde{x}_t / \varsigma_t]|_{s_k^+}^{t\to\tau_k^{-o}} / s_k^{+o} - 4S_-[\tilde{x}_t / \varsigma_t]|_{s_k^+}^{t\to\tau_k^{-o}} / s_k^{+o} = -\bar{u}(s_k^{+o}) - \bar{u}(s_k^{+o}) = -(2\times 1_{s_k^{+o}}) - (2\times 1_{s_k^{+o}}) = (2\times 1_{s_k^{+o}})^2 \,. \quad (2.32b)$$

These entropy expressions at any *current* moment $t$ within $\Delta_t = (t - s_k^{+o})$ do not comply with (1.1A, B).

The same results hold true for the entropy functional increments under functions

$$u_+ = +1_{s_k^{+o}}\bar{u}, u_- = -1_{\tau_k^{-o}}\bar{u} \,. \qquad (2.33)$$

Actually, for this function on $\Delta_t = (t - s_k^{+o})$ we have

$$\Delta S[\tilde{x}_t / \varsigma_t]|_{s_k^+}^{t} = -1/2(u_-(t) - u_+(s_k^{+o}))(t - s_k^{+o})^{-1}(s_k^{+o})^2 \qquad (2.34)$$

which for $t \to \tau_k^{-o}$ holds

$$\Delta S[\tilde{x}_t / \varsigma_t]|_{s_k^+}^{t\to\tau_k^{-o}} = -1/2(u_-(\tau_k^{-o}) - u_+(s_k^{+o}))(\tau_k^{-o} - s_k^{+o})^{-1}(s_k^{+o})^2,$$

and satisfies relations

$$S_+[\tilde{x}_t / \varsigma_t]|_{s_k^+}^{t\to\tau_k^{-o}} - S_-[\tilde{x}_t / \varsigma_t]|_{s_k^+}^{t\to\tau_k^{-o}} = \Delta S[\tilde{x}_t / \varsigma_t]|_{s_k^+}^{t\to\tau_k^{-o}}, \qquad (2.34a)$$

$$S_+[\tilde{x}_t / \varsigma_t]|_{s_k^+}^{t\to\tau_k^{-o}} = -S_-[\tilde{x}_t / \varsigma_t]|_{s_k^+}^{t\to\tau_k^{-o}} \qquad (2.34b)$$

which determine

$$S_+[\tilde{x}_t / \varsigma_t]|_{s_k^+}^{t\to\tau_k^{-o}} = -1/4(u_-(\tau_k^{-o}) - u_+(s_k^{+o}))(\tau_k^{-o} - s_k^{+o})^{-1}(s_k^{+o})^2 \qquad (2.35a)$$

$$S_-[\tilde{x}_t / \varsigma_t]|_{s_k^+}^{t\to\tau_k^{-o}} = 1/4(u_-(\tau_k^{-o}) - u_+(s_k^{+o}))(\tau_k^{-o} - s_k^{+o})^{-1}(s_k^{+o})^2 \,. \qquad (2.35b)$$

We get the entropy expressions through opposite directional discrete functions in (2.35a, b):

$$S_+[\tilde{x}_t / \varsigma_t]|_{s_k^+}^{t\to\tau_k^{-o}} \, 4(\tau_k^{-o} - s_k^{+o})^{-1}(s_k^{+o})^2 = -(u_-(\tau_k^{-o}) - u_+(s_k^{+o})),$$

$$S_-[\tilde{x}_t / \varsigma_t]|_{s_k^+}^{t\to\tau_k^{-o}} \, 4(\tau_k^{-o} - s_k^{+o})(s_k^{+o})^{-2} = u_-(\tau_k^{-o}) - u_+(s_k^{+o}),$$

which satisfy additivity at

$$-2(u_-(\tau_k^{-o}) - u_+(s_k^{+o})) = -2(-1_{\tau_k^{-o}}\bar{u} - 1_{s_k^{+o}}\bar{u})] = 2\bar{u}[1_{\tau_k^{-o}} + 1_{s_k^{+o}}] = 4[1_{\tau_k^{-o}} + 1_{s_k^{+o}}] \,. \quad (2.36)$$

While for each

$$S_{+}[\tilde{x}_t / \varsigma_t]|_{s_k^+}^{t\to\tau_k^{-o}} \; 4(\tau_k^{-o}-s_k^{+o})(s_k^{+o})^{-2} = -\bar{u}[-1_{\tau_k^{-o}} - 1_{s_k^{+o}}], \quad (2.36a)$$

$$S_{-}[\tilde{x}_t / \varsigma_t]|_{s_k^+}^{t\to\tau_k^{-o}} \; 4(\tau_k^{-o}-s_k^{+o})(s_k^{+o})^{-2} = \bar{u}[-1_{\tau_k^{-o}} - 1_{s_k^{+o}}] \quad (2.36b)$$

satisfaction of both 1.1A, B :

$$-(u_{-}(\tau_k^{-o}) - u_{+}(s_k^{+o})) \times (u_{-}(\tau_k^{-o}) - u_{+}(s_k^{+o})) = -[u_{-}(\tau_k^{-o}) - u_{+}(s_k^{+o})]^2,$$

requires

$$\bar{u} = -2j, \quad (2.37)$$

when holds equality

$$-[u_{-}(\tau_k^{-o}) - u_{+}(s_k^{+o})]^2 = (-2j)^2[-1_{\tau_k^{-o}} - 1_{s_k^{+o}}]^2. \quad (2.37a)$$

Simultaneous satisfaction of both 1.1.A, B leads to

$$\Delta S\;[\tilde{x}_t / \varsigma_t]|_{s_k^+}^{t\to\tau_k^{-o}} \; 2(\tau_k^{-o}-s_k^{+o})(s_k^{+o})^{-2} = -2\bar{u}[-1_{\tau_k^{-o}} - 1_{s_k^{+o}}] = 4j[1_{\tau_k^{-o}} + 1_{s_k^{+o}}],$$

$$-(-1_{\tau_k^{-o}}\bar{u} - 1_{s_k^{+o}}\bar{u}) \times (-1_{\tau_k^{-o}}\bar{u} - 1_{s_k^{+o}}\bar{u}) = (-2j)^2(-1_{\tau_k^{-o}} + 1_{s_k^{+o}})^2. \quad (2.37b)$$

At $o(t) \to 0$, the above conditions admit an instant existence of both $(-1_{\tau_k^{-o}}\bar{u}, +1_{s_k^{+o}}\bar{u})$.

Thus, under function (2.35), the entropy expressions are imaginary:

$$S_{+}[\tilde{x}_t / \varsigma_t]|_{s_k^+}^{t\to\tau_k^{-o}} \; 4(\tau_k^{-o}-s_k^{+o})(s_k^{+o})^{-2} = -2j[-1_{\tau_k^{-o}} + 1_{s_k^{+o}}] = -2j[1_{s_k^{+o}}^{\tau_k^{-o}}], \quad (2.38a)$$

$$S_{-}[\tilde{x}_t / \varsigma_t]|_{s_k^+}^{t\to\tau_k^{-o}} \; 4(\tau_k^{-o}-s_k^{+o})(s_k^{+o})^{-2} = 2j[-1_{\tau_k^{-o}} + 1_{s_k^{+o}}] = 2j[1_{s_k^{+o}}^{\tau_k^{-o}}], \quad (2.38b)$$

at their multiplicative and additive relations hold:

$$S_{-}[\tilde{x}_t / \varsigma_t]|_{s_k^+}^{t\to\tau_k^{-o}} \; 4(\tau_k^{-o}-s_k^{+o})(s_k^{+o})^{-2} \times S_{+}[\tilde{x}_t / \varsigma_t]|_{s_k^+}^{t\to\tau_k^{-o}} \; 4(\tau_k^{-o}-s_k^{+o})(s_k^{+o})^{-2} = 4, \quad (2.39)$$

$$S_{+}[\tilde{x}_t / \varsigma_t]|_{s_k^+}^{t\to\tau_k^{-o}} \; 4(\tau_k^{-o}-s_k^{+o})(s_k^{+o})^{-2} - S_{-}[\tilde{x}_t / \varsigma_t]|_{s_k^+}^{t\to\tau_k^{-o}} \; 4(\tau_k^{-o}-s_k^{+o})(s_k^{+o})^{-2} = -j2[1_{\tau_k^{-o}} + 1_{s_k^{+o}}] = -j2[1_{s_k^{+o}}^{\tau_k^{-o}}],$$

$$S_{+}[\tilde{x}_t / \varsigma_t]|_{s_k^+}^{t\to\tau_k^{-o}} - S_{-}[\tilde{x}_t / \varsigma_t]|_{s_k^+}^{t\to\tau_k^{-o}} = 1/2j[1_{s_k^{+o}}^{\tau_k^{-o}}]. \quad (2.40)$$

Relations (2.36), (2.36a, b) satisfy additivity only at points $\tau_k^{-o}, s_k^{+o}$.

Between these points, within $\Delta_t = (t - s_k^{+o}) \to o(t)$, the entropy expressions (2.38a, b) and (2.40) are imaginary.

Time direction may go back within this interval until an interaction occurs. •

These examples concur with *(2.5), (2.6) and illustrate it. Results show that a window of interaction with an environment opens only on the impulse border twice: at the beginning between moments* $\delta_k^{\tau+}/4$ *and* $\tau_k^{-o}$ *when the entropy flow with energy accesses impulse, and at the end of a gap when an entangled entropy with accesses of energy converts to equivalent Information.*

**4. THE RELATION BETWEEN THE CURVED TIME AND EQUIVALENT SPACE LENGTH WITHIN AN IMPULSE**

Let us have a two-dimensional rectangle impulse with plain $p$ measured in time length $[\tau]$ unit and orthogonal $h$ measured in space length $[l]$ unit, with the rectangle measure

$$M_i = p \times h. \quad (2.41)$$

The problem: Having a measure of the plain part of the impulse $\mathrm{M}_p$ to *find* high $h$ at equal measures of both parts:

$$\mathrm{M}_p = \mathrm{M}_h \text{ and } \mathrm{M}_p + \mathrm{M}_h = M_i. \quad (2.42)$$

From (2.42) it follows

$\mathrm{M}_h = 1/2M_i = 1/2p \times h$.  (2.43)

Assuming the impulse has only equal plain parts $1/2p$, it measures $\mathrm{M}_p = (1/2p)^2$.

Then, from $\mathrm{M}_p = (1/2p)^2 = \mathrm{M}_h = 1/2p \times h$ it follows that

$h/p = 1/2$.  (2.44)

Let us find a length unit $[l]$ of the curved time unit $[\tau]$ rotating on angle $\pi/2$ using relations

$2\pi h[l]/4 = 1/2p[\tau]$,  (2.45a)

$[\tau]/[l] = \pi h/p$.  (2.45)

Substitution (2.44) leads to ratio of the measured units:

$[\tau]/[l] = \pi/2$.  (2.46)

Relation (2.46) sustains orthogonality of these units in a time-space coordinate system, but since initial relations (2.42) are linear, ratio (2.46) represents a linear connection of time-space units (2.45). The impulse-jumps curve the time unit in (2.8). According to Proposition 1.3, the impulse' invariant entropy implies the multiplication, starting the rotation.

The microprocess, built in rotation movement curving the impulse time, adjoins the initial orthogonal axis of time and space coordinates (Figure 1a). The curving impulse illustrates Figure 1b.

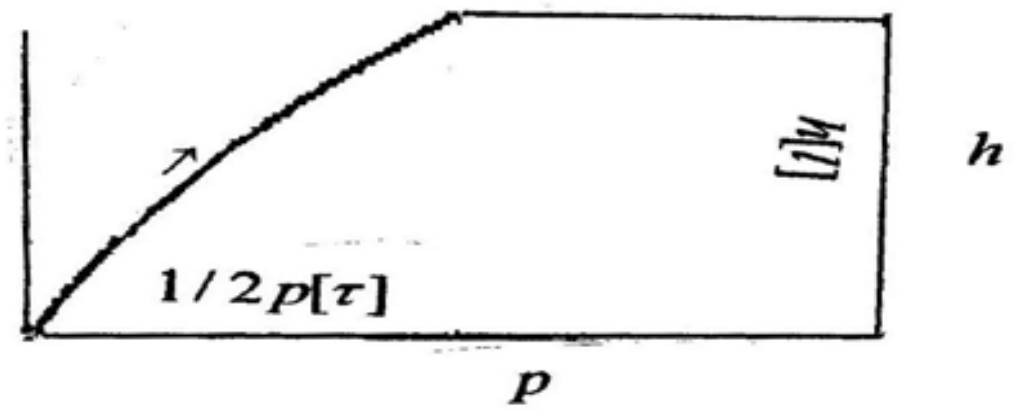

Figure 1(a). Illustration of origin the impulse space coordinate measure $h[l]$ at curving time coordinate measure $1/2p[\tau]$ in transitional movement.

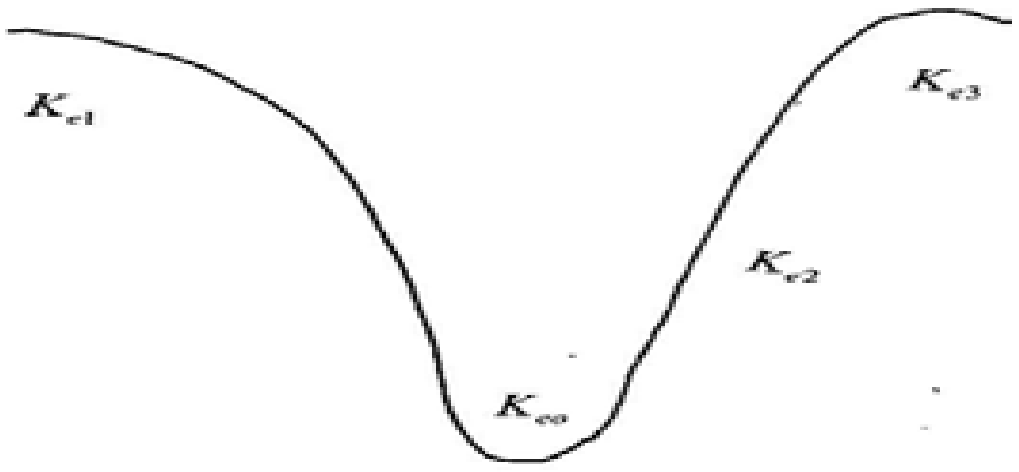

Figure 1(b). Curving impulse with curvature $K_{e1}$ of the impulse step-down part, curvature $K_{eo}$ of the cutting part, curvature $K_{e2}$ of impulse transferred part, and curvature $K_{e3}$ of the final part cutting all impulse entropy.

The impulses, preserving the multiplicative and additive measures, have common ratio of $h/p = 1/2$, whose curving part $p = 1/2$ brings universal ratio (2.46), which concurs with Lemma 2.1, (1.2a).

At the above assumption, measure $\mathrm{M}_h$ does not exist until the impulse-jump curves its only time plain $1/2p$ at transition of the impulse. This transition is measured only in time. The following impulse transition is measured both in time $1/2p$ and space coordinate $h$. According to (3.43), measure $\mathrm{M}_h$ emerges only on a half of that impulse' total measure $M_i$.

The transitional impulse could start on border of the virtual impulses $\downarrow\uparrow$, where the transition, curving time $\delta t_p = 1/2p$ under impulse-jump during $\delta t_p \to 0$, leads to

$$M_p \to 0 \text{ at } M_h \Rightarrow M_i = p \times h \ . \qquad (2.47)$$

If a virtual impulse $\downarrow\uparrow$ has equal opposite functions $u_-(t), u_+(t+\Delta)$, at $\bar{u}_+ = \bar{u}_-$, the additive condition for measure (1.2a): $\mathrm{U}_a(\Delta) = 0$ is violated, and the impulse holds only multiplicative measure $\mathrm{U}_m(\Delta) \neq 0$ in relation (1.2C): $U_m(\Delta) = \mathrm{U}_{am}$ which is finite only at $\bar{u}_+ = \bar{u}_- \neq 0$.

If any of $\bar{u}_+ = 0$, *or* $\bar{u}_- = 0$, both multiplicative $\mathrm{U}_m(\Delta) = 0$ and additive $\mathrm{U}_a(\Delta) = 0$ disappear.

At $\bar{u}_- \neq 0,$ measure $\mathrm{U}_a(\Delta)$ is a finite and positive, specifically, at $\bar{u}_- = 1$ it leads to $\mathrm{U}_a(\Delta) = 1$ preserving measure $\mathrm{U}_{amk} = |\mathrm{U}_a|_k$.

Existence of the transitional impulse has shown in (Secs. 2.3.2, 2.3.3).

An impulse-jump at $o[t_o^{\mp}] \to \delta t_p \to 0$ curves a "needle pleat" space at the transition to the finite form of the impulse.

The Bayes probabilities measure may overcome this transitive gap.

Since entropy (1.3.2.1) is proportional to the correlation time interval, whose impulse curvature $K_s = h[l]^{-1}$ is positive, this curved entropy is positive. The curving needle cut changes the curvature sign converting this entropy to Information.

**4.1. Curvature of the Impulse**

An external step-control carries entropy which evaluates:

$$\delta_{ue}^{i} = 1/4(\mathrm{u}_{io} - \mathrm{u}_i), \qquad (2.48)$$

where $\mathrm{u}_{io} = \ln 2 \cong 0.7Nat$ is the total cutoff entropy of the impulse and $\mathrm{u}_i \cong 0.5Nat$ is its cutting part.

The same entropy-Information carries each impulse step-down and step-up control, while both controls carry $\delta_{ueo}^{i} \cong 0.1Nat$.

That evaluates Information wide of each single impulse control's cut which the impulse carries:

$$\delta_{ue}^{i} \cong 0.05Nat. \qquad (2.48a)$$

To create Information, the starting step-down part and the step-up part transfer entropy to the final killing part generating Information. These three parts carry the entropy measures accordingly:

$$\delta_{ue1}^{i} \cong 0.025Nat, \delta_{ue2}^{i} \cong 0.02895Nat, \delta_{ue3}^{i} \cong 0.01847Nat. \qquad (2.48b)$$

The first relation in (2.48b) allows estimate Euclid's curvature $K_{e1}$ of the impulse step-down part, related to currying entropy $0.25Nat$ and its increment $\delta K_{e1}$:

$$K_{e1} = (r_{e1})^{-1}, r_{e1} = \sqrt{1 + (0.025/0.25)^2} = \mp 1.0049875,$$
$$K_{e1} \cong -0.995037, \delta K_{e1} \cong -0.004963. \qquad (2.49)$$

The cutting part's curvature estimates relations

$$K_{eo} = (r_{eo})^{-1}, r_{eo} = \mp\sqrt{1+(0.1/0.5)^2} = 1.0198,$$
$$K_{eo} \cong -0.98058, \delta K_{eo} \cong -0.01942. \quad (2.49a)$$

The transferred part's curvature estimates relations

$$K_{e2} = (r_{e2})^{-1}, r_{e2} = \sqrt{1+(0.02895/0.25)^2} \cong 1.0066825,$$
$$K_{e2} \cong +0.993362, \delta K_{e2} \cong 0.006638 \quad (2.49b)$$

which is opposite to the step-down part.

The final part cutting all impulse entropy estimates curvatures

$$K_{e3} = (r_{e2})^{-1}, r_{e3} = \sqrt{1+(0.01847/\ln 2)^2} \cong \pm 1.014931928,$$
$$K_{e3} \cong +0.99261662, \delta K_{e3} \cong -0.00738338. \quad (2.49c)$$

Thus, the entropy impulse is curved with three different curvature values (Figure 1b).

These values estimate each impulse' curvature holding the invariant entropies.

The entropies emerge in minimax cutoff of the impulse carrying entropy $S_{ki} = 0.5$ and *a priori* probability $p_{a\pm} = \exp(-0.5) = 0.6015$ after multiple numbers $m_p$ of probing impulses observe this probability.

Since the rectangle impulse, cutting a time correlation, has measure $\mathrm{M} = |1|_M$, the curving impulse, cutting the curving correlation, determines measure

$$r_{iM} = \mathrm{M} \times K_{ei}. \quad (2.50)$$

The rectangle impulse, not cutting time-correlations, possess Euclid's curvature $K_{iM} = 1$.

Accordingly, the impulse with both time and space measure $|M_{io}| = \pi$, which could appear in transitional impulse curvature of cutting part $K_{eo}$, determines correlation measures

$$r_{icM} = \mathrm{M}_{io} \times K_{eo}. \quad (2.50a)$$

At appearance of the impulse with emerging space coordinate, the increment of the curved impulse correlations measure ratio of the measures for the curved correlation to one with only time correlation:

$$r_{icM} / r_{iM} = \pi / |1| K_{ei} / K_{eio}. \quad (2.50b)$$

Counting (2.50b) leads to

$$r_{icM} / r_{iM} \cong 3.08.$$

Relative increment of correlation:

$$\Delta r_{iM} / r_{iM} = (r_{iM} + r_{icM}) / r_{iM} = 1 + r_{icM} / r_{iM} \cong 4$$

in limit:

$$\lim_{\Delta r(\Delta t), \Delta t \to 0} [\Delta r_{iM} / r_{iM}) = \dot{r}_{icM} / r_{iM}$$

brings contribution (1.3.4a) to entropy functional (I.3.4).

Measure $|M_{io}| = |[\tau] \times [l]| = \pi$ satisfies relations

$$[\tau] = \pi / \sqrt{2}, [l] = \sqrt{2} \quad (2.50c)$$

at

$$[\tau]/[l] = \pi / 2.$$

Shortening the cutting time intervals triples density (1.25) of each invariant curving correlation for the minimax impulse, preserving its measure (2.50).

Since any virtual cutting impulse preserves its virtual measure (2.50b), the related virtual time correlation is able to create the space during the entanglement that triple density measures.

For the invariant impulse that compresses the impulse curvature, the probability of both cutting time interval and emerging space coordinate increases.

The impulse measure $\mathrm{M}_{io} = r_{icM} \times (K_{eo})^{-1}$ defines the correlation multiplied on the inverse impulse curvature. But since $\mathrm{M}_{io} = inv = \pi$, the correlation with curvature $r_{icM} = \pi K_{eo}$ follows, with a growing correlation curvature increase, and vice versa.

Growing the impulse density accompanied with the shortening of cutting time intervals increases the space interval for the invariant impulse measure, but it changes the correlation only with the changing curvature.

Since the increasing IPF with growing density accompanies the increasing curvature of rotating impulses, the correlations also grow.

After accumulating energy, these Information curvatures evaluate the impulse Information gravity.

## 5. How the Observation's Cutting Jump Rotates the Microprocess Time and Creates Space Interval

Each observation, processing the interactive impulses, cuts the correlation of random distributions.

The virtual impulse's curved cutting correlations evaluates the entropy measure of the curvature, which with growing probability eventually brings Information-physical curvature to a real impulse.

The curved jump of the cutting correlation rotates the impulse time interval starting the impulse microprocess.

The edge of interval $\tau_k^{-o}$ determines both the jump width-displacement and the curvature forming in the rotation.

The curved time interval $\delta t_{\pm}^{ko} / \tau_k \cong 0.03125$ relative to the impulse time, formed during *the entanglement, turns on beginning a space before the entanglement ends at angle* $\pi / 2$. That illustrates Fig.1a.

Thus, the time and then space intervals emerge in the interacting impulse as a phase interval, whose probabilistic functions of frequencies enclose a fractional probability of the field available for the observation.

The negative curvature of the curved impulse (Figure 1b) attracts an observing positive curvature of an interacting impulse. The attraction in the interacting virtual impulses measures the entropy increment of the interacting curvatures as an analogy of a virtual gravitation.

A real impulse' negative curvature attracts energy from the random field necessary to create Information, which causes gravitational attraction. Hence, the attracting gravitation starts with the creation of space at the entanglement.

We detail it below.

The interactive impulse microprocess rotates in a transitive movement holding transitive action ↑.

This action, starting on angle of rotation |π/4|, initiates entanglement of the conjugated entropies.

The rotation movement, rotating action ↑ on additional angle, approaching |π/4|, conveys action ↓ that settles a transitional impulse, which finalizes the entanglement at angle| π/2|.

The transitional impulse holds temporal actions ↑↓ opposite to the primary impulse ↓↑ which intends to generate the conjugated entanglement, involved, for example in left and rights rotations ($\mp$).

The transitional impulse, interacting with the opposite correlated entanglements $\mp$ , reverses it on $\pm$ .

The interacting movement along the impulse boundary ends with cutting the impulse correlation, which carries the potential erasure, becoming a real with delivering an external energy.

Since the entropy' impulse is virtual, transition action within this impulse ↑↓ is also virtual and its interaction with the forming correlating entanglement is reversible, as well as the space and attractive entropy gravitation.

*Comments* 2.5

The time-interactions are emerging actions of the initial probability field of interacting events at the beginning of the random microprocesses. From this field emerges the first time-correlation and then space coordinates nearby the middle of the impulse, making it possible to deliver the field's energy.

Within the probability field, the emerging initial time has a discrete probability measure, satisfying the Kolmogorov Law •

Thus, the time-interactions hold a discrete sequence of impulses carrying entropy, from which emerges a space in the sequence: interactions-correlations–time-space.

The sequence of the impulses replicates the frequencies of observation, creating a wave function. The Information form of the Schrodinger equation (Sec.1.3) was established in [5] and published in [6].

Let us connect these results with the considered microprocess emerging from observations. That implies applying the emerging between impulses jump to the evolving between the impulses probability spot reaching the probability (2.18), when the wave function emerges. Thereafter, the wave function emerges during the probabilistic impulse observations when the between impulse jump has probability (2.18). Under evolving probabilistic wave function also emerges *Stochastic microprocess.*

Below the wave function's equation, rising in the observing probability evolution, is introduced.

## 6. The Interacting Curvatures of Step-Up and Step-Down Actions, and Memorizing a Bit

Each impulse (Figure 1a) step-down action has negative curvature (2.49, 2.49a) corresponding attraction, step-up reaction has positive curvature (2.49b) corresponding repulsion, the middle part of the impulse having negative curvature transfers the attraction between these parts.

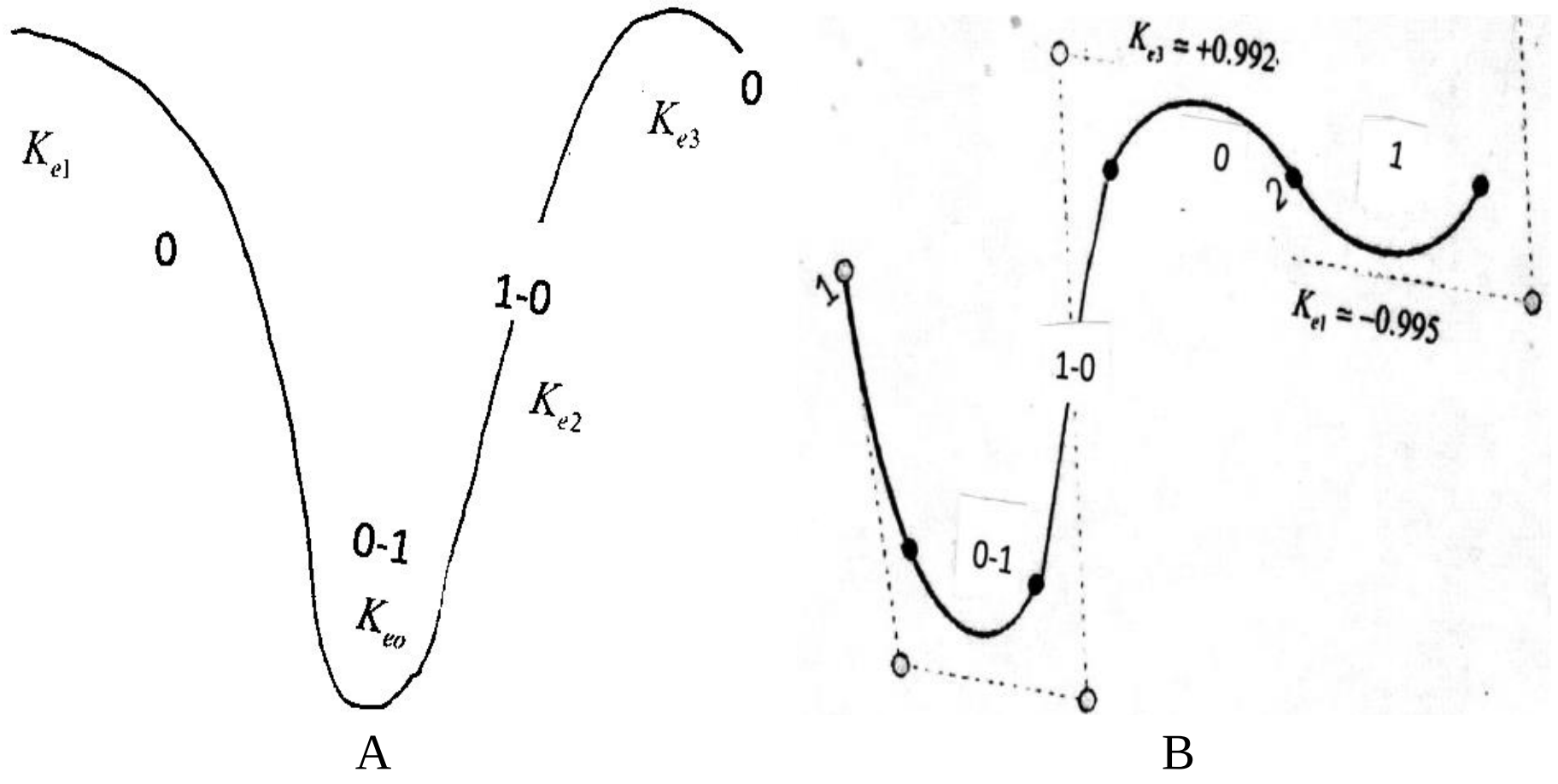

Figure 2. A virtual impulse (Figure 2A) starts step-down action with probability 0 of its potential cutting part; the impulse middle part has a transitional impulse with transitive logical 0-1; the step-up action changes it to 1-0 holding by the end interacting part 0, which, after the inter-active step-down cut, transforms the impulse entropy to Information Bit.

In Figure 2B, the impulse Figure 2A, starting from instance 1 with probability 0, transits at instance 2 during interaction to the interacting impulse with negative curvature $-K_{e1}$ of this impulse step-down action, which is opposite to curvature $+K_{e3}$ of ending the step-up action ( $-K_{e1}$ is analogous to that at beginning the impulse Fig.2A). The opposite curved interaction provides a time–space difference (a barrier) between 0 and 1 actions, necessary for creating the Bit. When the interactive process provides Landauer's energy [7] with maximal probability (certainty) 1, the interactive impulse' step-down action ending state memorizes the Bit. Such certain interaction injects the energy overcoming the transitive gap including the barrier toward creation the Bit.

The step-up action of an external (natural) process' curvature $+K_{e3}$ is equivalent of potential entropy $e_o = 0.01847 Nat$ which carries entropy $\ln 2$ of the impulse total entropy 1 Nat.

The interacting step-down part of internal process impulse' invariant entropy 1 Nat has potential entropy $1 - \ln 2 = e_1$. Actually, this step-down opposite interacting action brings entropy $-0.25 Nat$ with anti-symmetric impact $-0.025 Nat$ which carries the impulse wide $e_w \cong -0.05 Nat$ (Sec. 2.3.5) with total entropy $-0.3 Nat$ that equivalent to $-e_1$.

Thus, during the impulse interaction, the initial energy-entropy $W_o = k_B \theta_o e_o$ changes to $W_1 = -k_B \theta_1 e_1$, since the interacting parts of the impulses have opposite-positive and negative curvatures accordingly; the first one repulses, the second attracts the energies. For erasing the Bit equivalent to the internal impulse minimal entropy $e_{10} = \ln 2$, the needed Landauer energy is $W = k_B \theta \ln 2$.

If the interactive internal process accepts this Bit by memorizing (through erasure), the above Landauer energy should compensate the difference of these energies-entropy: $W_o - W_1 = W_B$ in balance form

$$k_B \theta_o e_o + k_B \theta_1 e_1 = k_B \theta \ln 2. \qquad (5.1)$$

Assuming the interactive process supplies the energy $W_B$ at moment $t_1$ of appearance of the interacting Bit, we get $k_B \theta_1(t_1) = k_B \theta(t_1)$. That brings (4.1) to form

$$k_B \theta_o 0.01847 + k_B \theta (1 - \ln 2) = k_B \theta \ln 2, \theta_o / \theta = (2 \ln 2 - 1) / 0.01847 = 20.91469199. \quad (5.2)$$

The opposite curved interaction decreases the ratio of above temperatures on increment $\ln 2 / 0.0187 - (2 \ln 2 - 1) / 0.01847 = 16.61357983$, with ratio

$$(2 \ln 2 - 1) / \ln 2 \cong 0.5573. \qquad (5.2a)$$

Natural impulse with maximal entropy density $e_{do} = 1 / 0.01847 = 54.14185$ interacting with internal curved impulse transfers minimal entropy density $e_{d1} = \ln 2 / 0.01847 = 37.52827182$.

Ratio of these densities $k_d = e_{do} / e_{d1} = 1.44269041$ equals

$$k_d = 1 / \ln 2 \qquad (5.3)$$

which identifies single impulse 1 measured by $k_d$ Bits-or $1 Nat$.

Hence, that ratio enables the production of the Information impulse $1 Nat = 1.44 bit$.

Here the interacting curvature, enclosing entropy density (4.3), lowers the initial energy and the related temperatures in the above ratio. From that follow

### 6.1. Conditions Creating a Bit in Interacting Curved Impulse

1) The opposite curving impulses in the interactive transition require keeping entropy ratio 1/ln2.

2) The interacting process should possess the Landauer energy by the moment ending the interaction.
3) The interacting impulse should hold invariant measure $M = [1]$ of entropy 1 Nat whose topological metric preserves the opposite curvatures. •

The last follows from the impulse's max-min mini-max law under its stepdown-stepup actions. They generate an invariant [1] Nat's time-space measure topological metric π(1/2circle), preserving opposite curvatures.

Results [8] prove that a physical process which holds the invariant entropy measure for each phase space volume (for example, at minimal phase volume $v_{eo} \cong 1.242$ per a process dimension [4]), the above topological invariant characterizes and satisfies the Second Thermodynamic Law. It also shows that "decreasing entropy and negative entropy production arises in arbitrary coordinates," applied to self-organizing systems.

Energy $W_B$ that delivers the external (natural) process will erase the entropy of both attracting and repulsive movements, covering energy of the both movements, which are ending at the impulse stopping states. The erased impulse total cutoff entropy is memorized as equivalent Information, encoding the impulse Bit in the impulse ending state.

The ending logic of natural step-up action captures its entropy, moving along the action positive curvature, transits to interacting step-down action' negative curvature, and by overcoming entropy-Information gap [4,25] acquires the equal Information that compensates for the movements logical cost.

Thus, the attractive logic of an invariant impulse, converting its entropy to Information within the impulse, performs the function of a *logical Maxwell's Demon* (MD) in the microprocess.

**6.2. Topological Transitivity at the Curving Interactions**

The impulse of the external process holds its $1Nat$ transitive entropy until its ending curved part interacts, creating an Information Bit during the interaction.

Theoretically, when a cutting maximum of entropy reaches a minimum at the end of the impulse, the interaction can occur, converting the entropy to Information by getting energy from the interactive process.

The invariant's topological transitivity has a duplication point (a transitive base) where one dense form changes to its conjugated form during orthogonal transition of hitting time.

During the transition, the invariant holds its measure (Figure 1b) preserving its total energy, while the densities of these energies are changing.

The topological transition separates (on the transitive base) both primary dense form and its conjugate dense form, while this transition turns the conjugated form to orthogonal.

At the transition-turning moment, a jump of the time curvature switches to a space curvature (Fig.1a) with potential rising space waves in a microprocess above. •

As a distinction from the traditional Maxwell's Demon, which uses an energy difference in temperature form [9], this approach reveals a Maxwell's Demon through the naturally created difference of the curvatures.

Forming a transitional impulse with the entangled qubits leads to possibility of memorizing them as a quantum Bit.

That requires first to provide the asymmetry of the entangled qubits, which starts the anti-symmetric impact through the main impulse step-down action $\downarrow$ interacting with opposite action $\uparrow$ of starting the transitional impulse.

This primary anti-symmetric impact $-0.025\times 2=-0.05Nat$ starts curving both the main observing and transitional impulses with curvature $K_{e1}\simeq -0.995037,$ enclosing $0.025Nat$, while the starting step-up action of the transitional impulse generates curvature $K_{e2}\simeq +0.993362$ enclosing $e_o=0.01847Nat$.

Difference $(0.025\text{-}0.01847)Nat$ estimates entropy measuring total asymmetry of main impulse $0.00653Nat=S_{as}$.

Entangled qubits in the transitional impulse evaluate the entropy volume 0.0636 Nat, which spends its entropy on transfer the minimal entangle phase volume $v_{eo}\cong 1.242$ to the entropy-Information gap; while the primary impact brings minimal entropy $0.05Nat$ starting the entangled curved correlation. Thus, the correlated curved entanglement can memorize $(0.05-0.0656)Nat$ in the equivalent Information of two qubits.

The middle part of the main impulse generates curvature $K_{e2}\simeq +0.993362$ which encloses entropy $0.02895Nat$.

The difference $0.02895-0.025=0.00395Nat$ adds asymmetry to the starting transitional entropy, while $0.02895-0.01845=0.0105Nat$ estimates the difference between the final asymmetry of the main impulse and the ended asymmetry of transitional impulse.

With the starting entropy of the curved transitional impulse $0.05Nat$, the ending entropy of the transitional impulse asymmetry is estimated:

$$0.0653\text{-}0.0105\text{-}0.00395\text{=}0.05085Nat.$$

Memorizing this asymmetry requires compensation from a source of equivalent energy. It could be supplied by opposite actions of the transitional step-down $\downarrow$ and main step-up interacting action $\uparrow$ ending the transitional impulse.

That action will create the needed curvature at the end of the main impulse, adding $0.0653-0.05085\text{=}0.01445, 0.01445\text{-}0.0105\text{=}0.00395Nat$ to entropy of transitional impulse curvature sum $0.05085$.

Another part 0.0105 will bring the difference of entropy' curvature $0.02895-0.01845=0.0105$ with total 0.0653.

Thus, $0.05085Nat=s_{as}$ is entropy of asymmetry of entropy volume $s_{ev}=0.0636$ of transitional impulse, whereas $0.0653Nat=S_{as}$ is the entropy of the asymmetry of the main impulse.

This asymmetry generates the same entangled entropy volume that the step-action of the main impulse transfers for interacting with the external impulse.

Thus, $s_{as}$ is the Information "demon cost" for the entangled correlation, which the curvature of the transitional impulse encloses.

The asymmetrical curvature of transitional impulse, holding the entangled volume, encloses the entangled correlation. Instead of direct evaluation of this correlation, allows memorizing the Information of two qubits in impulse measure 1 Nat.

That evaluation is closed to [10], obtained differently, and confirmed experimentally.

When the posteriori probability is closed to reality, the impulse positive curvature of step-up action, interacting with the merging impulse' negative curvatures of step-down action, transits a real interactive energy, which the opposite asymmetrical curvatures actions enfolds.

During curved interaction this primary virtual asymmetry compensates the asymmetrical curvature of a real external impulse, and that real asymmetry is memorized through the erasure by the supplied external Landauer's energy.

The ending action of the external impulse creates a classical Bit with probability

$P_k = \exp-(0.0636^2) = 0.99596321$.

Since the entanglement in the transitional impulse creates an entropy volume $0.0636$, the potential memorized pair of qubits has the same probability.

Therefore, both the memorized classical Bit and pair of qubits occur in a probabilistic process with high probability but less than 1.

The question is how to memorize the entropy enclosed in the correlated entanglement, which naturally holds this entropy and therefore has the same probability?

If a transitional impulse, created during interaction, has such a high probability, then its curvature holds the needed asymmetry, and it should be preserved for multiple encoding with the identified difference of the locations of both entangled qubits.

Information of the memorized qubits can be produced through interaction, which generates the qubits within a material-device (a conductor-transmitter) that preserves the curvature of the transitional impulse in a Black Box, by analogy with [11].

At such an invariant interaction, the multiple connected conductors memorize the qubits' code.

The needed memory of the transitional curved impulse encloses entropy $0.05085Nat$.

### 6.3. The Time Intervals of the Curved Interaction

If the natural space action curves the internal interactive part, the joint interactive time-space curved action measures its interactive impact.

If the interaction at moment $t_o$ creates internal curvature $K_{e1} \simeq -0.995037$ enclosing $-0.025Nat$ by moment $t_{o1} = 0.01845Nat$, the interacting time-space interval measures the difference of these intervals

$$|t_{o1}| - t_o \triangleq 0.0250 - 0.01847 = 0.00653Nat. \quad (5.4)$$

For that case, the internal curved interaction attracts the energy of natural interactive action.

By the moment $t_1$ of appearance of the interacting Bit, ratio (4.3) selects part of the Information impulse $i_{11} \cong (1.44 - \ln 2) \times 0.5573 \cong 0.2452bit$ which the curve interaction deducts from the internal impulse's Bit.

The anti-symmetric interaction involves middle part of the internal impulse with the asymmetry of curvature $K_{e2} \cong +0.993362$ which encloses entropy $0.02895Nat$.

Difference $0.02895 - 0.025 = 0.00395Nat$ adds asymmetry to the starting transitional entropy, while $0.02895 - 0.01845 = 0.0105$ estimates the difference between the final asymmetry of the main impulse and ended asymmetry of transitional impulse.

Taking into account the asymmetry Information $i_{13} \cong 0.0105 \times 1.44 = 0.015bit$, we get Information

$$i_f \cong 0.2452 - 0.015 \cong 0.23bit \quad (4.\ 4a)$$

evaluating the total asymmetrical increment of the curved interaction. This is free Information created in addition to the Bit, which measures the attracting action of the asymmetrical interaction. Its amount simply evaluates ~1/3/Bit.

If the No part of the interacting impulse emerges at $t_o$ and the Yes part arises by $t_1$, then the invariant interacting impulse will spend $1 - \ln 2 + \ln 2 = 1$ Nat creating the Bit ($\ln 2 Nat$).

If interaction of the natural process on the internal process delivers energy $W_B$ by moment $t_1$, this energy will erase the Bit and memorize it according to the balance relations.

The interacting impulse spends ~1 Nat on creating and memorizing Bit $\ln 2$ holding free Information $(1 - \ln 2) \cong 0.3Nat$.

The curved topology of interacting impulses decreases the needed energy ratio, according to the balance relation above.

Thus, the time interval $t_o - t_1$ creates the Bit and performs the *Maxwell's Demon* (MD) function.

Since the movement within the internal impulse ends at the impulse step-up stopping states, the thermodynamic process delivering this energy should stop in that state. Hence, the erased impulse cutoff entropy memorizes the equivalent Information $1.44bit$ in the impulse ending state. It includes $1.44-1.23=0.21$ where $0.21\times 1.44 \cong 0.3Nat$ is transferred to the next interacting impulse as the equivalent to is $-e_1$ .

In the ending Observer's probing logic, such a curving interaction moving along the negative curvature of its last *a priori* step-up action, overcomes the gap by moving along the positive curvature of the *a posteriori* step-down. It acquires Information (2.53a) that compensates for the logical cost of the movement.

Thus, the attractive free Information logics of an invariant impulse, converting its entropy to Information within the impulse, performs the function of a *logical* (MD) in the microprocess. (More details are in [15,4].)

Coordination of an Observer's external time-space scale with its internal time-space scale happens when an external step-down jump action interacts with the Observer's inner thermodynamic time-space interval, which, in the curved interaction, measures the difference of the time (4.4).

Ratio $[\tau]/[l]=\pi/2$ leads to $\Delta l_{10} = 2\Delta t_{10}/\pi, \Delta l_{10} \cong 0.00415Nat$ .

*Thus, the curvature of the rotating impulse encloses its time and space.*

The interacting jump injects energy, capturing the entropy of impulse's ending step-up action. This interaction models the 0-1 Bit. The opposite curved interaction provides a time–space difference (an asymmetrical barrier) between 0 and 1 actions, necessary for creating the Bit. The interactive impulse' step-down ending state memorizes the Bit when the Observer interactive process provides Landauer's energy with maximal probability.

**6.4. Applying the Jarzynski Equality (JE) to observing microprocess**

The Jarzynski Equality of irreversible thermodynamic transition [3] using the results of its experimental verification [12], has form

$$< e^{(\Delta \mathrm{F} - W)/k_B\theta} >= \gamma ,$$

or

$$e^{\Delta \mathrm{F}/k_B\theta} - < e^{W/k_B\theta} >= \gamma, 0 \le \gamma \le 2. \quad (4.5)$$

Here $\Delta F$ is the increment of free energy needed to produce energy $\mathrm{W};\gamma$ is the parameter of the verification, which measures the sum of the probabilities the inverse trajectory observed. At $\gamma=1$, the JE satisfies exactly:

$$e^{\Delta \mathrm{F}/k_B\theta} - < e^{W/k_B\theta} >=1. \quad (4.5a)$$

A thermodynamic process, satisfying the JE for all its states in sequence, evolves irreversibly.

Quantity of Information $I_\delta$ at the curved transtion $\delta_{to}$ we assume compensates the increment of free energy $\Delta F = \Delta F_{\delta t}$ satisfying relation

$$\Delta F_{\delta t} / k_B\theta = I_\delta . \quad (4.6)$$

Average thermodynamic energy $< \mathrm{W} >= W$ , which produces the multiple impulse dissipations (modeling by Markovian diffusion), integrates the EF equivalent entropy.

Since EF counts also curved observing probabilistic impulses, the average energy includes that in the curved impulses.

That allows measuring W by averaging exponential energy $<\exp W / k_b \theta>$ in the JE collected during observation of multiple impulses.

The dissipative energy has high entropy value compared with the considered natural source energy.

Erasing that entropy' energy by natural source, delivering high quality energy, for example $\Delta F = \Delta F_{\delta t}$ in (4.6), brings equal a non-random Information $I_{\delta}$.

The EF entropy $\Delta S_{\delta t}$, collected the average exponential energy during the observing random impulses' time intervals $\delta_t$ gets form

$$<\exp W / k_b \theta >= \exp \Delta S_{\delta t}, \quad (4.6a)$$

The influx of energy $\Delta F = \Delta F_{\delta t}$ at $\delta_{to}$ enables erasing entropy $\Delta S_{\delta t}$ of the curved impulse converting it to Information $I_{\delta t}$.

This entropy covers the observing microprocess.

According to formulas Sec.2, the microprocess evolves during observation of minimum number of $m_p = 5$ impulses on the interval $\delta_t$. After that a single elementary unit of information $I_{\delta t} = [1]$ may appear.

(Averaging the energy during multiple observations of evolving the microprocess and generation of information take different time intervals: $\delta_t$ and $\delta_{to}$ accordingly at $\delta_t > \delta_{to}$.)

Substituting in (4.5), Information $I_{\delta t}$ from (4.6) and entropy (4.6a), the JE acquires form:

$$\exp(I_{\delta t}) - \exp(\Delta S_{\delta t}) = \gamma \quad (4.7)$$

To get information $I_{\delta t} = [1]$ the five random impulses are need, whose entropy averages $\Delta S_{\delta t}$ satisfying the JE for the microprocess.

Writing (4.7) for $\gamma = 1$: $\exp([1]) - 1 = \exp(\Delta S_{\delta t})$ we find $\exp \Delta S_{\delta t} = 1.7$ or $\Delta S_{\delta t} \cong 0.54[1]$, which the influx of energy erases, providing the elementary unit of information $I_{\delta t} = [1]$.

Total entropy, observing during the evolving microprocess: $\Delta S_{\delta} \cong 0.54[1] \times 5 = 2.7[1]$, which $\Delta S_{\delta t}$ averages. The exponential energy $e^{\Delta F_{\delta t} / k_B \theta}$ erases $\Delta S_{\delta t}$ to produce a single unit $I_{\delta t} = [1]$ equivalent to unit of entropy $\Delta S_{\delta to} = [1]$ spent on transition during $\delta_{to}$. Or the microprocess, averaging exponential energy $< e^{W / k_B \theta} >$ produces entropy $\exp \Delta S_{\delta t} = 1.7$ and satisfies the JE balance energy at $e^{\Delta F_{\delta t} / k_B \theta} = 2.7$.

Thereafter, the equivalence of the JE for the evolving microprocess enables generating information $I_{\delta t} = [1]$ requires the MD energy $\Delta F_{\delta t} / k_B \theta = [1]$ to compensate for the unit of Information at the time of transmission of this Information.

Independently, assuming that the evolving microprocess holds relative entropy $S^*_{\mp a} = 2$ plus entropy $\Delta S_{\delta o} = \ln 2 \cong 0.7$ equivalent to the interacting impulse minimal entropy measure $| e_{10} | = [1]$. Then, the irreversible thermodynamic transition applies to microprocess with total entropy $\Delta S_{\delta} = 2.7[1]$ which evolves from five random impulses satisfying the minimax. Therefore the transitional entropy, whose erasures compensates appearance of irreversible Bit $I_{\delta t} = [1]$, equals $2.7/5 = 0.54$. Substituting, we come to JE balance (4.7) at $\gamma = 1$ for $\Delta S_{\delta t} \cong 0.54$.

From that, the JE (4.7) balance at $\gamma = 1$ is satisfied identically for the observing microprocess evolving according to minimax.

It confirms that the impulse minimax extreme principle (EP) satisfies the JE for impulse Information transition, or vice versa.

The EP determines invariant measure of every cutting impulse.

Each cutting impulse time interval enables encoding invariant unit of Information. Or, the EP follows from the JE in the physical process whose interactive time interval is an equivalent of the impulse Information cutting from the correlation carrying the energy.

The cutting correlation's time intervals hold the Information equivalent of this energy, and any real time interval of interaction brings the entropy equivalent of energy $\Delta F_{\delta t}$ which compensates for the MD while producing Information during the interaction. In an interactive random process whose sequence of cuts satisfy the EP, each impulse encodes the cutting correlation, and all Information of the process cutoff correlations encodes the Information process, fulfilling the minimax law which is independent on size of any impulse.

Thus, to satisfy the Maxwell's Demon, the Information produced by each impulse time interval should be invariant, holding constant the unit (Bit, Nat) in $I_{\delta t}$.

Moreover, the sum of probabilities of the inverse trajectories of the interacting impulses in the microprocess is part of the observing process, which exactly satisfies the JE initial conditions [12, 13]. The evolving microprocess starts with probability (Section 2.3.3) and relational entropy of inverse states $S^*_{\mp a} = 2$.

The 0-1 entropy units (potential Bit) of the microprocess impulse connect the impulse inner correlation, while 01-0-1 entropy entities (a potential qubit) bind the microprocess entanglement.

Such multiple microprocesses, which the observation generates, hold the statistical thermodynamic process where the JE automatically measures the energy of these impulse discrete units.

*The JE for the first time was applied in [15] for measuring energy within the impulse microprocess (quantum) connecting the JE with encoding this process' Information measure at the cutting correlation. The random interactions on the path to the generation of Information naturally average the impulse microprocess' dissipative work in the JE thermodynamics.*

*The curved impulse thermodynamics on the rotating microprocess trajectories describe the forming physical micro units encoding qubits, Bits.*

The Information process's last cutting impulse encodes the process's total Information integrated in its IPF.

Thus, the applied JE enables measuring energy of the entropy-Information unit in both statistical thermodynamic microprocess and encoding thermodynamic macroprocess.

This approach is distinct from other JE applications by averaging the work in the JE during the evolving observations naturally, while others need multiple experiments and specific procedures of averaging their results.

### 6.5.Multiple Interactions Generate a Code of the Interacting Process at the Following Conditions

1) Each impulse holds an invariant probability-entropy measure, satisfying the Bit conditions.
2) The impulse interactive process which delivers the code is a part of a real physical process that maintains this invariant entropy-energy measure. That process memorizes the Bit and creates an Information process of multiple encoded Bits. By attracting free Information, they build the process's Information dynamic structure.

For example, water, cooling interacting drops of hot oils in the found ratio of temperatures, enables spending energy of its chemical components to encode the components chemical structures. Or the water kinetic energy will carry the multiple drops' Bits as an arising Information dynamic flow.

Such a physical -chemical process supplies the needed energy to generate the code.

3) Building the multiple Bits code requires increasing the impulse information density three times with each following impulse acting on the interacting process (Section 2.1.).

To create a code of the Bits, each interactive impulse, producing a Bit, should follow three impulses measure π, i.e., frequency of interactive impulse should be f=1/3 π=~0.1061. •

The interval 3π provides the opportunity to join the impulses of three Bits in a triplet as an elementary macro unit. It combats the noise and redundancies from both the internal and external processes.

Natural encoding merges memory with the time of memorization, compensating the cutting cost by running time intervals of encoding.

The encoding process, preserving the invariant cutting Information, connects its multiple Bits or qubits in the invariant irreversible thermodynamics where each such discrete Information unit's energy measures the JE.

Multiplication mass M on curvature $K_{e2}$ of the impulse equals to relative density Nat/ Bit=1.44 which determines M=1.44/ $K_{e2}$ . At $K_{e2}$ =0.993362, we get a relative mass M=1.452335645.

The opposite curved interaction lowers the potential energy, compared to other interactions for generating a Bit.

The multiple curving interactions create topological Bits code, which sequentially forms a moving spiral structure [4].

## 7. FINDING AN INVARIANT ENERGY MEASURE WHICH EACH BIT ENCLOSES, STARTING MAXWELL'S DEMON

Since its minimal energy is $W = k_B \theta \ln 2$ , it's possible to find such temperature $\theta_1^o$ that is equal to inverse value of $k_B$. If the interacting process carries this temperature, then its minimal energy holds

$$W_1^o = \ln 2 \text{ at } \theta_1^o = 1/k_B ,$$

which becomes equal to the Bits' time-space Nat measure' entropy invariant.

Let us evaluate $\theta_1^o$ at $k_B = 8617 \times 10^{-5} \text{eV}/\text{K}$ at Kelvin temperature $K = 20/293 = 0.0682259386^{oC/K}$ equivalent to $20^{oC}$ .

Then $\theta_1^o = 588.19 \times 10^5 / \text{eV}$ .

If we assume that this primary natural energy brings eV amount equivalent to quanta of light: $e_q = 1240\text{eVnm}$, $1\text{nm}=10^{-9} m$ , then we get

$$\theta_1^o = 588.19 \times 10^5 \times 1.240 \times 10^3 / e_q \times 10^{-9} m \cong 72.9356^{oC/m} / e_q.$$

Or each quant brings temperature' density $\theta_1^o = 72.9356^{oC/m}$ , which is reasonably real.

With this $\theta_o^o$, the interacting impulse will bring energy $W_1^o = \ln 2$ to create its Bit.

Following the balance relation, the external process at this $\theta_o^o$ holds temperature $\theta_o^o = 20.914691 99\theta_1^o = 1525,42^{oC/m}$ brought by a quant.

This temperature energy holds an invariant impulse measure $|1|_M = 1Nat$ with metric π, or each such impulse has entropy density $1Nat/\pi$. At temperature $\theta_1^o$, the interacting impulse' Bit has minimal density energy equivalent to $\ln 2/\pi = 0.22$.

In cognitive dynamics [16], it allows spending energy ln2 for erasure of the observing Bit and memorizes the equivalent cognitive quantity equal to Landauer's Bit by the neuron Information Bits.

*With such energy, the Information attraction-gravitation imitates* $0.23bit$ *enables* attracting actions. Therefore, the curving interaction dynamically encodes Bits in a *natural process* of the interacting Information process.

The growing curvature of the impulses during rotation increases the density of the Bit.

*The rotating thermodynamic process with minimal Landauer energy performs the natural memorizing of each natural Bit.*

The invariant Information-cutting process holds the invariant irreversible thermodynamics measured through piece-wise Hamiltonian and diffusion-kinetic matrix equations [17]:

$$I_f = L_t X_t, L_t = 2b_t$$

where $I_f$ is diffusion-kinetic flows, $X_t$ is thermodynamic forces, $b_t$ and is diffusion matrix.

At $L_t \geq 2b_t$ kinetic flow transfers to diffusion $2b_t$ at $L_t \leq 2b_t$ the diffusion flow transfers to kinetics, where the transformation applies on a small $\varepsilon$-localities of the bordered impulse.

At these conditions, the Hamiltonian includes increments of chemical potentials of interacting physical-chemical entities.

Hence, the JE with both connections to irreversible thermodynamic and kinetics describes increments of temperature, entropy, energy, diffusion, and physical-chemical components in the variety of thermodynamic processes within interacting impulses and their cooperative macrodynamics.

The Information Macrodynamics describes all these through the equivalent Information parameters.

**REFERENCES TO PART II**

**REFERENCES TO PRINCIPLES OF OBSERVATION AND OBSERVER**

[1] Bohr N. *Atomic physics and human knowledge*, Wiley, New York, 1958.

[2] Dirac P. A. M. *The Principles of Quantum Mechanics*, Oxford University Press (Clarendon), London/New York, 1947.

[3] Von Neumann J. *Mathematical foundations of quantum theory*, Princeton University Press, Princeton, 1955.

[4] Wigner E. Review of the quantum mechanical measurement problem. In *Quantum Optics, Experimental Gravity, and Measurement Theory*. NATO AS1 Series: Physics, Series B, **94,** 58, Eds. P. Meystre & M. 0. Scully, 1983.

[5] Wigner E. The unreasonable effectiveness of mathematics in the natural sciences, *Communications in Pure and Applied Mathematics*, **13**(1), 1960.

[6] Wheeler J. A. and Feynman R. P. Interaction with the absorber as the mechanism of radiation. Reviews of Modern Physics, **17**(2-3):157, 1945.

[7] Wheeler J. A. and Feynman R. P. Classical electrodynamics in terms of direct interparticle action. *Reviews of Modern Physics*, **21**(3):425, 1949.

[8] Bohm D. J. “Information and Meaning,” in The Search for Meaning, Paavo Pylkkanen (ed.), Crucible, 1989; Bohm D. J. A new theory of the relationship of mind to matter. *Journal Am. Soc. Psychic. Res*. **80**, 113–135, 1986.

[9] Eccles J. C. Do mental events cause neural events analogously to the probability fields of quantum mechanics? *Proceedings of the Royal Society*, B**277**: 411–428, 1986.

[10] Wheeler J. A. On recognizing “law without law.” *Am. J. Phys.,* **51**(5), 398–404, 1983.

[11] Wheeler J. A., Zurek W. Editor. Information, physics, quantum: *The search for links, Complexity, Entropy, and the Physics of Information*, Redwood, California, Wesley, 1990.

[12] Wheeler J. A. The Computer and the Universe, *International Journal of Theoretical Physics,* **21**(6/7): 557-572, 1982.

[13] Wheeler J. A. and Ford K. It from bit. In Geons, *Black Holes & Quantum Foam: A life in Physics,* New York, Norton, 1998.

[14] Wheeler J. A. Include the Observer in the Wave Function?. In: Lopes J. L., Paty M. (eds.) *Quantum Mechanics, A Half Century Later.* Episteme, vol 5. Springer, Dordrecht, 1977.

[15] Einstein A., Podolsky B., and N. Rosen N. Can Quantum-Mechanical Description of Physical Reality be Considered Complete? *Phys. Rev*. **47**(10), 777-780. 1935.

[16] Penrose’s R. *Fashion, Faith and Fantasy in the New Physics of the Universe,* Princeton University Press, 2016.

[17] Weinberg S. *The Trouble with Quantum Mechanics,* The New York Review of Books, January 19, 2017.

[18] Tong D. *Quantum Fields: The real building blocks of the Universe,* The Royal Institution, Cambridge, 2017.

[19] Misner C. W., Thorne K. S., and Zurec W. H. John Wheeler, relativity, and quantum information, *Physics Today,* 2009.

[20] Kolmogorov A. N. Foundations of the Theory of Probability, Chelsea, New York, 1956. See also Kolmogorov in Perspective, *American Mathematical Society*, 2006.

[21] Kolmogorov A. N. Logical basis for information theory and probability theory, *IEEE Trans. Inform. Theory,* **14** (5): 662–664, 1968.

[22] Shannon C. E. A Mathematical theory of communication, *The Bell System Technical Journal,* **27**: 379–423, 623–656, 1948.

[23] Kullback S. *Information theory and statistics*, Wiley, New York, 1959.

[24] Jaynes E. T. *Information Theory and Statistical Mechanics in Statistical Physics,* Benjamin, New York, 1963.

[25] Jaynes E. T. *How Does the Brain do Plausible Reasoning?* Stanford University, 1998.

[26] Landauer R. Irreversibility and heat generation in the computing process, *IBM Journal Research and Development,* **5**(3):183–191, 1961.

[27] Lerner V. S. Optimal control of superimposing macroprocesses on the basis of a physical approach, *Radiophysics,* **25**(11):1608-1626,1972,link.springer.com/article/10.1007/BF01031152.

[28] Lerner V. S. Macrosystemic Approach to Solution of Control Problems under Condition of Indeterminacy, trans. by Scripta Technical, Inc. 1989, from *Journal Automatics,* **5:**43-52 Kiev, 1988.

[29] Lerner V. S. Mathematical Foundation of Information Macrodynamics, *J. Systems Analysis-Modeling-Simulation,* **26**:119-184,1996.

[30] Lerner V. S. Information Systems Theory and Informational Macrodynamics: Review of the Main Results, *IEEE Transactions on systems, man, and cybernetics—Part C: Applications and reviews,* **37** (6):1050-1066, 2007.

[31] Lerner V. S. *Information Path Functional and Informational Macrodynamics,* Nova Science, New York, 2010.

[32] Lerner V. S. An observer's information dynamics: Acquisition of information and the origin of the cognitive dynamics, *J. Information Sciences*, **184**: 111-139, 2012.
[33] Lerner V. S. The boundary value problem and the Jensen inequality for an entropy functional of a Markov diffusion process, *Journal of Mathematical Analysis and Applications*, **353** (1): 154–160, 2009.
[34] Lerner V. S., Solution to the variation problem for information path functional of a controlled random process functional, *Journal of Mathematical Analysis and Applications*, **334**: 441-466, 2007.
[35] Lerner V. S. The Impulse Interactive Cuts of Entropy Functional Measure on Trajectories of Markov Diffusion Process, Integrating in Information Path Functional, Encoding and Application, British Journal of Mathematics & Computer *Science*, **20**(3): 1-35, 2017.
[36] Lerner V. S. The impulse observations of random process generate information binding reversible micro and irreversible macro processes in Observer: regularities, limitations, and conditions of self-creation, arXiv: 1204.5513.
[37] Lerner V. S. Arising information regularities in an observer *arXiv: 1307.0449*.
[38] Jeremy Norman's History of Information, http://www.historyofinformation. com/index.php on December 19, 2018.
[39] Perkovac M. Maxwell's Equations as the Basis for Model of Atoms. *Journal of Applied Mathematics and Physics*, **2**, 235-251, 2014.
[40] Lerner V. S. Information Path from Randomness and Uncertainty to Information, Thermodynamics, and Intelligence of Observer, *arXiv:1401.7041*.
[41] Chu Shu-Yuan. Time-Symmetric Approach to Gravity, *arXiv:gr-qc/98020v1*, 1998.
[42] Wolchover N. How Space and Time Could Be a Quantum Error-Correcting Code, *Quanta Magazine*, January, 2019.
[43] Lerner V. S. Natural Encoding of Information through Interacting Impulses, *arXiv*: 1701.04863, *IEEE Xplore*: http://ieeexplore.ieee.org/xpl/Issue7802033,p.103-115,2016.

## REFERENCES TO SECTIONS 2-7

[1] Le Jan, Yves. Markov paths, loops and fields, *Lecture Notes in Mathematics*, vol. 2026, Springer, Heidelberg, 2011, Lectures from the *38th Probability Summer School* held in Saint-Flour, 2008.
[2] Axencott R. Behavior of diffusion semi-groups at infinity, *Bulletin de la S.M.F.*, tome **102**:193-240, 1974.
[3] Jarzynski C. Nonequilibrium Equality for Free Energy Differences, *Phys. Rev. Lett*., **78**, 2690. 1997.
[4] Lerner V. S. Information Path from Randomness and Uncertainty to Information, Thermodynamics, and Intelligence of Observer, *arXiv*:1401.7041.
[5] Lerner V. S. Hidden stochastic, quantum and dynamic Information of Markov diffusion process and its evaluation by an entropy integral measure under the impulse control's actions, applied to Information observer, *arXiv:*1207.3091.
[6] Lerner V. S. *The Information Hidden in Markov Diffusion*, Lambert Academic Publisher, 2017.
[7] Landauer R. Irreversibility and heat generation in the computing process, *IBM Journal Research and Development*, **5**(3):183–191, 1961.
[8] Sato N. and Yoshida Z. Up-Hill Diffusion Creating Density Gradient-What is the Proper Entropy? *arXiv*:1603.04551.
[9] Bennett C. H. Demons, Engines and the Second Law, *Scientific American*, 108-116, 1987.
[10] Palsson M. et al. Experimentally modeling stochastic process with less memory by use of a quantum processor, *Sci. Adv.* e1601302, **3**, 2017.
[11] Acin A., Quantum Information Theory with Black Boxes, *The Zurich Physics Collloquium*, 2015 https://www.video.ethz.ch/speakers/zurich_physics_colloquium/ 23db0d4b-ef08-4644-8c8b-7fe09adbc632.html;

[12] Weber, S. J. et al. Mapping the optimal route between two quantum states, *arXiv*:1403.4992, 2014.

[13] Bérut A., Arakelyan A, Petrosyan A., Ciliberto S., Dilenshneider R., Lutz E. Experimental verification of Landauer's principle linking Information and thermodynamics, *Nature*, **484**, 187-189, 2012.

[14] Toyabe S., Sagawa T., Ueda M., Muneyuki E. and Sano M. Experimental demonstration of Information-to-energy conversion and validation of the generalized Jarzynski Equality, *Nature Physics*, **6**: 988-991, 2010.

[15] Lerner V. S. Natural Encoding of Information through Interacting Impulses, *arXiv: 1701.04863.*

[16] Lerner V. S. *How Information creates its Observer. The Emergence of the Information Observer with Regularities*, Nova Science, 2019.

[17] Lerner V. S. About the biophysical conditions initiating cooperative complexity, Letter to the Editor, *Journal of Biological Systems,* **14(**2):315-332, 2006.